\documentclass[letterpaper,preprintnumbers,english,floats,floatfix,showpacs,amssymb,prd,twocolumn,nofootinbib]{revtex4}
\usepackage{amssymb,amsmath}
\usepackage{epsfig,hyperref}
\usepackage[svgnames]{xcolor}
\usepackage{pgf,tikz}
\usepackage{epstopdf}
\usepackage[british]{babel}
\usepackage{enumerate}

\makeatletter
\DeclareRobustCommand*{\bfseries}{%
  \not@math@alphabet\bfseries\mathbf
  \fontseries\bfdefault\selectfont
  \boldmath
}
\makeatother

\def\be{\begin{equation}}
\def\ee{\end{equation}}
\def\beq{\begin{eqnarray}}
\def\eeq{\end{eqnarray}}

\makeatletter

\newcommand{\arXiv}[2][]{\href{http://arxiv.org/abs/#2}{\texttt{arXiv:#2\@ifempty{#1}{}{ [#1]}}}}
\makeatother

\begin{document}
\title{Interior dynamics of neutral and charged black holes}

\author{Jun-Qi Guo}
\email{junqi.guo@tifr.res.in}
\author{Pankaj S. Joshi}%
\email{psj@tifr.res.in}
\affiliation{Department of Astronomy and Astrophysics, Tata Institute of Fundamental Research, Homi Bhabha Road, Mumbai 400005, India}

\date{\today}

\begin{abstract}
  In this paper, we explore the interior dynamics of neutral and charged black holes. Scalar collapses in flat, Schwarzschild, and Reissner-Nordstr\"{o}m geometries are simulated. We examine the dynamics in the vicinities of the central singularity of a Schwarzschild black hole and of the inner horizon of a Reissner-Nordstr\"{o}m black hole. In simulating scalar collapses in Schwarzschild and Reissner-Nordstr\"{o}m geometries, Kruskal and Kruskal-like coordinates are used, respectively, with the presence of a scalar field being taken into account. It is found that, besides near the inner horizons of Reissner-Nordstr\"{o}m and Kerr black holes, mass inflation also takes place near the central singularity in neutral scalar collapse. Approximate analytic expressions for dif{}ferent types of mass inflation are partially obtained via a close interplay between numerical and analytical approaches and an examination of the connections between Schwarzschild black holes, Reissner-Nordstr\"{o}m black holes, neutral collapse, and charge scattering. We argue that the mass inflations near the central singularity and the inner horizon are related to the localness of the dynamics in strong gravity regions. This is in accord with the Belinskii, Khalatnikov, and Lifshitz conjecture.
\end{abstract}
\maketitle

\section{Introduction\label{sec:introduction}}
The internal structure of black holes and spacetime singularities have been intriguing and basic topics in gravitation and cosmology, theoretical and realistic~\cite{Burko_1997_book,Brady_1999,Berger_2002,Joshi_2007,Henneaux_2008}. Exploring the internal structure of black holes and spacetime singularities inside black holes can lead to a better understanding on black hole physics, gravitation, and quantum physics. Because of similarities between the singularities inside black holes and the singularity in the early Universe, this study may also shed light on cosmology.

It is widely believed that in reality, gravitational collapses may produce rotating black holes. Price's theorem states that gravitational radiation, produced on the surface of a collapsing star, carries away all the initial features of the star's gravitational field, except the mass, charge, and angular momentum parameters~\cite{Price}. As a next step, there is naturally the question of what the final state of the internal collapses might be. A simpler version of the question is how, in reality, inside a rotating black hole, particles from the accretion disk can af{}fect the internal geometry of the black hole.

On the spacelike singularity curve inside a Schwarzschild black hole, the Kretschmann scalar curvature diverges. Then the maximal globally hyperbolic region defined by initial data is inextendible. However, inside charged (Reissner-Nordstr\"{o}m) and rotating (Kerr) black holes, the central singularity is timelike. The globally hyperbolic region is up to the Cauchy horizon, and the spacetime is extendible beyond this horizon to a larger manifold. The Reissner-Nordstr\"{o}m inner (Cauchy) horizon is a surface of infinite blueshift, which in turn may cause the inner horizon unstable~\cite{Simpson_1973}. This motivated the strong cosmic censorship conjecture, which states that for generic asymptotically flat initial data, the maximal Cauchy development is future inextendible. For mathematical explorations of the internal structures of charge black holes, see Refs.~\cite{Dafermos_2003,Dafermos_2014}. For reviews on the Cauchy problem in general relativity and strong cosmic censorship, see Refs.~\cite{Ringstrom_2015} and \cite{Isenberg_2015}, respectively.

As the singularities are approached, the tidal force diverges, and classical general relativity does not apply. It is nevertheless important to explore the dynamics in extremely strong-gravity regions in classical general relativity. Belinskii, Khalatnikov, and Lifshitz (BKL) investigated the asymptotic behaviors in the vicinity of a spacelike singularity and found that the generic solution has chaotic oscillations of the Kasner axes as the singularity is approached~\cite{Belinskii_1970,Belinskii_1973,Belinski_1404}. When a massless scalar field is present, the oscillations will be destroyed, and {\lq\lq}the collapse is described by monotonic (but anisotropic) contraction of space along all directions{\rq\rq}~\cite{Belinskii_1970,Belinskii_1973,Belinski_1404}. Moreover, in the vicinity of the singularity, {\lq\lq}the variation of the gravitational field from one location to the next can be neglected {---} what is much more important is the way gravity changes over time\rq\rq~\cite{einstein_online}. The BKL conjecture was verified numerically in the singularity formation in a closed cosmology in Refs.~\cite{Berger,Garfinkel_1}, and in the dynamics of a test scalar field approaching the singularity of a black hole in Ref.~\cite{Garfinkel_2}. In Ref.~\cite{Ashtekar}, the BKL conjecture in the Hamiltonian framework was examined under the concern of loop quantum gravity. In Ref.~\cite{Guo_1312}, the BKL conjecture was tested in spherical scalar collapse in dark energy $f(R)$ gravity in the Einstein frame. Note that in the Einstein frame, in the vicinity of the central singularity, the scalar degree of freedom $\phi[\equiv(\sqrt{3/2}\ln{f'})/\sqrt{8\pi G}]$ dominates the physical scalar field and the potential term for $\phi$, where $f'{\equiv}df/dR$ and $R$ is the Ricci scalar. Therefore, the dynamics in this case is essentially the same as that in general relativity, and it is expected that the BKL conjecture is also valid in spherical scalar collapse in general relativity. (In fact, this is also confirmed in the current paper.) Compared to such ef{}forts that have been made on verifying this conjecture near spacelike singularities, little work has been done on examining whether the BKL conjecture is valid in the vicinities of the inner horizons of charged and rotating black holes.

\subsection{Mass inflation}
The backreaction of the radiative tail from a gravitational collapse on the inner horizon of a Reissner-Nordstr\"{o}m black hole was investigated by Poisson and Israel~\cite{Poisson_1989,Poisson_1990}. It was shown that due to the divergence of the tail's energy density occurring on the inner horizon, the ef{}fective internal gravitational-mass parameter becomes unbounded. This phenomenon is usually called mass inflation. These arguments were extended to the rotating black hole case in Ref.~\cite{Barrabes_1990}.

In Refs.~\cite{Poisson_1989,Poisson_1990}, approximate analytic expressions were obtained by considering a simplified model in which the perturbations were modeled by cross-flowing radial streams of infalling and outgoing lightlike particles. To get more information, some numerical simulations in more realistic models have been performed. The dynamics of a spherical, charged black hole perturbed nonlinearly by a self-gravitating massless scalar field was numerically studied in Refs.~\cite{Gnedin_1991,Gnedin_1993,Brady_1995,Burko_1997,Burko_1997b,Hansen_2005}. Under the influence of the scalar field, the inner horizon of a charged black hole contracts to zero volume, and the central singularity is converted from timelike into spacelike. The mass inflation phenomenon was observed. In Refs.~\cite{Hod_1997,Oren_2003}, with regular initial data, spherical collapse of a charged scalar field was simulated. An apparent horizon was formed. A null, weak mass-inflation singularity along the Cauchy horizon and a final, spacelike, central singularity were obtained. The same process was investigated rigorously in Ref.~\cite{Kommemi_2011}. In addition, gravitational collapses in some modified gravity theories have been studied numerically. Spherical scalar collapse in $f(R)$ gravity was simulated in Ref.~\cite{Guo_1312}. Asymptotic analysis was implemented in the vicinity of the singularity of a formed black hole. Spherical collapse of a neutral scalar field in a given spherical, charged black hole in Brans-Dicke theory was investigated in Ref.~\cite{Avelino_2009}. Spherical collapses of a charged scalar field in dilaton gravity and $f(R)$ gravity were explored in Refs.~\cite{Borkowska_2011} and~\cite{Hwang}, respectively.

In numerical relativity, it is important to connect approximate analytic candidate expressions with numerical results. In Refs.~\cite{Burko_1998,Burko_1999}, the features of the Cauchy horizon singularity in charge scattering were studied. Analytic and numerical results were compared at some steps.

Despite many ef{}forts that have been spent on the mass function near the inner horizons of charged and rotating black holes, little work has been done in an even simpler case: mass function in the vicinity of the central singularity of a Schwarzschild black hole. So far this has remained an unexplored area of work.

In this paper, we use the following notations:
\begin{enumerate}[(i)]
  \item Neutral collapse: neutral scalar collapse toward a black hole formation.
  \item Neutral scattering: neutral scalar collapse in a (neutral) Schwarzschild geometry.
  \item Charge scattering: neutral scalar collapse in a (charged) Reissner-Nordstr\"{o}m geometry. In this process, the scalar field is scattered by the inner horizon of a Reissner-Nordstr\"{o}m black hole.
\end{enumerate}

\subsection{New results}
In this paper, we explore neutral collapse, neutral scattering, dynamics in Schwarzschild and Reissner-Nordstr\"{o}m geometries, and charge scattering. The connections between such processes will also be examined.

We simulate neutral collapse and investigate the asymptotic dynamics in the vicinity of the central singularity of the formed black hole via mesh refinement. Approximate analytic solutions near the central singularity are obtained. We find that, because of the backreaction of the scalar field on the geometry, mass inflation also takes place near the central singularity. In neutral scattering, similar results are obtained.

We explore the dynamics in a Reissner-Nordstr\"{o}m geometry and charge scattering. By comparing these two processes, we investigate the causes of mass inflation. We seek further approximate analytic solutions with the following improvements. Usually, double-null coordinates are used in studies of mass inflation in spherical symmetry. In the line element of double-null coordinates, the two null coordinates $u$ and $v$ are present in the form of product $dudv$. In the equations of motion, mixed derivatives of $u$ and $v$ are present quite often. In this paper, we use a slightly modified line element, in which one coordinate is timelike and the rest are spacelike. In this case, in the equations of motion, spatial and temporal derivatives are usually separated. This simplifies the numerical formalism and helps to obtain approximate analytic solutions. In addition, we compare numerical results and approximate analytic solutions closely at each step. We compare the dynamics for Schwarzschild black holes, Reissner-Nordstr\"{o}m black holes, neutral collapse, neutral scattering, and charge scattering. With the enlightenment of high-resolution numerical results, we treat the system as a mathematical dynamical system rather than a physical one, examining the contributions from all the terms in the equations of motion.

According to the strength of the scalar field, charge scattering can be classified into five types as follows:
\begin{enumerate}[(i)]
  \item Type I: spacelike scattering. When the scalar field is very strong, the inner horizon can contract to zero volume rapidly, and the central singularity becomes spacelike. The dynamics near the central singularity is similar to that in neutral collapse.
  \item Type II: null scattering. When the scalar field is intermediate, the inner horizon can contract to a place close to the center or reach the center. In this type of mass inflation, for each quantity, the spatial and temporal derivatives are almost equal. In the case of the center being reached, the center is null. This type has two stages: early/slow and late/fast. In the early stage, the inner horizon contracts slowly, and the scalar field also varies slowly. In the late stage, the inner horizon contracts quickly, and the dynamics is similar to that in the spacelike scattering case.
  \item Type III: critical scattering. This case is on the edge between the above two cases. The central singularity becomes null.
  \item Type IV: weak scattering. When the scalar field is very weak, the inner horizon contracts but not much.
  \item Type V: tiny scalar scattering. When the scalar field is very tiny, the influence of the scalar field on the internal geometry is negligible.
\end{enumerate}
In this paper, we will explore the dynamics of the first four types of mass inflation, and obtain approximate analytic solutions for the first two.

Some similarities between neutral and charged mass inflations are obtained. The gravitational/electric field(s) is(are) strong in the vicinities of the central spacelike singularity of a Schwarzschild black hole and the inner horizon of a Reissner-Nordstr\"{o}m black hole. Therefore, as verified by numerical results, the BKL conjecture applies at both cases. It is found that the BKL conjecture can interpret well how the mass inflation happens. The dynamics at strong gravity regions is local. Then at such regions, the Misner-Sharp mass function does not provide global information on the black holes.

We explore scalar collapses in both general relativity and $f(R)$ gravity. The results are similar. For simplicity, we focus on scalar collapses in general relativity in this paper, and present the results in $f(R)$ gravity in a separate paper~\cite{Guo_1508}.

This paper is organized as follows. In Sec.~\ref{sec:framework}, we build the framework for charge scattering, including action for charge scattering and the coordinate system. In Sec.~\ref{sec:set_up}, we set up the numerical formalism for charge scattering. In Sec.~\ref{sec:neutral_collapse}, neutral collapse will be studied. In Sec.~\ref{sec:neutral_scattering}, we discuss neutral scalar scattering in a Schwarzschild geometry. In Sec.~\ref{sec:dynamics_RN}, the dynamics in a Reissner-Nordstr\"{o}m geometry will be examined. In Sec.~\ref{sec:results_scattering}, we explore charge scattering, obtaining approximate analytic solutions. In Sec.~\ref{sec:weak_scattering}, we consider charge scattering with a weak scalar field. In Sec.~\ref{sec:summary}, the results will be summarized.

In this paper, we set $G=c=4\pi\epsilon_0=1$.

\section{Framework for charge scattering\label{sec:framework}}
In this section, we build the framework for charge scattering, in which a self-gravitating massless scalar field collapses in a Reissner-Nordstr\"{o}m geometry. For comparison and verification considerations, we use Kruskal-like coordinates, and set up the initial conditions by modifying those in a Reissner-Nordstr\"{o}m geometry with a physical scalar field.

\subsection{Action}
The action for the charge scattering system in general relativity is
\be S=\int d^{4}x \sqrt{-g}\left(\frac{R}{16\pi G} + \mathcal{L}_{\psi} + \mathcal{L}_{F}\right), \label{GR_action} \ee
with
\begin{align}
\mathcal{L}_{\psi}&=-\frac{1}{2}g^{\alpha\beta}\psi_{,\alpha}\psi_{,\beta},\\
\nonumber\\
\mathcal{L}_{F}&=-\frac{F_{\mu\nu}F^{\mu\nu}}{4}.
\end{align}
$R/(16{\pi}G)$, $\mathcal{L}_{\psi}$, and $\mathcal{L}_{F}$ are the Lagrange densities for gravity, a physical scalar field $\psi$, and the electric field for a Reissner-Nordstr\"{o}m black hole, respectively. $R$ is the Ricci scalar, and $G$ is the Newtonian gravitational constant. $F_{\mu\nu}$ is the electromagnetic-field tensor for the electric field of a Reissner-Nordstr\"{o}m black hole.

The energy-momentum tensor for a massless scalar field $\psi$ is
\be T^{(\psi)}_{\mu\nu}
\equiv-\frac{2}{\sqrt{|g|}}\frac{\delta(\sqrt{|g|}\mathcal{L}_{\psi})}{\delta g^{\mu\nu}}
=\psi_{,\mu}\psi_{,\nu}-\frac{1}{2}g_{\mu\nu}g^{\alpha\beta}\psi_{,\alpha}\psi_{,\beta}.\ee
The dynamics for the scalar field $\psi$ is described by the Klein-Gordon equation,
\be
\Box\psi=0.
\label{Box_psi}
\ee

The electric field of a Reissner-Nordstr\"{o}m black hole can be treated as a static electric field of a point charge of strength $q$ sitting at the origin $r=0$. In the Reissner-Nordstr\"{o}m metric~(\ref{RN_metric}), the only nonvanishing components of $F_{\mu\nu}$ are $F_{tr}=-F_{tr}=-q/r^2$. The corresponding energy-momentum tensor for the electric field is~\cite{Poisson_2004}
\be
\begin{split}
{T^{(F)}}^{\mu}_{\nu}
&\equiv-\frac{2}{\sqrt{|g|}}\frac{\delta(\sqrt{|g|}\mathcal{L}_{F})}{\delta{g_{\mu}^{\nu}}}\\
&=\frac{1}{4\pi}\left(F^{\mu\rho}F_{\nu\rho}-\frac{1}{4}\delta^{\mu}_{\nu}F^{\alpha\beta}F_{\alpha\beta}\right)\\
&=\frac{q^2}{8\pi r^4}\mbox{diag}(-1,-1,1,1).
\end{split}
\label{em_tensor}
\ee
Although Eq.~(\ref{em_tensor}) is obtained in the Reissner-Nordstr\"{o}m metric, it is valid in any coordinate system, since as seen by static observers, the electromagnetic field should be purely electric and radial~\cite{Poisson_1990,Poisson_2004}. We denote the total energy-momentum tensor for the source fields as
\be T^{\scriptsize{(\mbox{total})}}_{\mu\nu}=T^{(\psi)}_{\mu\nu}+T^{(F)}_{\mu\nu}.\ee
\begin{figure}
  \epsfig{file=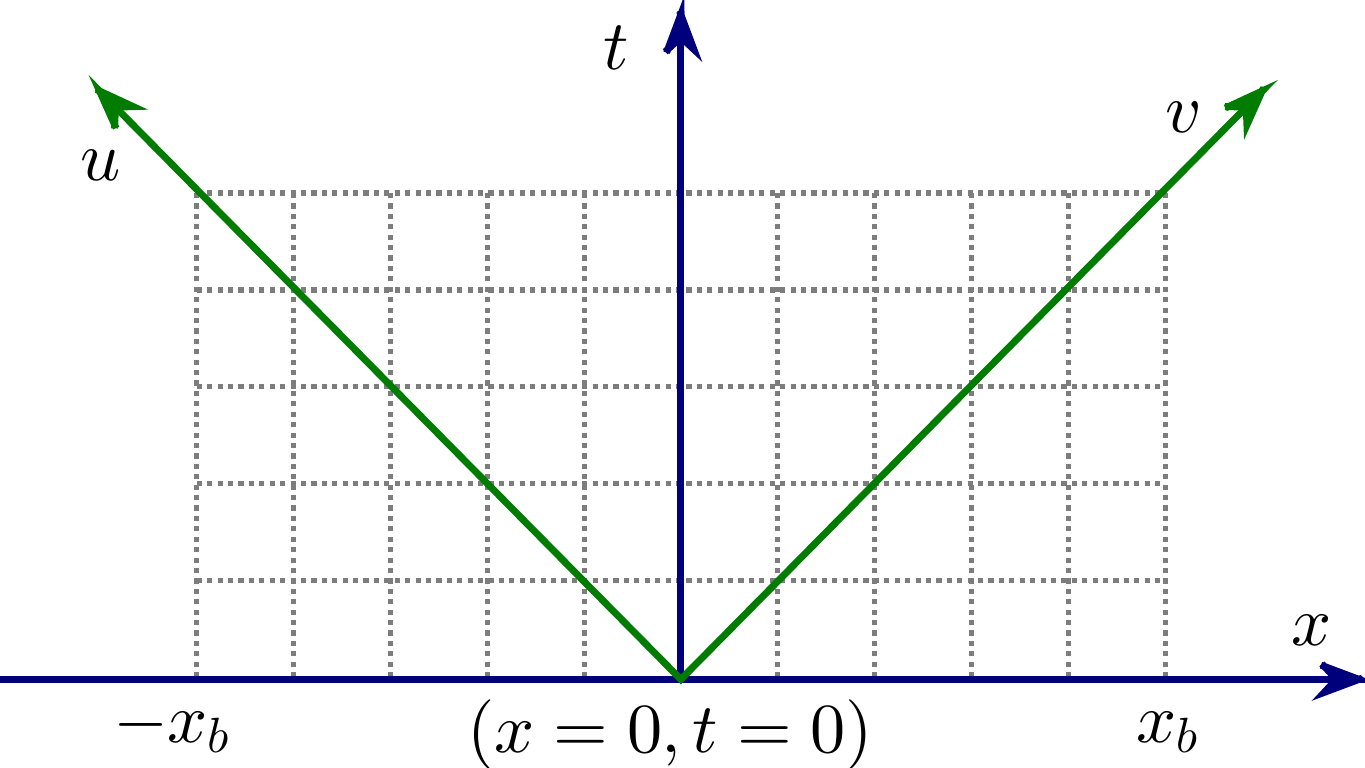, height=4cm}
  \caption{Initial and boundary conditions for charge scattering. Initial slice is at $t=0$. Definition domain for $x$ is $[-x_b~x_b]$.
  $u=(t-x)/2$ and $v=(t+x)/2$.}
  \label{fig:ic_and_bc}
\end{figure}

\begin{figure}
  \epsfig{file=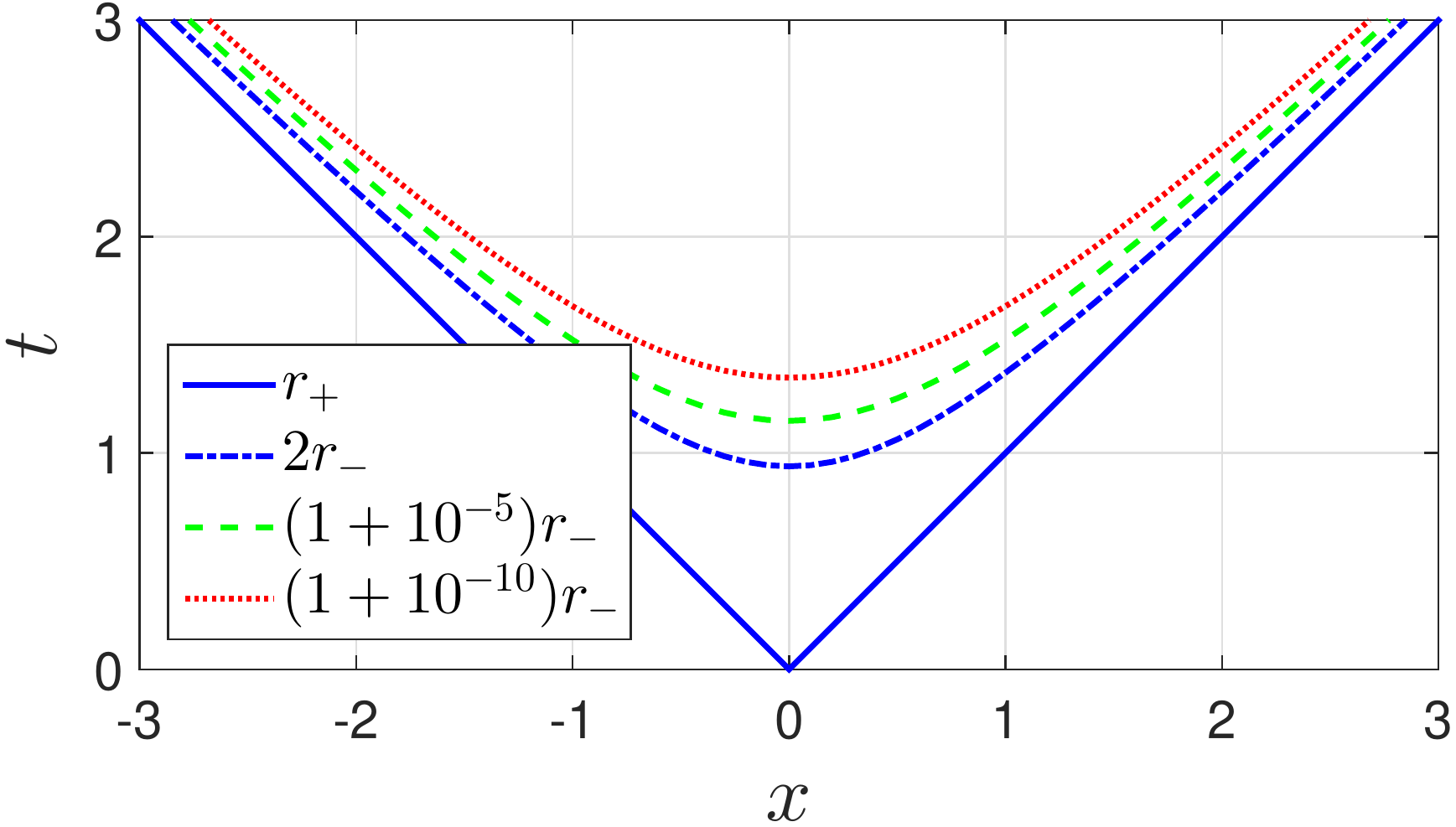, height=4cm}
  \caption{Contour lines for $r$ defined by Eq.~(\ref{r_RN_metric}) in a Reissner-Nordstr\"{o}m geometry with $m=1$ and $q=0.7$. Although the exact inner horizon is at regions where $uv$ and $(t^2-x^2)$ are infinite, $r$ can be very close to the inner horizon $r=r_{-}$ even when $uv$ and $(t^2-x^2)$ take moderate values.}
  \label{fig:contour_lines_r}
\end{figure}

\subsection{Coordinate system}
In the studies of mass inflation, the double-null coordinates described by Eq.~(\ref{double_null_metric_dudv}) are usually used,
\be
ds^{2} = -4e^{-2\sigma}dudv+r^2d\Omega^2,
\label{double_null_metric_dudv}
\ee
where $\sigma$ and $r$ are functions of the coordinates $u$ and $v$. $u$ and $v$ are outgoing and ingoing characteristics (trajectories of photons), respectively. For convenience, in this paper, we use a slightly modified form described by Eq.~(\ref{double_null_metric_dtdx}), obtained by defining $u=(t-x)/2$ and $v=(t+x)/2$~\cite{Frolov_2004},
\be
ds^{2} = e^{-2\sigma}(-dt^2+dx^2)+r^2d\Omega^2.
\label{double_null_metric_dtdx}
\ee
This set of coordinates is illustrated in Fig.~\ref{fig:ic_and_bc}. Similar to the Schwarzschild metric, the Reissner-Nordstr\"{o}m metric can be expressed in Kruskal-like coordinates~\cite{Graves_1960} (also see Refs.~\cite{Poisson_1989,Poisson_1990,Poisson_2004,Reall_2015}). So for ease and intuitiveness, we set the initial conditions close to those of the Reissner-Nordstr\"{o}m metric in Kruskal-like coordinates, taking into account the presence of a physical scalar field $\psi$.

In the form of Ref.~\cite{Reall_2015}, the Reissner-Nordstr\"{o}m metric in Kruskal-like coordinates in the region of $r>r_{-}$ can be expressed as
\be ds^2=\frac{r_{+}r_{-}}{k_{+}^2r^{2}}e^{-2k_{+}r}\left(\frac{r}{r_{-}}-1\right)^{1+\frac{k_{+}}{|k_{-}|}}(-dt^2+dx^2)+r^2d\Omega^2,
\label{RN_metric_text}\ee
where $r_{\pm}(=m\pm\sqrt{m^2-q^2})$ and $k_{\pm}[=(r_{\pm}-r_{\mp}]/(2r_{\pm}^2)]$ are the locations and surface gravities for the outer and inner horizons of a Reissner-Nordstr\"{o}m black hole, respectively. $m$ and $q$ are the mass and charge of the black hole, respectively. $r(t,x)$ is defined implicitly below~\cite{Reall_2015},
\be 4uv=t^2-x^2=e^{2k_{+}r}\left(1-\frac{r}{r_{+}}\right)\left(\frac{r}{r_{-}}-1\right)^{-\frac{k_{+}}{|k_{-}|}}.\label{r_RN_metric}\ee
Some details for this set of coordinates are given in Appendix~\ref{sec:appendix_Kruskal_RN}. In this set of coordinates, as implied by Eq.~(\ref{r_RN_metric}), the exact inner horizon is at regions where $uv$ and $(t^2-x^2)$ are infinite; however, it is found that, even when $uv$ and $(t^2-x^2)$ take moderate values, $r$ still can be very close to the inner horizon, e.g., $r=(1+10^{-10})r_{-}$. (See Fig.~\ref{fig:contour_lines_r}.) Therefore, at such places, the interaction between the scalar field and the inner horizon still can be very strong, and then we can investigate mass inflation numerically.

This formalism has several advantages as follows:
\begin{enumerate}[(i)]
  \item In the line element (\ref{double_null_metric_dtdx}), one coordinate is timelike and the rest are spacelike. This is a conventional setup. It is more convenient and more intuitive to use this set of coordinates. For the set of coordinates described by Eq.~(\ref{double_null_metric_dudv}), in the equations of motion, many terms are mixed derivatives of $u$ and $v$; while for the set of coordinates described by Eq.~(\ref{double_null_metric_dtdx}), in the equations of motion, spatial and temporal derivatives are usually separated.
  \item We set initial conditions close to those in the Reissner-Nordstr\"{o}m metric. Consequently, with the terms related to the scalar field being removed, we can test our code by comparing the numerical results to the analytic ones in the Reissner-Nordstr\"{o}m case conveniently. Moreover, comparing dynamics for scalar collapse to that in the Reissner-Nordstr\"{o}m case helps us to obtain intuitions on how the scalar field af{}fects the geometry.
  \item The interactions between a scalar field and the geometry are local ef{}fects. In Refs.~\cite{Gnedin_1991,Gnedin_1993}, the space between the inner and outer horizons are compactified into finite space. This overcompactification, at least to us, makes it a bit hard to understand the dynamics. In the configuration that we choose, the space is partially compactified, and the picture of charge scattering turns out to be simpler.
\end{enumerate}

\section{Numerical setup for charge scattering\label{sec:set_up}}
In this section, we set up the numerical formalisms for charge scattering, including field equations, initial conditions, boundary conditions, discretization scheme, and tests of numerical codes.

\subsection{Field equations\label{sec:field_eqs}}
In this subsection, we list the field equations for charge scattering. Details for the Einstein tensor and the energy-momentum tensor of a scalar field are given in Appendix~\ref{sec:appendix_tensors}. In double-null coordinates (\ref{double_null_metric_dtdx}), using
\be G^{t}_{t}+G^{x}_{x}=8\pi\left[{T^{\scriptsize{(\mbox{total})}}}^{t}_{t}+{T^{\scriptsize{(\mbox{total})}}}^{x}_{x}\right],\nonumber\ee
one obtains the equation of motion for $r$,
\be r(-r_{,tt}+r_{,xx})-r_{,t}^2+r_{,x}^2 = e^{-2\sigma}\left(1-\frac{q^2}{r^2}\right), \label{equation_r}\ee
where $r_{,t}\equiv dr/dt$ and other quantities are defined analogously. For simplicity, we define $\eta\equiv r^2$ and integrate the equation of motion for $\eta$, instead. The equation of motion for $\eta$ can be obtained by rewriting Eq.~(\ref{equation_r}) as~\cite{Frolov_2004}
\be -\eta_{,tt}+\eta_{,xx}=2e^{-2\sigma}\left(1-\frac{q^2}{r^2}\right).\label{equation_r_2}\ee
$G^{\theta}_{\theta}=8\pi{T^{\scriptsize{(\mbox{total})}}}^{\theta}_{\theta}$ provides the equation of motion for $\sigma$,
\be
-\sigma_{,tt}+\sigma_{,xx} + \frac{r_{,tt}-r_{,xx}}{r}+4\pi(\psi_{,t}^2-\psi_{,x}^2)+e^{-2\sigma}\frac{q^2}{r^4} = 0.
\label{equation_sigma}
\ee
In double-null coordinates, the equation of motion for $\psi$~(\ref{Box_psi}) becomes
\be
-\psi_{,tt}+\psi_{,xx}+\frac{2}{r}(-r_{,t}\psi_{,t}+r_{,x}\psi_{,x})=0.
\label{equation_psi}
\ee

The $\{uu\}$ and $\{vv\}$ components of the Einstein equations yield the constraint equations,
\be r_{,uu}+2r_{,u}\sigma_{,u}+4\pi r\psi_{,u}^2=0, \label{constraint_eq_uu}\ee
\be r_{,vv}+2r_{,v}\sigma_{,v}+4\pi r\psi_{,v}^2=0. \label{constraint_eq_vv}\ee
Via the definitions of $u=(t-x)/2$ and $v=(t+x)/2$, the constraint equations can be expressed in $(t,x)$ coordinates.
Equations $(\ref{constraint_eq_vv})-(\ref{constraint_eq_uu})$ and $(\ref{constraint_eq_vv})+(\ref{constraint_eq_uu})$ generate the constraint equations
for the $\{tx\}$ and $\{tt\}+\{xx\}$ components, respectively,
\be r_{,tx}+r_{,t}\sigma_{,x}+r_{,x}\sigma_{,t}+4\pi r\psi_{,t}\psi_{,x}=0,
\label{constraint_eq_xt}
\ee
\be
r_{,tt}+r_{,xx}+2(r_{,t}\sigma_{,t}+r_{,x}\sigma_{,x})+4\pi r(\psi_{,t}^2+\psi_{,x}^2)=0.
\label{constraint_eq_xx_tt}
\ee

\subsection{Initial conditions\label{sec:ic}}
We set the initial data to be time symmetric,
\be r_{,t}=\sigma_{,t}=\psi_{,t}=0 \hphantom{ddd} \mbox{at} \hphantom{d} t=0. \label{ic_t_0}\ee
Therefore, in this configuration, the constraint equation (\ref{constraint_eq_xt}) is satisfied identically. Note that, in this configuration, the values of $r_{,t}$ and $\sigma_{,t}$ at $t=0$ are the same as those in the Reissner-Nordstr\"{o}m metric case.

We set the initial value for $\psi$ as
\be \psi|_{\scriptsize{t=0}}=a\exp\left[-\frac{(x-x_{0})^2}{b}\right],\ee
where $a$, $b$, and $x_{0}$ are parameters. The initial value for $\sigma$ is defined to be the same as the corresponding one in the Reissner-Nordstr\"{o}m metric case~(\ref{RN_metric_text}),
\be e^{-2\sigma}\big|_{\scriptsize{t=0}}
=e^{-2\sigma}\big|^{\scriptsize{\mbox{RN}}}_{\scriptsize{t=0}}
=\frac{r_{+}r_{-}}{k_{+}^2r^{2}}e^{-2k_{+}r}\left(\frac{r}{r_{-}}-1\right)^{1+\frac{k_{+}}{|k_{-}|}},
\label{sigma_RN_metric}\ee
where $r$ is defined by Eq.~(\ref{r_RN_metric}) with $t=0$. We obtain the initial value for $r$ in charge scattering at $t=0$ by combining Eqs.~(\ref{equation_r}) and (\ref{constraint_eq_xx_tt}),
\be
\begin{split}
r_{,xx}=
&-r_{,t}\sigma_{,t}-r_{,x}\sigma_{,x}+\frac{r_{,t}^2-r_{,x}^2}{2r}\\
&-2{\pi}r(\psi_{,t}^2+\psi_{,x}^2)+\frac{1}{2r}e^{-2\sigma}\left(1-\frac{q^2}{r^2}\right).
\end{split}
\label{r_xx_ic}\ee

We set $r_{,x}=\sigma_{,x}=0$ at the origin $(x=0,t=0)$ as in the Reissner-Nordstr\"{o}m metric case. The definition domain for the spatial coordinate $x$ is $[-x_{b}~x_{b}]$. Then $r(x=x,t=0)$ can be obtained by integrating Eq.~(\ref{r_xx_ic}) via the fourth order Runge-Kutta method from $x=0$ to $x={\pm}x_{b}$, respectively.

In this paper, we employ the finite dif{}ference method. The leapfrog integration scheme is implemented, which is a three-level scheme and requires initial data on two dif{}ferent time levels. With the initial data at $t=0$, we compute the data at $t=\Delta t$ with a second-order Taylor series expansion as done in Ref.~\cite{Pretorius}. Take $\psi$ as an example,
\be \psi|_{\scriptsize{t={\Delta}t}}=\psi|_{\scriptsize{t=0}}+\psi_{,t}|_{\scriptsize{t=0}}\Delta t
+\frac{1}{2}\psi_{,tt}|_{\scriptsize{t=0}}(\Delta t)^2.\label{ic_taylor_expansion}\ee
The values of $\psi|_{t=0}$ and $\psi_{,t}|_{t=0}$ are set up as discussed above, and the value of $\psi_{,tt}|_{t=0}$ can be obtained from the equation of motion for $\psi$ (\ref{equation_psi}).

Up to this point, the initial conditions are fixed with all the field equations being taken into account. The first-order time derivatives of $r$, $\sigma$, and $\psi$ at $t=0$ described by Eq.~(\ref{ic_t_0}) ensure that the constraint equation (\ref{constraint_eq_xt}) is satisfied. The equation for $r_{,xx}$ at $t=0$ expressed by (\ref{r_xx_ic}) implies that the constraint equation (\ref{constraint_eq_xx_tt}) is satisfied. Computations of $r$, $\sigma$, and $\psi$ at $t={\Delta}t$ via a second-order Taylor series expansion, as expressed by Eq.~(\ref{ic_taylor_expansion}) for the case of $\psi$, satisfy all the equations of motion.

\subsection{Boundary conditions\label{sec:bc}}
The values of $r$, $\sigma$, and $\psi$ at the boundaries of $x={\pm}x_{b}$ are obtained via extrapolations. In fact, we are mainly concerned with the dynamics around $x=0$. Therefore, as long as $x_b$ is large enough, the boundary conditions will not af{}fect the dynamics that we are interested in.

\begin{figure}
\hspace{-15.0pt}
\includegraphics[width=9cm,height=7.2cm]{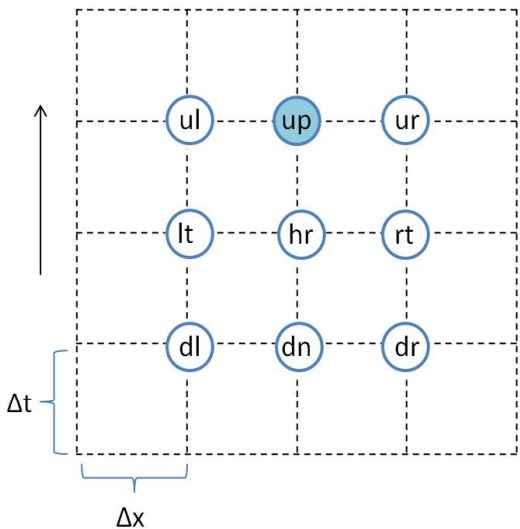}
\caption{Numerical evolution scheme.}
\label{fig:grid_scheme}
\end{figure}

\subsection{Discretization scheme}
In this paper, we implement the leapfrog integration scheme, which is second-order accurate and nondissipative. With the demonstration of Fig.~\ref{fig:grid_scheme} and using the quantity $\psi$ as an example, our discretization scheme is expressed below,
\be
\begin{aligned}
& \psi_{,t}=\frac{\psi_{\text{up}}-\psi_{\text{dn}}}{2\Delta t},
& \psi_{,x}=\frac{\psi_{\text{rt}}-\psi_{\text{lt}}}{2\Delta x},\\
& \psi_{,tt}=\frac{\psi_{\text{up}}-2\psi_{\text{hr}}+\psi_{\text{dn}}}{(\Delta t)^{2}},
& \psi_{,xx}=\frac{\psi_{\text{lt}}-2\psi_{\text{hr}}+\psi_{\text{rt}}}{(\Delta x)^{2}},\\
& \psi_{,xt}=\frac{\psi_{\text{ur}}-\psi_{\text{ul}}-\psi_{\text{dr}}+\psi_{\text{dl}}}{4{\Delta}x\cdot{\Delta}t},
& \psi_{,u}=\psi_{,t}-\psi_{,x},\\
& \psi_{,v}=\psi_{,t}+\psi_{,x}.\\
\end{aligned} \nonumber
\ee
We let the temporal and spatial grid spacings be equal, $\Delta t=\Delta x$.

\begin{figure*}[t!]
  \epsfig{file=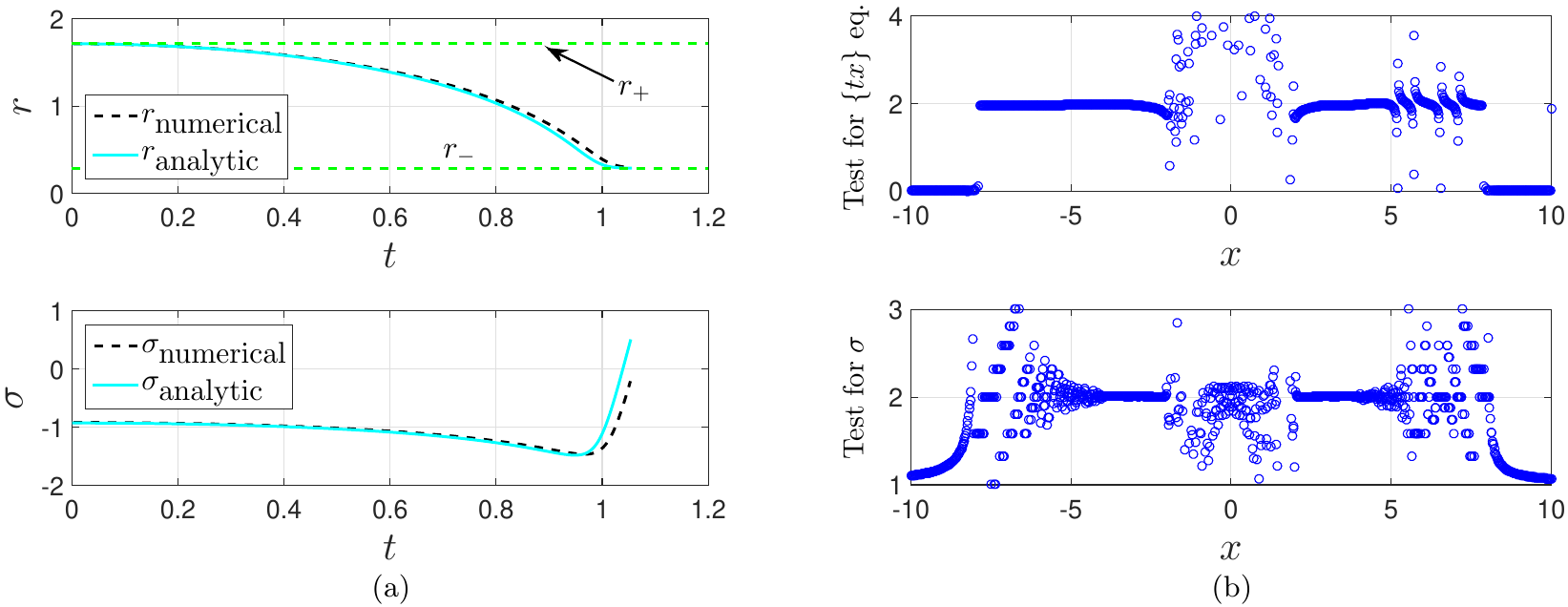, width=1\textwidth}
  \caption{Tests of numerical code for charge scattering. (a) Numerical vs analytic results for a Reissner-Nordstr\"{o}m black hole. $m=1$, $q=0.7$, and ${\Delta}x={\Delta}t=10^{-4}$. The slice is for $(x=3{\Delta}x,t=t)$. This is a special case of charge scattering with the contribution of the scalar field being set to zero. Numerical and analytic results match well at an early stage, while at a later stage where the gravity and electric field become stronger, the numerical evolutions have a time delay, compared to analytic solutions. (b) Numerical tests for the $\{tx\}$ constraint equation (\ref{constraint_eq_xt}) and the evolution of $\sigma$ on the slice $(x=x,t=0.65)$. They are both second-order convergent.}
  \label{fig:numerical_tests}
  \end{figure*}

\subsection{Tests of numerical code}
It is necessary to test the numerical code before we explore the results further. We compare the numerical results obtained by the code with the analytic ones for the dynamics in a Reissner-Nordstr\"{o}m geometry, and examine the convergence of the constraint equations and dynamical equations in charge scattering.

For charge scattering, the dynamics in a Reissner-Nordstr\"{o}m geometry in Kruskal-like coordinates is a special case, in which the contributions from the scalar field are set to zero. This special case has analytic solutions expressed by Eqs.~(\ref{RN_metric_text}) and (\ref{r_RN_metric}). Therefore, we can test our code by comparing the numerical and analytic results in the Reissner-Nordstr\"{o}m metric. Set $m=1$, $q=0.7$, and ${\Delta}x={\Delta}t=10^{-4}$. We plot the evolutions of $r$ and $\sigma$ along the slice $(x=3{\times}10^{-4},t=t)$ in Fig.~\ref{fig:numerical_tests}(a). As shown in Fig.~\ref{fig:numerical_tests}(a), numerical and analytic results match well at an early stage, while at a late stage where gravity and electric field become strong, the numerical evolutions have a time delay compared to analytic solutions.

When the numerical results are obtained, we substitute the numerical results into the disretized equations of motion and constraint equations, and find that they are well satisfied. Moreover, the convergence of the constraint equations (\ref{constraint_eq_xt}) and (\ref{constraint_eq_xx_tt}) is examined. We assume one constraint equation is $n$th-order convergent: residual=$\mathcal{O}(h^{n})$, where $h$ is the grid size. Therefore, the convergence rate of the discretized constraint equations can be obtained from the ratio between residuals with two dif{}ferent step sizes,
\be n=\log_{2}\left[\frac{\mathcal{O}(h^{n})}{\mathcal{O}\left(\left(\frac{h}{2}\right)^{n}\right)}\right].\label{accuracy_test}\ee
Our numerical results show that both of the constraint equations are about second-order convergent. As a representative, we plot the results for the $\{tx\}$ constraint equation (\ref{constraint_eq_xt}) in Fig.~\ref{fig:numerical_tests}(b) for the slice $(x=x,t=0.65)$.

Convergence tests via simulations with dif{}ferent grid sizes are also implemented~\cite{Sorkin,Golod}. If the numerical solution converges, the relation between the numerical solution and the real one can be expressed by
\be F_{\mbox{real}}=F^{h}+\mathcal{O}(h^{n}),\nonumber\ee
where $F^{h}$ is the numerical solution. Then for step sizes equal to $h/2$ and $h/4$, we have
\be F_{\mbox{real}}=F^{\frac{h}{2}}+\mathcal{O}\left[\left(\frac{h}{2}\right)^{n}\right],\nonumber\ee
\be F_{\mbox{real}}=F^{\frac{h}{4}}+\mathcal{O}\left[\left(\frac{h}{4}\right)^{n}\right].\nonumber\ee
Defining $c_1\equiv F^{h}-F^{\frac{h}{2}}$ and $c_2\equiv F^{\frac{h}{2}}-F^{\frac{h}{4}}$, one obtains the convergence rate
\be n=\log_{2}\left(\frac{c_1}{c_2}\right). \label{convergence_test}\ee
The convergence tests for $\eta\equiv r^{2}$, $\sigma$, and $\psi$ are investigated. They are all second-order convergent. As a representative, the results for $\sigma$ are plotted in Fig.~\ref{fig:numerical_tests}(b) for the slice $(x=x,t=0.65)$. The values of the parameters in charge scattering in this section are described at the beginning of Sec.~\ref{sec:results_scattering}. We use the spatial range of $x\in[-10~10]$ and the grid spacings of $h={\Delta}x={\Delta}t=0.02$.

\section{Neutral scalar collapse\label{sec:neutral_collapse}}
In Ref.~\cite{Guo_1312}, neutral scalar collapse in $f(R)$ gravity was explored. In this section, we consider a simpler case: neutral scalar collapse in general relativity. We examine the dynamics in the vicinity of the central singularity of the formed black hole. Mass inflation in the vicinity of the central singularity will be discussed.

\subsection{Numerical setup}
The numerical setup is a simpler version of the one in neutral scalar collapse in $f(R)$ gravity discussed in Ref.~\cite{Guo_1312}. The dynamical equations for $r$, $\eta$, $\sigma$, and $\psi$ can be obtained by setting the terms related to the electric field in the corresponding equations in Sec.~\ref{sec:field_eqs} to zero,
\be r(-r_{,tt}+r_{,xx})-r_{,t}^2+r_{,x}^2 = e^{-2\sigma},\label{equation_r_collapse}\ee
\be -\eta_{,tt}+\eta_{,xx}=2e^{-2\sigma},\label{equation_eta_collapse}\ee
\be -\sigma_{,tt}+\sigma_{,xx} + \frac{r_{,tt}-r_{,xx}}{r}+4\pi(\psi_{,t}^2-\psi_{,x}^2)=0,\label{equation_sigma_collapse}\ee
\be -\psi_{,tt}+\psi_{,xx}+\frac{2}{r}(-r_{,t}\psi_{,t}+r_{,x}\psi_{,x})=0. \label{equation_psi_collapse}\ee

In the equation of motion for $\sigma$~(\ref{equation_sigma_collapse}), the term $(r_{,tt}-r_{,xx})/r$ can create big errors near the center $x=r=0$. To avoid such a problem, we use the constraint equation (\ref{constraint_eq_uu}) alternatively~\cite{Frolov_2004}. Defining a new variable $g$
\be g\equiv-2\sigma-\ln(-r_{,u}), \label{g_definition}\ee
one can rewrite Eq.~(\ref{constraint_eq_uu}) as the equation of motion for $g$,
\be g_{,u}=4{\pi}\cdot\frac{r}{r_{,u}}\cdot\psi_{,u}^2. \label{equation_g}\ee
In the numerical integration, once the value of $r$ at the advanced level is obtained, the value of $\sigma$ at the current level will be computed using Eq.~(\ref{g_definition}).

We set the initial data as
\be r_{,tt}=r_{,t}=\sigma_{,t}=\psi_{,t}=0 \hphantom{ddd} \mbox{at} \hphantom{d} t=0.\ee
The initial value of $\psi(r)$ is defined as
\be \psi(r)|_{t=0}=a\cdot\tanh\left[\frac{(r-r_0)^2}{b}\right],\ee
with $a=0.1$, $b=1$, and $r_{0}=5$. The local Misner-Sharp mass $m$ is defined as~\cite{Misner}
\be g^{\mu\nu}r_{,\mu}r_{,\nu}=e^{2\sigma}(-r_{,t}^2+r_{,x}^2){\equiv}1-\frac{2m}{r}.\label{mass_Misner}\ee
(See Ref.~\cite{Hayward} for details on various properties of the Misner-Sharp mass/energy in spherical symmetry.) Then on the initial slice $(x=x,t=0)$, the equations for $r$, $m$, and $g$ are~\cite{Frolov_2004,Guo_1312}
\begin{align}
r_{,x}&=\left(1-\frac{2m}{r}\right)e^g,\label{r_ic_collapse}\\
\nonumber\\
m_{,r}&=4{\pi}r^2\left[V+\frac{1}{2}\left(1-\frac{2m}{r}\right)\psi_{,r}^2\right],\label{m_ic_collapse}\\
\nonumber\\
g_{,r}&=4{\pi}r\psi_{,r}^2.\label{g_ic_collapse}
\end{align}
Set $r=m=g=0$ at the origin $(x=0,t=0)$. Then the values of $r$, $m$, and $g$ on the initial slice $(x=x,t=0)$ can be obtained by integrating Eqs.~(\ref{r_ic_collapse})-(\ref{g_ic_collapse}) from the center $x=0$ to the outer boundary $x=x_{b}$ via the fourth-order Runge-Kutta method. The values of $r$, $\sigma$, and $\psi$ at $t={\Delta}t$ can be obtained with a second-order Taylor series expansion, as discussed in Sec.~\ref{sec:ic}. The value of $g$ at $t={\Delta}t$ can be obtained using Eq.~(\ref{g_definition}).

The range for the spatial coordinate is defined to be $x\in[0 \mbox{ } 10]$. At the inner boundary $x=0$, $r$ is always set to zero. The term $2(-r_{,t}\psi_{,t}+r_{,x}\psi_{,x})/r$ in Eq.~(\ref{equation_psi_collapse}) needs to be regular at $x=r=0$. Since $r$ is always set to zero at the center,  so is $r_{,t}$. Then we enforce $\psi$ to satisfy $\psi_{,x}=0$ at $x=0$. The value of $g$ at $x=0$ is obtained via extrapolation. We set up the outer boundary conditions via extrapolation. The temporal and spatial grid spacings are ${\Delta}t={\Delta}x=0.005$.

The numerical code used here is a simplified version of the second-order convergent one for spherical scalar collapse in $f(R)$ gravity developed in Ref.~\cite{Guo_1312}.

\subsection{Black hole formation}
\begin{figure*}[t!]
  \begin{tabular}{ccc}
  \epsfig{file=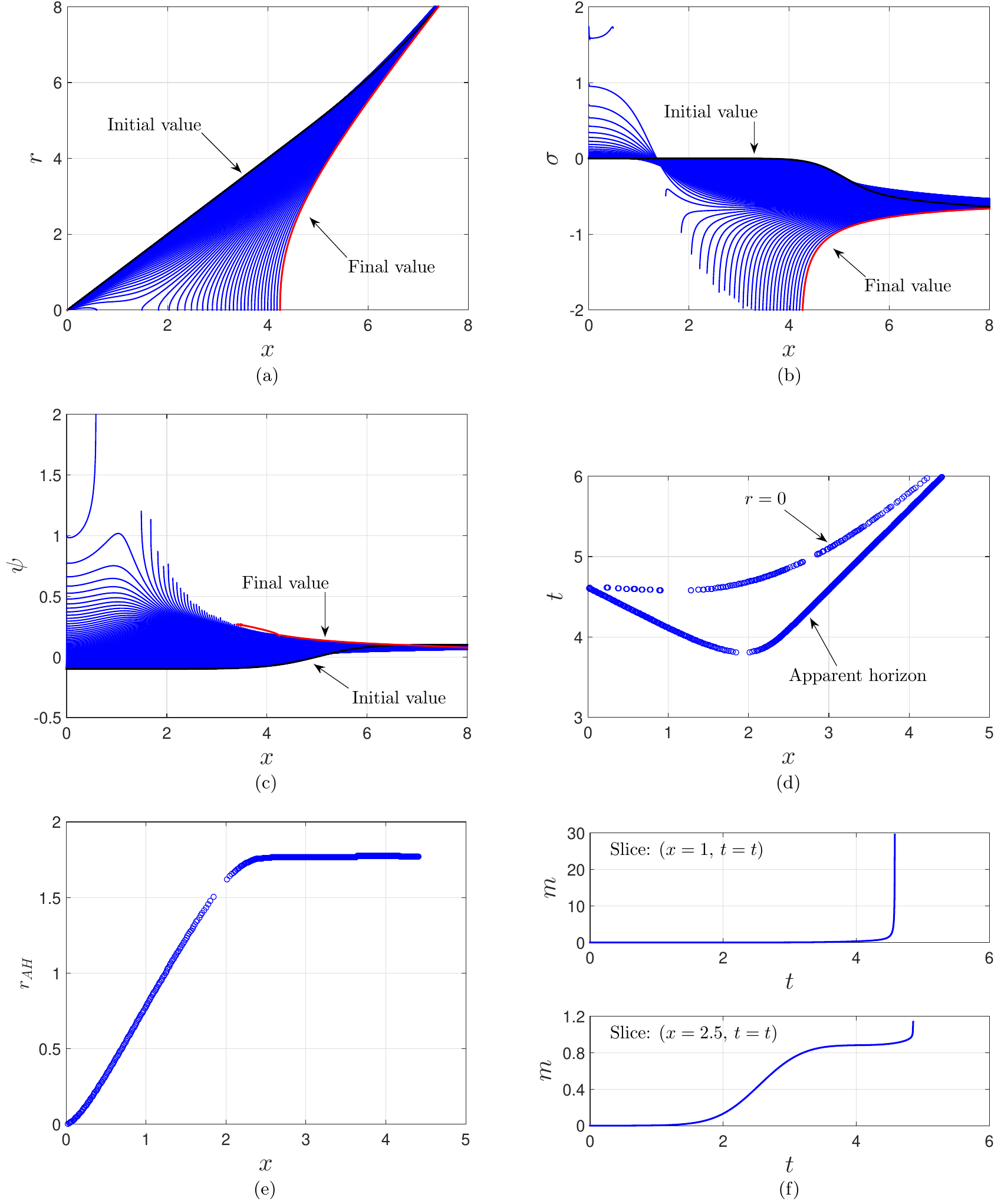, width=1\textwidth}
  \end{tabular}
  \caption{Evolutions in neutral scalar collapse. (a)-(c): evolutions of $r$, $\sigma$, and $\psi$. In (a) and (b), the time interval between two consecutive slices is $10{\Delta}t=0.05$. In (c), it is $5{\Delta}t=0.025$. (d) and (e) are for the apparent horizon and the singularity curve of the formed black hole. (f) The Misner-Sharp mass function along the slices $(x=1,t=t)$ and $(x=2.5,t=t)$.}
  \label{fig:evolutions_collapse}
\end{figure*}

When the scalar field is weak enough, the scalar field will collapse and then disperse. A flat space is left. However, when the scalar field carries enough energy, the scalar field can collapse into a black hole. In this paper, we are interested in the latter case. In Fig.~\ref{fig:evolutions_collapse}, we plot the evolutions of $r$, $\sigma$, and $\psi$ and the apparent horizon of the formed black hole. On the apparent horizon of a black hole, the expansion of the outgoing null geodesics orthogonal to the apparent horizon is zero~\cite{Baumgarte}. Then in double-null coordinates, on the apparent horizon, there is~\cite{Csizmadia}
\be g^{\mu\nu}r_{,\mu}r_{,\nu}=e^{2\sigma}(-r_{,t}^2+r_{,x}^2){\equiv}1-\frac{2m}{r}=0.\ee
Using such a property, the apparent horizon is found and plotted in Figs.~\ref{fig:evolutions_collapse}(d) and \ref{fig:evolutions_collapse}(e). As shown in Fig.~\ref{fig:evolutions_collapse}(e), the radius of the apparent horizon is $r_{\scriptsize{\mbox{AH}}}\approx1.8$. Therefore, a black hole is formed.

We plot the Misner-Sharp mass function~(\ref{mass_Misner}) along the slices $(x=1,t=t)$ and $(x=2.5,t=t)$ in Fig.~\ref{fig:evolutions_collapse}(f), from which one can see that, near the central singularity, the mass function diverges.

\subsection{\texorpdfstring{Asymptotic dynamics near the singularity curve of a Schwarzschild black hole}{Asymptotic dynamics near the central singularity of a Schw. BH}}
In this section, we are mainly interested in the dynamics near the central singularity of the formed black hole. As a preparation, in this subsection, we list some basic results on the dynamics near the central singularity of a (stationary) Schwarzschild black hole. We consider the Schwarzschild metric in Kruskal coordinates,
\be ds^2=\frac{32m^3}{r}e^{-\frac{r}{2m}}(-dt^2+dx^2)+r^2d\Omega^2,\label{Kruskal_coordinate}\ee
with
\begin{align}
e^{-2\sigma}&=\frac{32m^3}{r}e^{-\frac{r}{2m}},\label{sigma_schw_BH}\\
\nonumber\\
t^2-x^2&=\left(1-\frac{r}{2m}\right)e^{\frac{r}{2m}}.\label{radius_schw_BH}
\end{align}

As discussed in Appendix~\ref{sec:appendix_singularity_Schw}, in the vicinity of the singularity curve, the ratios between the spatial and temporal derivatives of $r$ can be expressed by the slope of the contour line $r=0$, $K$,
\be \Big|\frac{r_{,x}}{r_{,t}}\Big|=|K|,\hphantom{ddd}\Big|\frac{r_{,xx}}{r_{,tt}}\Big|\approx|K^2|.\ee

In the vicinity of the singularity curve, we rewrite $t$ as $t=t_0-\xi$, where $t_0$ is the coordinate time on the singularity curve and $\xi\ll t_0$. With the spatial coordinate $x$ being fixed, a perturbation expansion near the singularity curve directly yields
\be r\approx\left(16m^{2}t_0\xi\right)^{\frac{1}{2}}.\label{r_Kruskal}\ee
Moreover, combining Eqs.~(\ref{sigma_schw_BH}) and (\ref{r_Kruskal}), one obtains
\be \sigma\approx-\frac{1}{2}\ln\left(\frac{32m^3}{r}\right)\approx\frac{1}{2}\ln{r}\approx\frac{1}{4}\ln\xi.\ee

\newpage
\be \nonumber\ee
\be \nonumber\ee

As shown in Appendix~\ref{sec:appendix_singularity_Schw}, throughout the whole spacetime of a Schwarzschild black hole, inside and outside the horizon and also near the singularity, there is
\be g^{\mu\nu}r_{,\mu}r_{,\nu}=e^{2\sigma}(-r_{,t}^2+r_{,x}^2){\equiv}1-\frac{2m}{r}.\label{mass_function}\ee
Namely, for a (stationary) Schwarzschild black hole, the Misner-Sharp mass function is equal to the black hole mass everywhere.

\begin{figure}[h]
\hspace{-10pt}
\epsfig{file=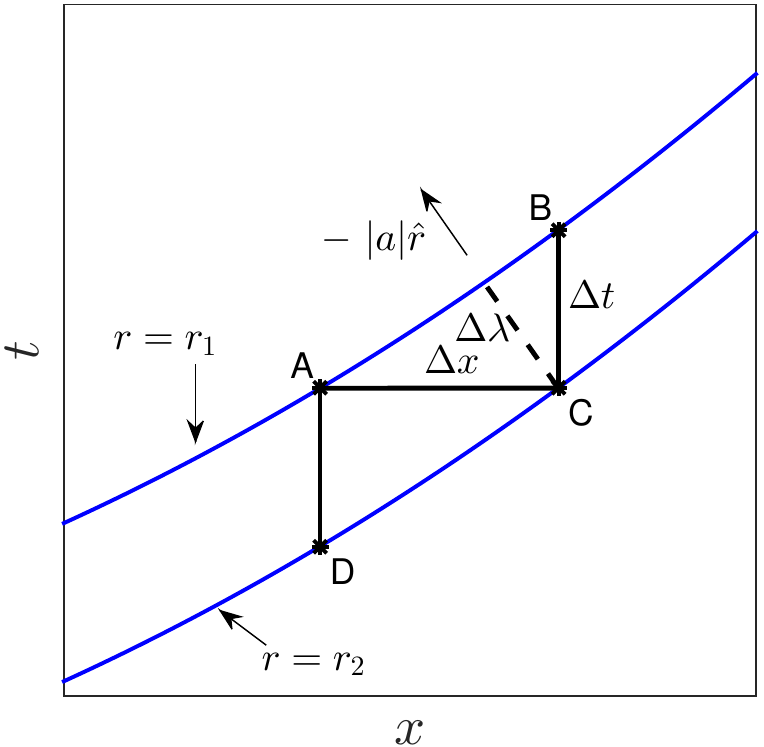, width=6.3cm, height=6.2cm}
\caption{Spatial derivative vs temporal derivative near the central spacelike singularity in neutral scalar collapse. Point $A$ and point $B$ are on one same hypersurface $r=\mbox{const}$, while point $C$ is on another one. At point $C$, in first-order accuracy, $r_{,x}\approx(r_{C}-r_{A})/\Delta x$ and $r_{,t}\approx(r_{B}-r_{C})/\Delta t$. Since $r_{A}=r_{B}$ and the absolute value of the slope of the singularity curve $dt/dx$ is less than $1$, there is $|r_{,x}/r_{,t}|\approx|\Delta t/\Delta x|<1$. $\Delta\lambda$ and $-|a|\hat{r}$ are perpendicular to the lines of $r=r_{1}$ and $r=r_{2}$. $a$ is a certain number, such that the magnitude of $-|a|\hat{r}$ has the length as shown in the figure.}
\label{fig:ratio_center}
\end{figure}

\subsection{\texorpdfstring{Asymptotic dynamics in the vicinity of the central singularity of the formed black hole in neutral collapse}{Asymptotic dynamics in the vicinity of the central singularity}\label{sec:asymptotics_collapse}}
In this subsection, we discuss the asymptotic dynamics in the vicinity of the central singularity of the formed black hole in neutral collapse, by connecting the reduced field equations to numerical results and comparing the dynamics of neutral collapse to that in Schwarzschild black holes.

We first explain why near the singularity curve of the formed black hole, the ratio between spatial and temporal derivatives are defined by the slope of the contour lines $r=\mbox{const}$. We take the quantity $r$ as an example. As illustrated in Fig.~\ref{fig:ratio_center}, point $A$ and point $B$ are on one same hypersurface $r=\mbox{const}$, while point $C$ is on another one. At point $C$, in first-order accuracy, $r_{,x}\approx(r_{C}-r_{A})/\Delta x$ and $r_{,t}\approx(r_{B}-r_{C})/\Delta t$. Since the singularity curve is spacelike, the absolute value of the slope of the singularity curve $K{\equiv}dt/dx$ is less than $1$. Consequently, there is
\be \Big|\frac{r_{,x}}{r_{,t}}\Big|\approx\Big|\frac{\Delta t}{\Delta x}\Big|\approx|K|<1.\label{ratio_spatial_temporal}\ee
Similar arguments yield
\be \Big|\frac{r_{,xx}}{r_{,tt}}\Big|\approx\Big|\frac{{\Delta}t}{{\Delta}x}\Big|^2{\approx}K^2.\label{ratio_spatial_temporal_2nd}\ee
This can also be interpreted in the following way. As shown in Fig.~\ref{fig:ratio_center}, in double-null coordinates, the time vector $\hat{t}$ is not normal to the hypersurface of $r=\mbox{const}$. Then the derivatives in the radial direction have nonzero projections on both hypersurfaces of $x=\mbox{const}$ and $t=\mbox{const}$. Along an arbitrary slice $(x=\mbox{const},t=t)$, near the singularity, the ratios between first- and second-order spatial and temporal derivatives are equal to $|K|$ and $K^2$, respectively.

The quantities $\sigma$ and $\psi$ have similar features as described by Eqs.~(\ref{ratio_spatial_temporal}) and (\ref{ratio_spatial_temporal_2nd}).
This can be explained as follows. Take the scalar field $\psi$ as an example. With the illustration of Fig.~\ref{fig:ratio_center}, as this scalar field moves toward the center, two neighboring points on this scalar wave $\psi$ should take close values when they respectively cross points $C$ and $D$ (which are on one same hypersurface $r=\mbox{const}$) at two consecutive moments, because these two points on the scalar wave are neighbors and the \lq\lq distances\rq\rq~$AD$ and $BC$ are more important for their values than the dif{}ference between these two neighboring points. In other words, in the vicinity of the singularity curve, gravity is more important than the dif{}ference between neighboring points on the scalar wave. Alternatively, to a large extent, later evolutions in a strong gravitational field largely erase away the initial information on the connections between neighboring points. These arguments are supported by numerical results. Near the singularity, the evolution of $\psi$ is described by $\psi\approx C\ln\xi$, where $\xi=t_0-t{\ll}t_0$ and $t_0$ is the coordinate time on the singularity curve. In Fig.~\ref{fig:ratio_center}, $\xi$ means $AD$ and $BC$. The parameter $C$ changes slowly along the singularity curve, compared to the dramatic running of $\ln\xi$~\cite{Guo_1312}. In other words, $\psi_{\scriptsize{A}}\approx\psi_{\scriptsize{B}}$ and $\psi_{\scriptsize{C}}\approx\psi_{\scriptsize{D}}$. Then as in the case of $r$, there is
\be \Big|\frac{\psi_{,x}}{\psi_{,t}}\Big|\approx\Big|\frac{\Delta t}{\Delta x}\Big|<1.\label{ratio_spatial_temporal_psi}\ee

\begin{figure*}[t!]
  \epsfig{file=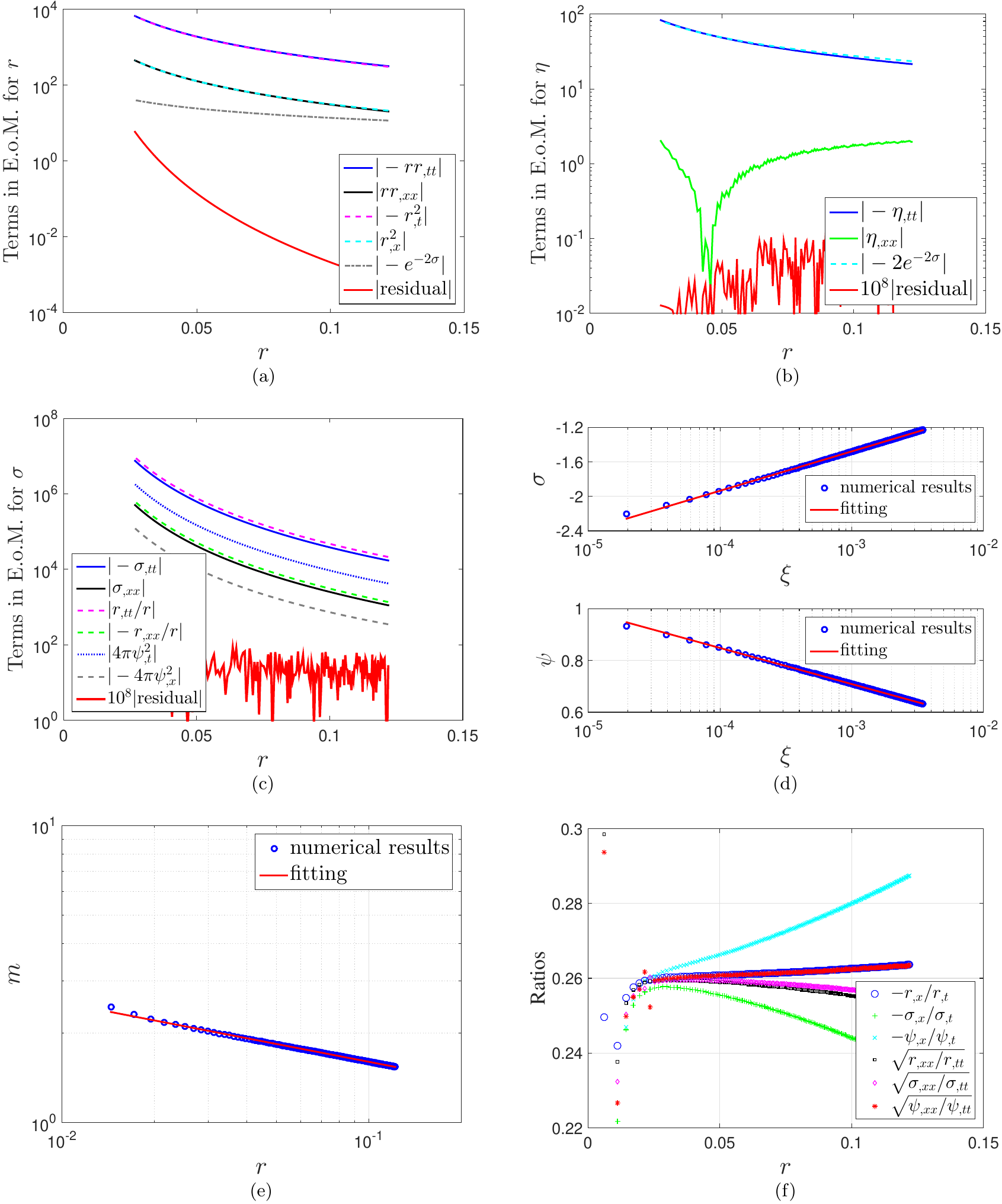, width=1\textwidth}
  \caption{(color online). Dynamics and solutions in the vicinity of the central singularity in neutral scalar collapse on the slice $(x=2,t=t)$.
  (a)-(c): dynamical equations for $r$, $\eta$, and $\sigma$.
  (d) $\sigma=a\ln\xi+b$, $a=0.1980\pm0.0004$, $b=-0.112\pm0.003$.
      $\psi=a\ln\xi+b$, $a=-0.0606\pm0.0001$, $b=0.2899\pm0.0008$.
  (e) $\ln m=a\ln r+b$, $a=-0.1993\pm0.0006$, $b=0.013\pm0.002$.
  (f) Ratios between spatial and temporal derivatives for $r$, $\sigma$, and $\psi$.}
  \label{fig:solutions_collapse_x_2}
\end{figure*}

Taking the slice $(x=2,t=t)$ as an example, we examine the dynamics along this slice via mesh refinement that was implemented in Refs.~\cite{Guo_1312,Garfinkel} and plot the results in Fig.~\ref{fig:solutions_collapse_x_2}. As shown in Fig.~\ref{fig:solutions_collapse_x_2}(f), in the vicinity of the singularity along this slice, the ratios between spatial and temporal derivatives are
\be
-\frac{r_{,x}}{r_{,t}}\approx-\frac{\sigma_{,x}}{\sigma_{,t}}\approx-\frac{\psi_{,x}}{\psi_{,t}}
\approx\sqrt{\frac{r_{,xx}}{r_{,tt}}}\approx\sqrt{\frac{\sigma_{,xx}}{\sigma_{,tt}}}\approx\sqrt{\frac{\psi_{,xx}}{\psi_{,tt}}}
\approx0.26.
\ee
As a comparison, the slope of the singularity curve, which is plotted in Fig.~\ref{fig:evolutions_collapse}(d), at $x=2$ is $K\approx0.25$.

\begin{figure*}[t!]
  \epsfig{file=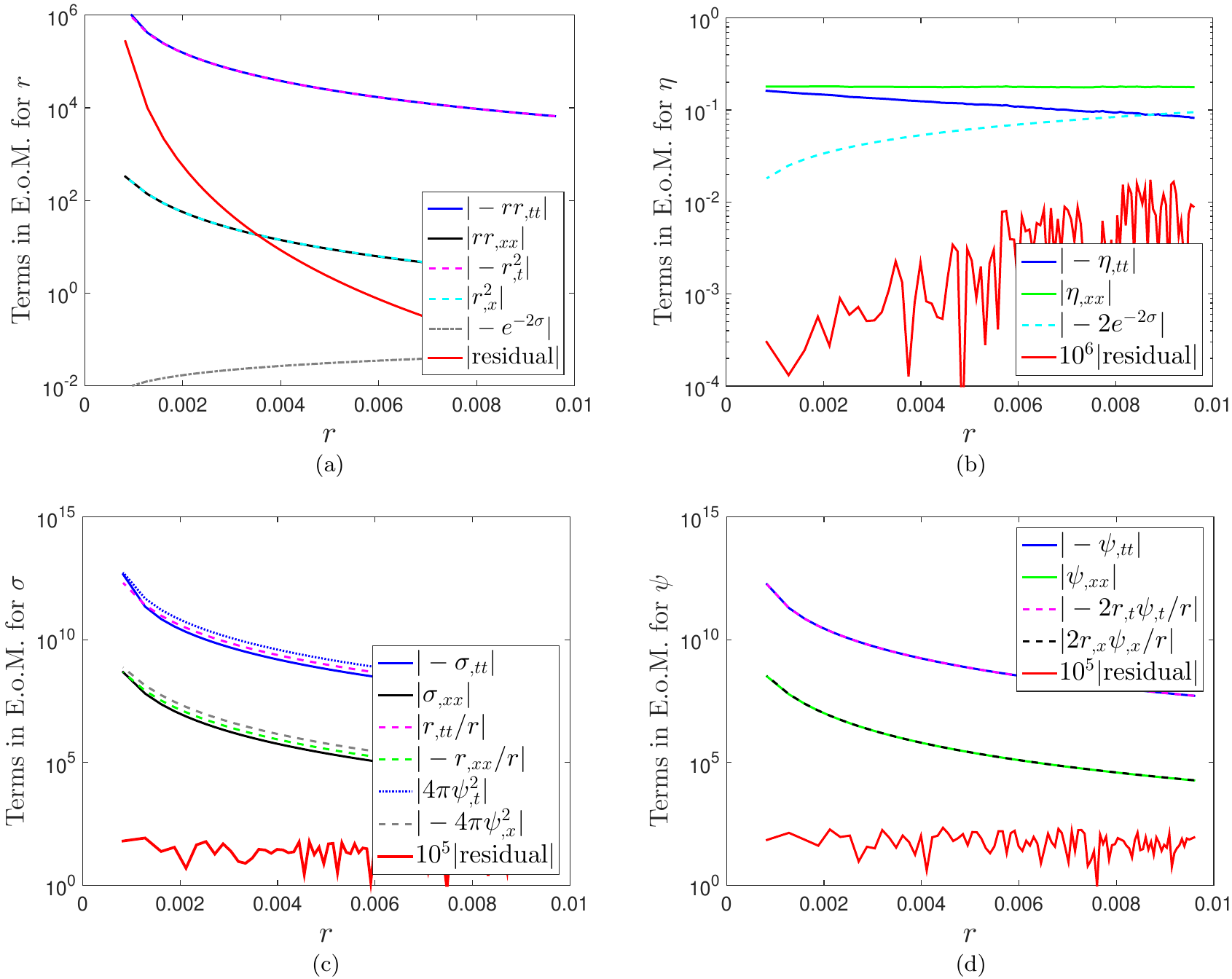, width=1\textwidth}
  \caption{(color online). Dynamics in the vicinity of the singularity in neutral collapse on the slice $(x=1,t=t)$. The numerical results show that in the vicinity of the singularity, the field equations for $r$~(\ref{equation_r_collapse}), $\eta$~(\ref{equation_eta_collapse}), $\sigma$~(\ref{equation_sigma_collapse}), and $\psi$~(\ref{equation_psi_collapse}) can be reduced to the following formats, respectively:
  (a) $rr_{,tt}\approx-r_{,t}^2$, (b) $\eta_{,tt}\approx\eta_{,xx}$, (c) $\sigma_{,tt}-r_{,tt}/r\approx4\pi\psi_{,t}^2$,
  and (d) $\psi_{,tt}\approx-2r_{,t}\psi_{,t}/r$. The big numerical errors near $r=0$ in (a) come from converting Eq.~(\ref{equation_r_2}) into Eq.~(\ref{equation_r}).}
  \label{fig:dynamics_collapse_x_1}
\end{figure*}

\begin{figure*}[t!]
  \epsfig{file=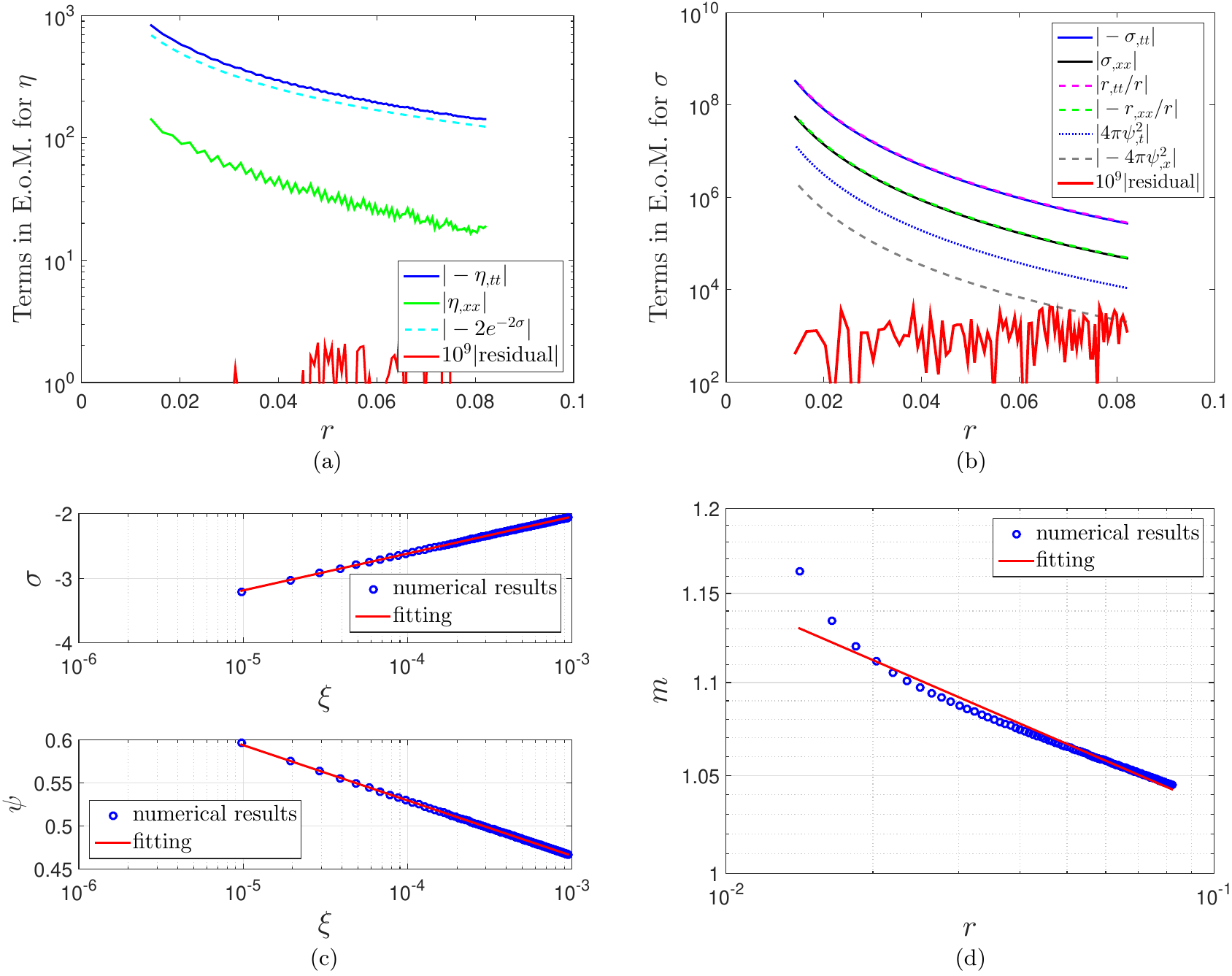, width=1\textwidth}
  \caption{(color online). Dynamics and solutions in the vicinity of the central singularity in neutral scalar collapse on the slice $(x=2.5,t=t)$.
  (a) and (b): dynamical equations for $\eta$ and $\sigma$.
  (c) $\sigma=a\ln\xi+b$, $a=0.2476\pm0.0002$, $b=-0.337\pm0.002$.
      $\psi=a\ln\xi+b$, $a=-0.02786\pm0.00003$, $b=0.2732\pm0.0003$.
  (d) $\ln m=a\ln r+b$, $a=-0.046\pm0.001$, $b=0.072\pm0.003$.}
  \label{fig:solutions_collapse_x_2_point_5}
\end{figure*}

Because of the fact that in the vicinity of the singularity, the ratios between spatial and temporal derivatives for $r$, $\sigma$, and $\psi$ take similar values, and with the numerical results plotted in Figs.~\ref{fig:dynamics_collapse_x_1}, \ref{fig:solutions_collapse_x_2},
and \ref{fig:solutions_collapse_x_2_point_5}, the field equations for $r$~(\ref{equation_r_collapse}), $\sigma$~(\ref{equation_sigma_collapse}),
and $\psi$~(\ref{equation_psi_collapse}) can be simplified as follows:
\begin{align}
rr_{,tt}&\approx-r_{,t}^2,\label{equation_r_asymptotic}\\
\nonumber\\
\sigma_{,tt}&\approx\frac{r_{,tt}}{r}+4\pi\psi_{,t}^2,\label{equation_sigma_asymptotic}\\
\nonumber\\
\psi_{,tt}&\approx-\frac{2}{r}r_{,t}\psi_{,t}.\label{equation_psi_asymptotic}
\end{align}
The asymptotic solutions to Eqs.~(\ref{equation_r_asymptotic})-(\ref{equation_psi_asymptotic}) are~\cite{Guo_1312}
\begin{align}
r&\approx A\xi^{\beta},\label{r_asymptotic_collapse}\\
\nonumber\\
\sigma&\approx B\ln\xi+\sigma_{0}\approx[\beta(1-\beta)-4{\pi}C^2]\ln\xi+\sigma_{0},\label{sigma_asymptotic_collapse}\\
\nonumber\\
\psi&\approx C\ln\xi.\label{psi_asymptotic_collapse}
\end{align}
Substituting Eq.~(\ref{r_asymptotic_collapse}) into Eq.~(\ref{equation_r_asymptotic}) yields
\be (1-\beta)\xi^{2(\beta-1)}\approx\beta\xi^{2(\beta-1)}.\nonumber\ee
Then we have
\be \beta\approx\frac{1}{2}.\label{beta_asymptotic_collapse}\ee

\begin{figure*}[t!]
  \epsfig{file=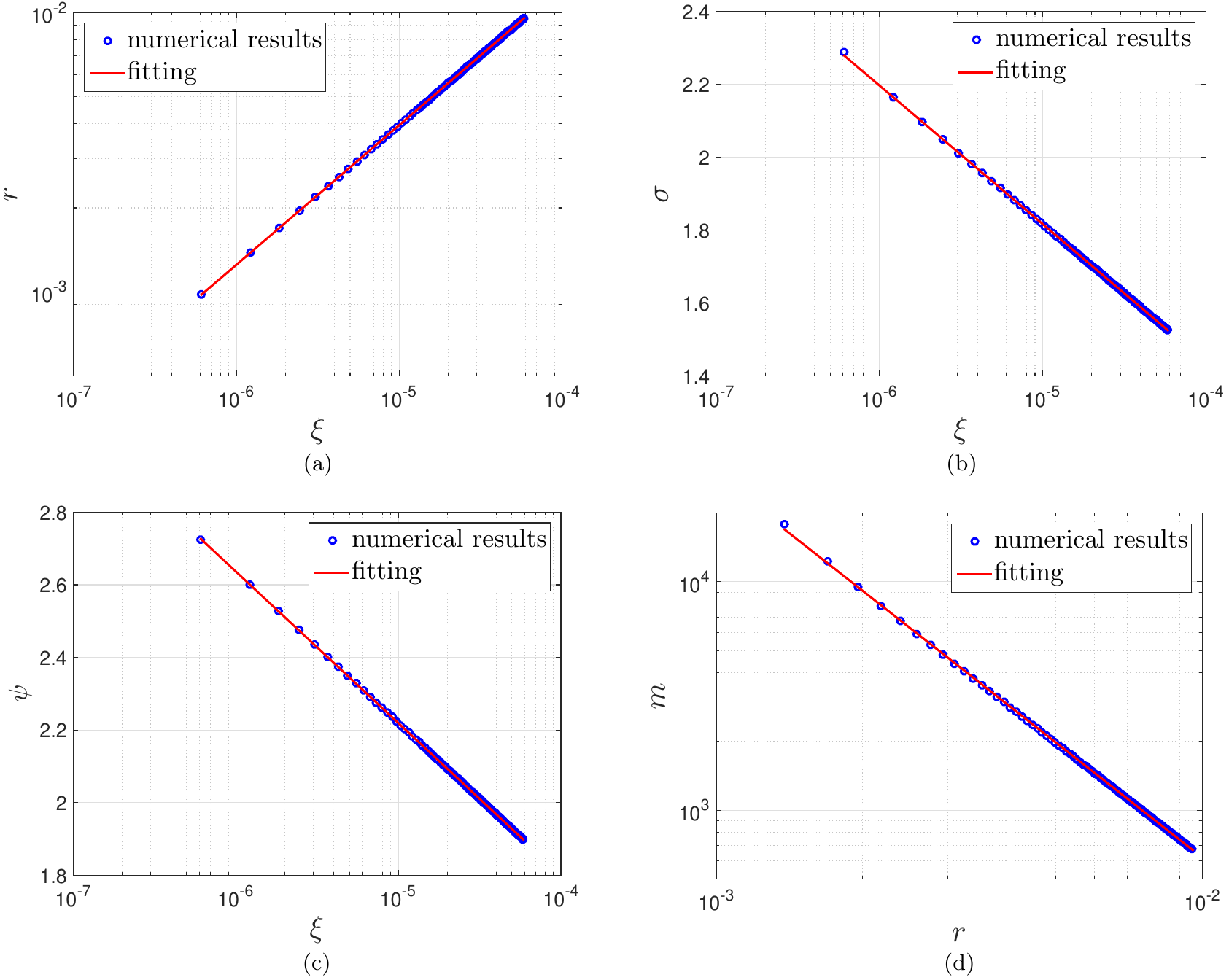, width=1\textwidth}
  \caption{Solutions in the vicinity of the central singularity in neutral scalar collapse on the slice $(x=1,t=t)$.
  (a) $\ln r=a\ln\xi+b$, $a=0.49948\pm0.00004$, $b=0.2172\pm0.0005$.
  (b) $\sigma=a\ln\xi+b$, $a=-0.1651\pm0.0001$, $b=-0.083\pm0.001$.
  (c) $\psi=a\ln\xi+b$, $a=-0.18129\pm0.00003$, $b=0.1322\pm0.0003$.
  (d) $\ln m=a\ln r+b$, $a=-1.670\pm0.002$, $b=-1.258\pm0.009$.}
  \label{fig:solutions_collapse_x_1}
\end{figure*}

\subsection{\texorpdfstring{Causes of mass inflation in neutral scalar collapse}{Causes of mass inflation in neutral scalar collapse}\label{sec:causes_mass_inflation}}
In neutral scalar collapse, the equation of motion for $\sigma$ is
\be
-\sigma_{,tt}+\sigma_{,xx} + \frac{r_{,tt}-r_{,xx}}{r}+4\pi(\psi_{,t}^2-\psi_{,x}^2)=0.
\ee
Then as discussed in the above subsection, the term $4\pi(\psi_{,t}^2-\psi_{,x}^2)$ is positive and can make $\sigma(x,t)$ greater than the corresponding value in the Schwarzschild black hole case. [See Eq.~(\ref{sigma_asymptotic_collapse}).] This makes the mass function divergent near the singularity as will be discussed below.

Near the singularity, using Eqs.~(\ref{r_asymptotic_collapse})-(\ref{beta_asymptotic_collapse}), there is
\be e^{2\sigma}(-r_{,t}^2+r_{,x}^2)\approx(K^{2}-1)\frac{A^{2}}{4}\xi^{-\frac{1}{2}-8{\pi}C^2}.\ee
Then with Eq.~(\ref{mass_Misner}), the mass function can be written as
\be
\begin{split}
m&=\frac{r}{2}[1+e^{2\sigma}(r_{,t}^2-r_{,x}^2)]\\
&\approx\left[\frac{1}{8}(1-K^{2})A^{3}e^{2\sigma_0}\right]\xi^{-8{\pi}C^2}\\
&\approx\left[\frac{1}{8}(1-K^{2})A^{3+16{\pi}C^2}e^{2\sigma_0}\right]r^{-16{\pi}C^2}.
\end{split}
\label{mass_analytic}
\ee
In the Schwarzschild black hole case, $C=0$. As shown in Appendix~\ref{sec:appendix_singularity_Schw}, the mass function is always constant and equal to the black hole mass. In neutral collapse, the parameter $\beta$ does not change much and remains about $1/2$. However, the parameter $C$ is not zero. Then the metric quantity $\sigma$ is modified. As a result, the delicate balance between $r$ and $e^{2\sigma}(-r_{,t}^2+r_{,x}^2)$ is broken. Consequently, as implied in Eq.~(\ref{mass_analytic}), near the singularity, the mass function diverges: mass inflation occurs. With numerical simulations, the divergence of $m$ near the central singularity was also reported in Ref.~\cite{Csizmadia}, while further explorations were absent.

As shown in Fig.~\ref{fig:evolutions_collapse}(b), at large- and small-$x$ regions, $\sigma$ approaches negative and positive infinities, respectively. We interpret this dif{}ference using Eqs.~(\ref{sigma_asymptotic_collapse})-(\ref{beta_asymptotic_collapse}), Fig.~\ref{fig:evolutions_collapse}(c), and three sample sets of results at $x=1$, $x=2$, and $x=2.5$, which are obtained via mesh refinement and are plotted in Figs.~\ref{fig:dynamics_collapse_x_1} and \ref{fig:solutions_collapse_x_1}, \ref{fig:solutions_collapse_x_2}, and \ref{fig:solutions_collapse_x_2_point_5}, respectively. As shown in Fig.~\ref{fig:evolutions_collapse}(c), at large-$x$ regions, the scalar field is very weak. This is also illustrated in Figs.~\ref{fig:solutions_collapse_x_2}(c) and \ref{fig:solutions_collapse_x_2_point_5}(b), which show the contributions of all the terms in the equation of motion for $\sigma$ at the slices $(x=2,t=t)$ and $(x=2.5,t=t)$, respectively. From these two figures, one can see that at large-$x$ regions, in the vicinity of the singularity, in the equation of motion for $\sigma$, the contribution of the scalar field $\psi$ is much less than those from metric quantities, $4\pi\psi_{,t}^2\ll|\sigma_{,tt}|\approx|r_{,tt}/r|$. As shown in Figs.~\ref{fig:solutions_collapse_x_2}(d) and \ref{fig:solutions_collapse_x_2_point_5}(c), on the slices $(x=2,t=t)$ and $(x=2.5,t=t)$, $C=-0.0606$ and $C=-0.02786$. Then $4{\pi}C^2<1/4$. Consequently, as implied in Eq.~(\ref{sigma_asymptotic_collapse}) and as plotted in Figs.~\ref{fig:solutions_collapse_x_2}(d) and \ref{fig:solutions_collapse_x_2_point_5}(c), in such a case, $\sigma$ is negative. We also note that because of the negativeness of $\sigma$, the contribution of the term $2e^{-2\sigma}$ in the equation of motion for $\eta$~(\ref{equation_eta_collapse}) is important. [See Figs.~\ref{fig:solutions_collapse_x_2}(b) and \ref{fig:solutions_collapse_x_2_point_5}(a).] This is quite dif{}ferent from the positive case, as plotted in Fig.~\ref{fig:dynamics_collapse_x_1}(b). Because of weakness of the scalar field at large-$x$ regions, $\sigma$ is not modified very much, compared to the corresponding value in the Schwarzschild black hole case. As a result, as implied in Eq.~(\ref{mass_analytic}), the mass function does not grow as fast as in the strong scalar field case. [See Figs.~\ref{fig:solutions_collapse_x_1}(d), \ref{fig:solutions_collapse_x_2}(e), and \ref{fig:solutions_collapse_x_2_point_5}(d).]

At small-$x$ regions, before the center $x=0$ converts from timelike into spacelike, the scalar field is reflected at $x=0$. In addition, new pulses of the scalar field from large-$x$ regions keep coming to the central region. Then some of the energy of the scalar field is accumulated around the center.
Then the scalar field becomes very strong in this region. [See Fig.~\ref{fig:evolutions_collapse}(c).] In this case, the contribution from the scalar field can be comparable to or even greater than those from the metric quantities, $4\pi\psi_{,t}^2\approx|\sigma_{,tt}|\approx|r_{,tt}/r|$. (See Fig.~\ref{fig:dynamics_collapse_x_1}.) As shown in Fig.~\ref{fig:solutions_collapse_x_1}(c), $C\approx-0.18$ in the vicinity of the singularity on the slice $(x=1,t=t)$. There is $4{\pi}C^2\approx0.41>1/4$. Consequently, $\sigma$ grows to be positive. [See Fig.~\ref{fig:solutions_collapse_x_1}(b).] Because of the positiveness of $\sigma$, the contribution from the term $2e^{-2\sigma}$ in the equation of motion for $\eta$~(\ref{equation_eta_collapse}) is small compared to those from $\eta_{,tt}$ and $\eta_{,xx}$. [See Fig.~\ref{fig:dynamics_collapse_x_1}(b).] Because of strongness of the scalar field, $\sigma$ is significantly modified. As a result, the mass function grows dramatically. [See Eq.~(\ref{mass_analytic}) and Fig.~\ref{fig:solutions_collapse_x_1}(d).]

\subsection{Differential form of the mass function near the central singularity\label{sec:inflation_vs_compression}}
As a further step of exploring mass inflation, we examine the dif{}ferential form of the mass function. Using Eq.~(\ref{mass_function}) and Einstein equations, one obtains~\cite{Poisson_1989,Poisson_1990},
\be \frac{\partial m}{\partial x^a}=4{\pi}r^{2}[T^{b}_{a}-\delta^{b}_{a}T^{(2)}]\frac{\partial r}{\partial x^b},\label{dm_dx}\ee
where $x^a$ is a coordinate, $t$ or $x$. $T^{(2)}({\equiv}T^{t}_{t}+T^{x}_{x})$ is zero for the massless scalar field $\psi$ with $V(\psi)=0$.
[See Eqs.~(\ref{T_tt}) and (\ref{T_xx}).] We denote $\hat{r}$, $\hat{t}$, and $\hat{x}$ as unit vectors along the radial, temporal, and spatial directions, respectively. As illustrated in Fig.~\ref{fig:ratio_center}, $\hat{r}$ is perpendicular to the contour lines $r=\mbox{const}$.
Combining Eqs.~(\ref{dm_dx}), (\ref{T_tt}), and (\ref{T_xx}), one obtains the gradient of $m$,
\be
\begin{split}
\nabla{m}
&=\frac{\partial m}{\partial t}\hat{t}+\frac{\partial m}{\partial x}\hat{x}\\
&=-4{\pi}r^{2}\cdot\frac{1}{2}e^{2\sigma}(\psi_{,t}^2+\psi_{,x}^2)[r_{,t}\hat{t}+r_{,x}(-\hat{x})]\\
&=-4{\pi}r^{2}\cdot\frac{1}{2}e^{2\sigma}(\psi_{,t}^2+\psi_{,x}^2)r_{,t}\sqrt{1+K^2}(-\hat{r})\\
&{\approx}\frac{{\Delta}m}{{\Delta}\lambda}(-\hat{r}).
\end{split}
\label{dm_dlambda}\ee
As implied in Fig.~\ref{fig:ratio_center},
\be \Delta\lambda
=\frac{\Delta t\cdot\Delta x}{\sqrt{(\Delta t)^2+(\Delta x)^2}}
\approx\frac{\Delta t}{\sqrt{1+K^2}}.\ee
Then using
\be \frac{\Delta r}{\Delta\lambda}
=\frac{\Delta r}{\Delta t}\cdot\frac{\Delta t}{\Delta \lambda}
{\approx}r_{,t}\sqrt{1+K^2},\ee
we have
\be m_{,r}\approx\frac{{\Delta}m}{{\Delta}r}
=\frac{{\Delta}m}{{\Delta}\lambda}\cdot\frac{{\Delta}\lambda}{{\Delta}r}
\approx-4{\pi}r^{2}\cdot\frac{1}{2}e^{2\sigma}(\psi_{,t}^2+\psi_{,x}^2).
\label{dm_dr}\ee

Equation~(\ref{dm_dr}) implies that in the vicinity of the central singularity, the ef{}fective density for the scalar field $\psi$ is
\be
\rho_{\scriptsize{\mbox{ef{}f}}}
=\frac{1}{2}e^{2\sigma}(\psi_{,t}^2+\psi_{,x}^2)
{\sim}r^{-3-16{\pi}C^2},
\ee
which is consistent with the integration form of the mass function~(\ref{mass_analytic}).

\subsection{Discussions}
\emph{1. In which cases is the mass function equal to the black hole mass?}

For stationary black holes, the Misner-Sharp mass function is always equal to the black hole mass. This is verified for the Schwarzschild black hole case in Appendix~\ref{sec:appendix_singularity_Schw} and for the Reissner-Nordstr\"{o}m black hole case in Sec.~\ref{sec:dynamics_RN}.

In spherical symmetry, at spatial infinity, the mass function describes the total energy/mass of an asymptotically flat spacetime~\cite{Hayward}. In a gravitational collapse case, it means the total mass of the collapsing system.

\emph{2. In which cases does the mass function not describe the black hole mass?}

In the vicinities of the central singularity of a Schwarzschild black hole and the inner horizon of a Reissner-
\\
\\
\\
Nordstr\"{o}m or Kerr black hole, the dynamics and some quantities are \emph{local}. The mass function is just a parameter which varies at each point, not giving \emph{global} information on the black hole mass.

\emph{3. How does one interpret the behaviors of the mass function at dif{}ferent circumstances?}

One may interpret this issue via an analog to Newtonian gravity. In Newtonian gravity, suppose we want to measure the mass $M$ of a source sphere with a pointlike test mass $m$. Denote the distance between the two masses by $r$. Without perturbations, the gravitational force between the two masses is simply $F=GMm/r^2$. However, if another mass $\delta M$ passes by the test mass, the gravitational field near the test mass can become very strong and local, and it is not able to provide accurate information on the source sphere. When the perturbation mass is gone or sticks to the source sphere, one can measure the mass of the source object accurately again.

Similarly, in gravitational collapse in general relativity, in the vicinity of the singularity of the formed black hole, because of the strong interaction between the scalar field and the geometry, the dynamics becomes local. Then the computed mass function does not provide global information on the mass of the black hole or of the collapsing system. As discussed in Sec.~\ref{sec:causes_mass_inflation}, at large-$x$ regions, most of the energy of the scalar field has been absorbed into the black hole, and the tail of the scalar field is very weak. As a result, although in this case the mass function does not seem to approach a fixed value, it grows very slowly. [See Fig.~\ref{fig:solutions_collapse_x_2_point_5}(d).]

\emph{4. Is there a suitable mass definition inside black holes?}

Although quite several local definitions of mass have been constructed for regular spacetime, little work has been done to provide more alternatives for global conservation laws. Considering the large amount of ef{}forts that have been spent on formulating proper quasi-local mass, it can be even more challenging to develop the ``ultimate'' mass definition inside black holes. The role of unavoidable spacetime singularities has also an ef{}fect in all of these definitions.

\section{Neutral scalar scattering\label{sec:neutral_scattering}}
In this section, we consider neutral scattering, in which a neutral scalar collapses in a Schwarzschild geometry. The numerical formalism is a simpler version of the one in charge scattering that has been constructed in Sec.~\ref{sec:set_up}, and it can be obtained by removing the electric terms in the field equations presented in Sec.~\ref{sec:field_eqs} and replacing the Reissner-Nordstr\"{o}m geometry with a Schwarzschild one.

The parameters are set as follows:
\begin{enumerate}[(i)]
  \item Schwarzschild geometry: $m=1$.
  \item Physical scalar field:

  $\psi(x,t)|_{\scriptsize{t=0}}=a\exp\left[-(x-x_{0})^2/b\right]$,

  $a=0.2$, $b=1$, and $x_{0}=1$.
  \item Grid. Spatial range: $x\in[-10~10]$. Grid spacings: ${\Delta}x={\Delta}t=0.005$.
\end{enumerate}

\begin{figure*}[t!]
  \epsfig{file=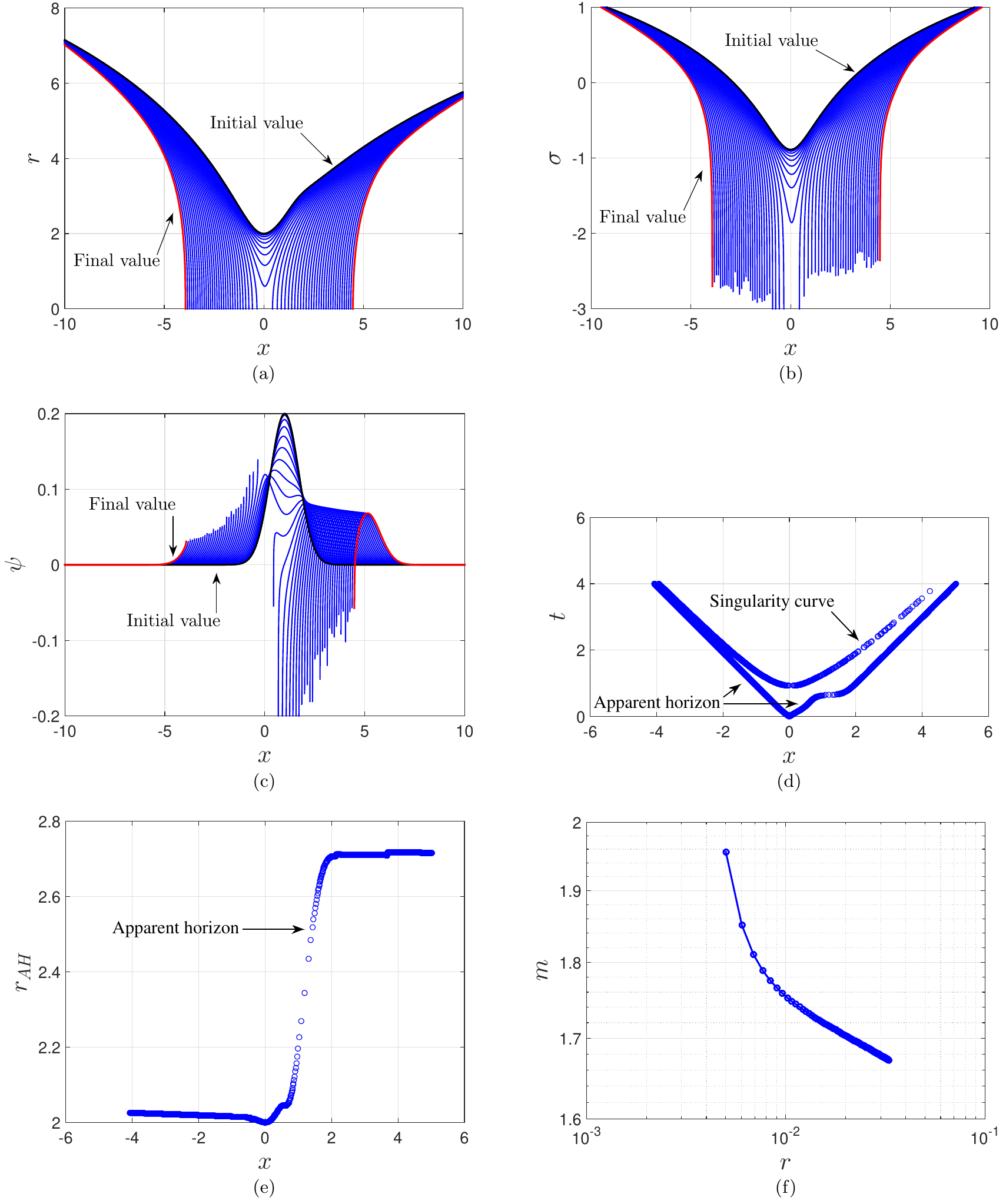, width=1\textwidth}
  \caption{Evolutions in neutral scattering. (a)-(c): evolutions of $r$, $\sigma$, and $\psi$. The time interval between two consecutive slices is $20{\Delta}t=0.1$. (d) and (e) are for the apparent horizon and the singularity curve of the formed black hole. (f) The Misner-Sharp mass function along the slice $(x=2,t=t)$.}
  \label{fig:evolutions_neutral_scattering}
\end{figure*}

The numerical results for neutral scattering are plotted in Fig.~\ref{fig:evolutions_neutral_scattering}. As shown in Fig.~\ref{fig:evolutions_neutral_scattering}(b), $\sigma$ is always negative. Our interpretations for this are the following. In neutral collapse, before the center $x=0$ changes from timelike into spacelike, because of the reflection at $x=0$ and the consecutive incoming of pulses of the scalar field from large $x$, some of the energy of the scalar field is accumulated at the central region. Then the scalar field becomes very strong and its contribution can be comparable to those from the metric quantities, i.e., $4\pi\psi_{,t}^2\sim|r_{,tt}/r|\sim|\sigma_{,tt}|$. As a result, $\sigma$ can be positive. However, in neutral scattering, there is a given Schwarzschild geometry, and the center is a spacelike singularity from the beginning. So the energy of the scalar field simply moves to the central singularity rather than being reflected. Then the energy of the scalar field has no opportunity of accumulating at the central region. Consequently, the scalar field is always weak compared to the metric quantities, and $\sigma$ remains negative.

We explore the dynamics in the vicinity of the central singularity, including the behaviors of the field equations, the metric quantities, the scalar field, and the mass function. The results are similar to those in the weak scalar field case in neutral collapse as plotted in Figs.~\ref{fig:solutions_collapse_x_2} and \ref{fig:solutions_collapse_x_2_point_5}. As a representative, we plot the Misner-Sharp mass function along the slice $(x=2,t=t)$ in Fig.~\ref{fig:evolutions_neutral_scattering}(f).

\begin{figure*}[t!]
  \epsfig{file=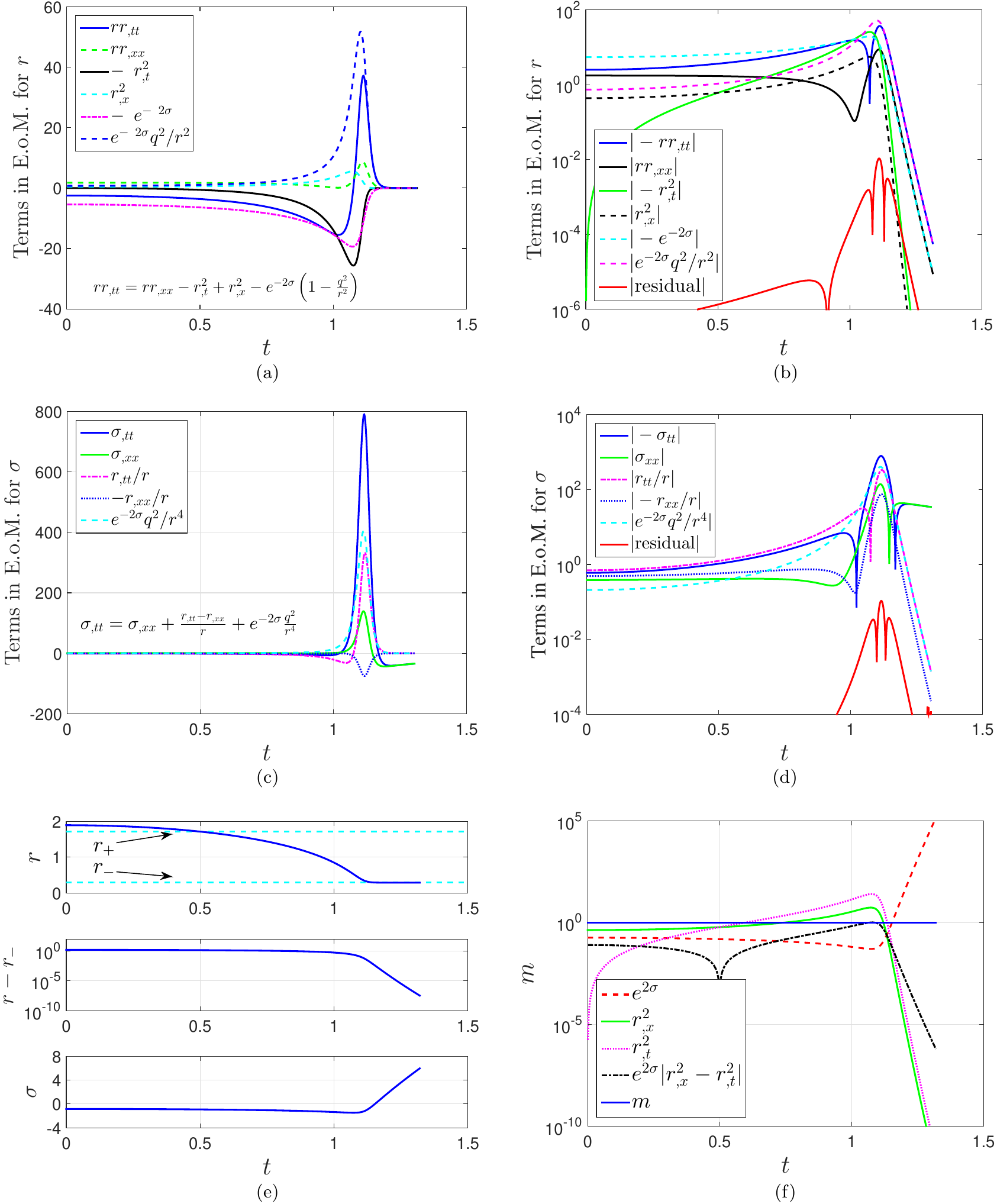, width=0.95\textwidth}
  \caption{(color online). Dynamics in a Reissner-Nordstr\"{o}m geometry on the slice $(x=0.5,t=t)$.
  (a) and (b): dynamical equation for $r$.
  (c) and (d): dynamical equation for $\sigma$.
  (e) and (f): evolutions of $r$, $\sigma$, and $m$.
  At the early stage, $r$ and $\sigma$ move toward zero and negative infinity, respectively, as in the Schwarzschild metric. Later on, due to the repulsive force from $2e^{-2\sigma}q^2/r^2$, $r$ decelerates and asymptotes to $r_{-}$. [See Figs.~(a) and (e).] Mainly because of $e^{-2\sigma}q^2/r^4$, $\sigma$ switches the sign and approaches positive infinity. [See Figs.~(c) and (e).] As the inner horizon is approached, the term $e^{2\sigma}(-r_{,t}^2+r_{,x}^2)$ asymptotes to zero, and the mass function $m$ remains equal to the mass of the Reissner-Nordstr\"{o}m black hole. [See Fig.~(f).]}
  \label{fig:dynamics_RN}
\end{figure*}

\section{Dynamics in the Reissner-Nordstr\"{o}m geometry\label{sec:dynamics_RN}}
As one more preparation for charge scattering, we discuss the dynamics in the Reissner-Nordstr\"{o}m geometry in the patch $r>r_{-}$ in Kruskal-like coordinates
\be ds^2=\frac{r_{+}r_{-}}{k_{+}^2r^{2}}e^{-2k_{+}r}\left(\frac{r}{r_{-}}-1\right)^{1+\frac{k_{+}}{|k_{-}|}}(-dt^2+dx^2)+r^2d\Omega^2,
\label{RN_metric_example}\ee
where $r$ and $\sigma$ are defined by
\begin{align}
t^2-x^2&=e^{2k_{+}r}\left(1-\frac{r}{r_{+}}\right)\left(\frac{r}{r_{-}}-1\right)^{-\frac{k_{+}}{|k_{-}|}},\label{r_RN_example}\\
\nonumber\\
e^{-2\sigma}&=\frac{r_{+}r_{-}}{k_{+}^2r^{2}}e^{-2k_{+}r}\left(\frac{r}{r_{-}}-1\right)^{1+\frac{k_{+}}{|k_{-}|}}.
\label{sigma_RN}
\end{align}
The equations of motion for $r$ and $\sigma$ are
\be r(-r_{,tt}+r_{,xx})-r_{,t}^2+r_{,x}^2 = e^{-2\sigma}\left(1-\frac{q^2}{r^2}\right),\label{eom_r_RN}\ee
\be
-\sigma_{,tt}+\sigma_{,xx}+\frac{r_{,tt}-r_{,xx}}{r}+e^{-2\sigma}\frac{q^2}{r^4}=0.
\label{eom_sigma_RN}
\ee

The mass and charge of the Reissner-Nordstr\"{o}m black hole are set as $m=1$ and $q=0.7$, respectively. We take the slice $(x=0.5,t=t)$ as a sample slice. The dynamics for $r$ and $\sigma$ and the evolutions of $r$, $\sigma$, and $m$ on this slice are plotted in Fig.~\ref{fig:dynamics_RN}. As shown in Fig.~\ref{fig:dynamics_RN}, in Eqs.~(\ref{eom_r_RN}) and (\ref{eom_sigma_RN}), around the outer horizon, the charge terms are negligible compared to other terms. The dynamics is close to that in the Schwarzschild black hole case. $\sigma$ is negative and $|\sigma|$ keeps decreasing until $r$ is close to the inner horizon. Then $|\sigma|$ decelerates and turns around. $\sigma$ approaches positive infinity, $r_{,x}$ and $r_{,t}$ approach zero, and $r$ asymptotes to $r_{-}$. The term $e^{2\sigma}(-r_{,t}^2+r_{,x}^2)$ asymptotes to zero, as plotted of in Fig.~\ref{fig:dynamics_RN}(f). Note that the Misner-Sharp mass function a charged black hole is defined as
\be g^{\mu\nu}r_{,\mu}r_{,\nu}=e^{2\sigma}(-r_{,t}^2+r_{,x}^2){\equiv}1-\frac{2m}{r}+\frac{q^2}{r^2}.\label{mass_definition_RN}\ee
Then as shown in Fig.~\ref{fig:dynamics_RN}(f), the mass function remains equal to the mass of the Reissner-Nordstr\"{o}m black hole as the inner horizon is approached as expected.

In the next section, we will show that in charge scattering, near the inner horizon, due to the additional contribution from a scalar field, $\sigma$ approaches positive infinity faster than in the Reissner-Nordstr\"{o}m black hole case. Consequently, $r$ decreases faster and then is able to cross the inner horizon. As a result, the term $e^{2\sigma}(-r_{,t}^2+r_{,x}^2)$ and the mass parameter diverge: mass inflation takes place.

\begin{figure*}[t!]
  \epsfig{file=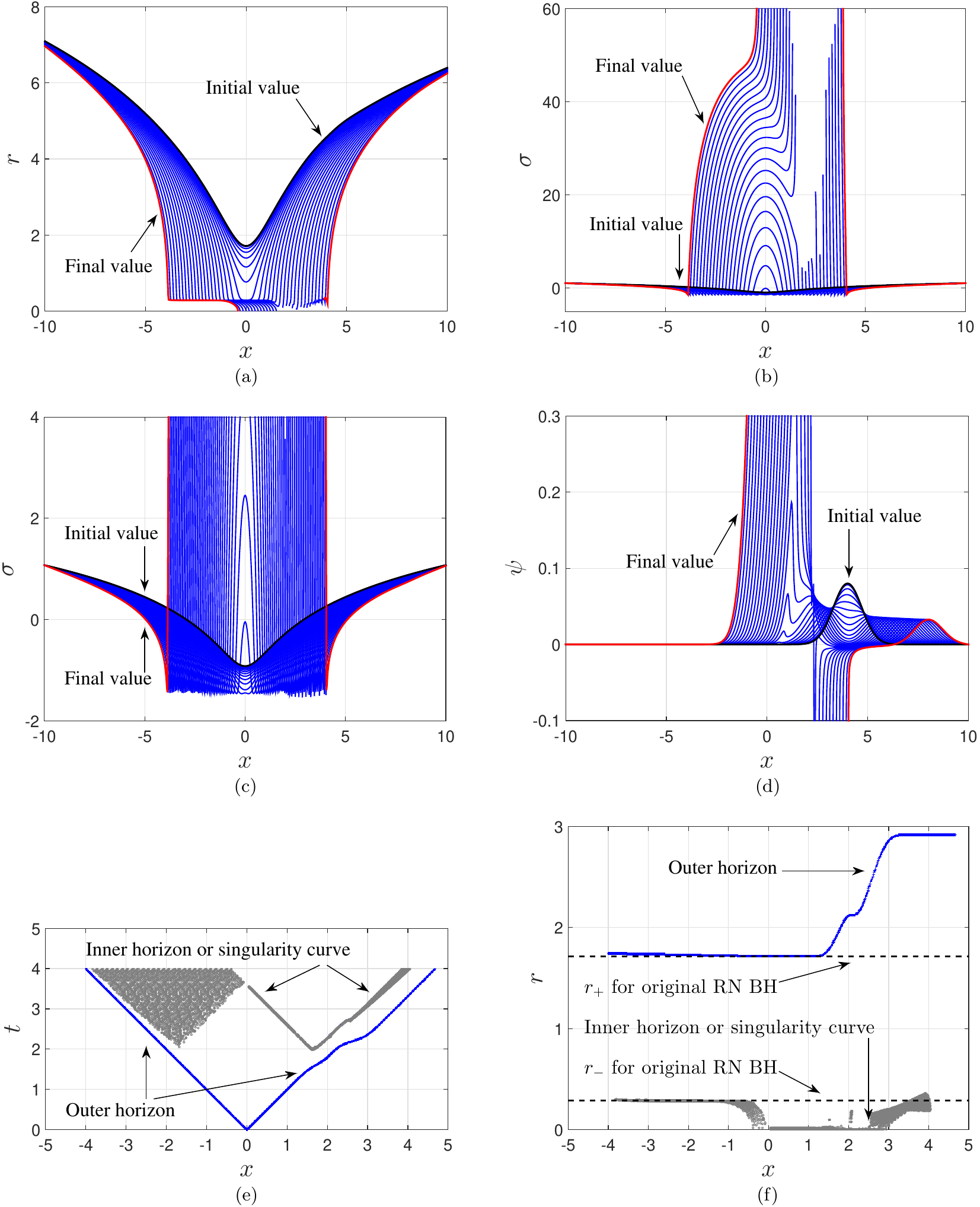, width=0.9\textwidth}
  \caption{Evolutions in charge scattering. In (a), (b), and (d), the time interval between two neighboring slices is $30{\Delta}t=0.15$. In (c), it is $15{\Delta}t=0.075$. (e) and (f) are for the horizons. When the scalar field is strong enough (around $x=2$), the inner horizon can be pushed to the center, and the central singularity becomes spacelike. When the scalar field is weak enough (e.g., $-4<x<-1$), the inner horizon does not change much, and the central singularity remains timelike. At the intermediate state (e.g., $0<x<1.5$), the inner horizon contracts to zero, and the central singularity becomes null. The results for the inner horizon, especially for $x>2$, are not that accurate. We are aware that the inner horizon is actually at infinity, while $r$ still can be very close to $r_{-}$ when $x$ and $t$ take moderate values.}
  \label{fig:evolutions}
\end{figure*}

\newpage
\be \nonumber\ee
\be \nonumber\ee
\be \nonumber\ee
\be \nonumber\ee
\be \nonumber\ee
\be \nonumber\ee

\section{Results for charge scattering\label{sec:results_scattering}}
In this section, we explore charge scattering: scalar field collapse in a Reissner-Nordstr\"{o}m geometry. We study the evolutions of the metric components and scalar field, and obtain approximate analytic solutions. We closely compare the dynamics in Schwarzschild black holes, Reissner-Nordstr\"{o}m black holes, neutral scalar collapse, and charge scattering. In addition, in obtaining the analytic solutions, we compare the analytic candidate solutions to the numerical results at each step.

In this section, the parameters are set as follows:
\begin{enumerate}[(i)]
  \item Reissner-Nordstr\"{o}m geometry: $m=1$, and $q=0.7$.
  \item Physical scalar field:

  $\psi(x,t)|_{\scriptsize{t=0}}=a\exp\left[-(x-x_{0})^2/b\right]$,

  $a=0.08$, $b=1$, and $x_{0}=4$.
  \item Grid. In Sec.~\ref{sec:evolutions}, we use the spatial range of $x\in[-10~10]$ and grid spacings of ${\Delta}x={\Delta}t=0.005$. In Secs.~\ref{sec:spacelike}-\ref{sec:critical_scattering}, we use the spatial range of $x\in[-5~5]$ and grid spacings of ${\Delta}x={\Delta}t=0.0025$.
\end{enumerate}

\subsection{Evolutions\label{sec:evolutions}}
\subsubsection{Outline}
In this subsection, we describe the evolutions of $r$, $\sigma$, and $\psi$ that are plotted in Fig.~\ref{fig:evolutions}. Examining the equations of motion (\ref{equation_r})-(\ref{equation_psi}), one can see that in the charge scattering dynamical system, there are three types of quantities as follows:
\begin{enumerate}[(i)]
  \item Metric components: $r$ and $\sigma$. They contribute as gravity.
  \item Scalar field: $\psi$. It contributes as a self-gravitating field.
  \item Electric field. It acts as a repulsive force. See Eq.~(\ref{equation_r}).
\end{enumerate}
Furthermore, these quantities can be separated into two sides: the gravitating side ($r$, $\sigma$, and $\psi$) and the repulsive side (electric field). The dynamics in charge scattering consists mainly of how these quantities interact, including how the gravitating and repulsive sides compete.

According to the strength of the scalar field, charge scattering can be classified as follows:
\begin{enumerate}[(i)]
  \item Type I: spacelike scattering. When the scalar field is very strong, the inner horizon can contract to zero volume rapidly, and the central singularity becomes spacelike. Sample slice: $(x=1.67,t=t)$ in Fig.~\ref{fig:evolutions}. See Sec.~\ref{sec:spacelike}.
  \item Type II: null scattering. When the scalar field is intermediate, the inner horizon can contract to a place close to the center or reach the center. In this type of mass inflation, for each quantity, the spatial and temporal derivatives are almost equal. In the case of the center being reached, the central singularity becomes null. This type of scattering has two stages: early/slow and late/fast. In the early stage, the inner horizon contracts slowly; while in the late stage, the inner horizon contracts rapidly. Sample slice: $(x=0.5,t=t)$ in Fig.~\ref{fig:evolutions}. See Secs.~\ref{sec:fast_stage} and \ref{sec:slow_stage}.
  \item Type III: critical scattering. This case is on the edge between the above two cases. When the central singularity is reached, it becomes null. Sample slice: $(x=1.53,t=t)$ in Fig.~\ref{fig:evolutions}. See Sec.~\ref{sec:critical_scattering}.
  \item Type IV: weak scattering. When the scalar field is very weak, the inner horizon does not contract much.
        Sample case: Figs.~\ref{fig:weak_scattering_evolution} and \ref{fig:weak_scattering_eom}. See Sec.~\ref{sec:weak_scattering}.
  \item Type V: tiny scalar scattering. When the scalar field is very tiny, the influence of the scalar field on the geometry is negligible. Sample slice: $(x=-3,t=t)$ in Fig.~\ref{fig:evolutions}.
\end{enumerate}
In this paper, we will discuss the first three types.

\subsubsection{Causes of mass inflation and evolutions of $r$, $\sigma$, and $\psi$}
In a Reissner-Nordstr\"{o}m geometry, in Kruskal-like coordinates expressed by Eq.~(\ref{RN_metric_example}), near the inner horizon, although $\sigma$ asymptotes to positive infinity, $(r_{,t}^2-r_{,x}^2)$ is much less than $e^{-2\sigma}$. Consequently, $e^{2\sigma}(r_{,t}^2-r_{,x}^2)$ approaches zero. Then as implied in Eq.~(\ref{mass_definition_RN}), $m$ takes a finite value.

\begin{figure*}[t!]
  \epsfig{file=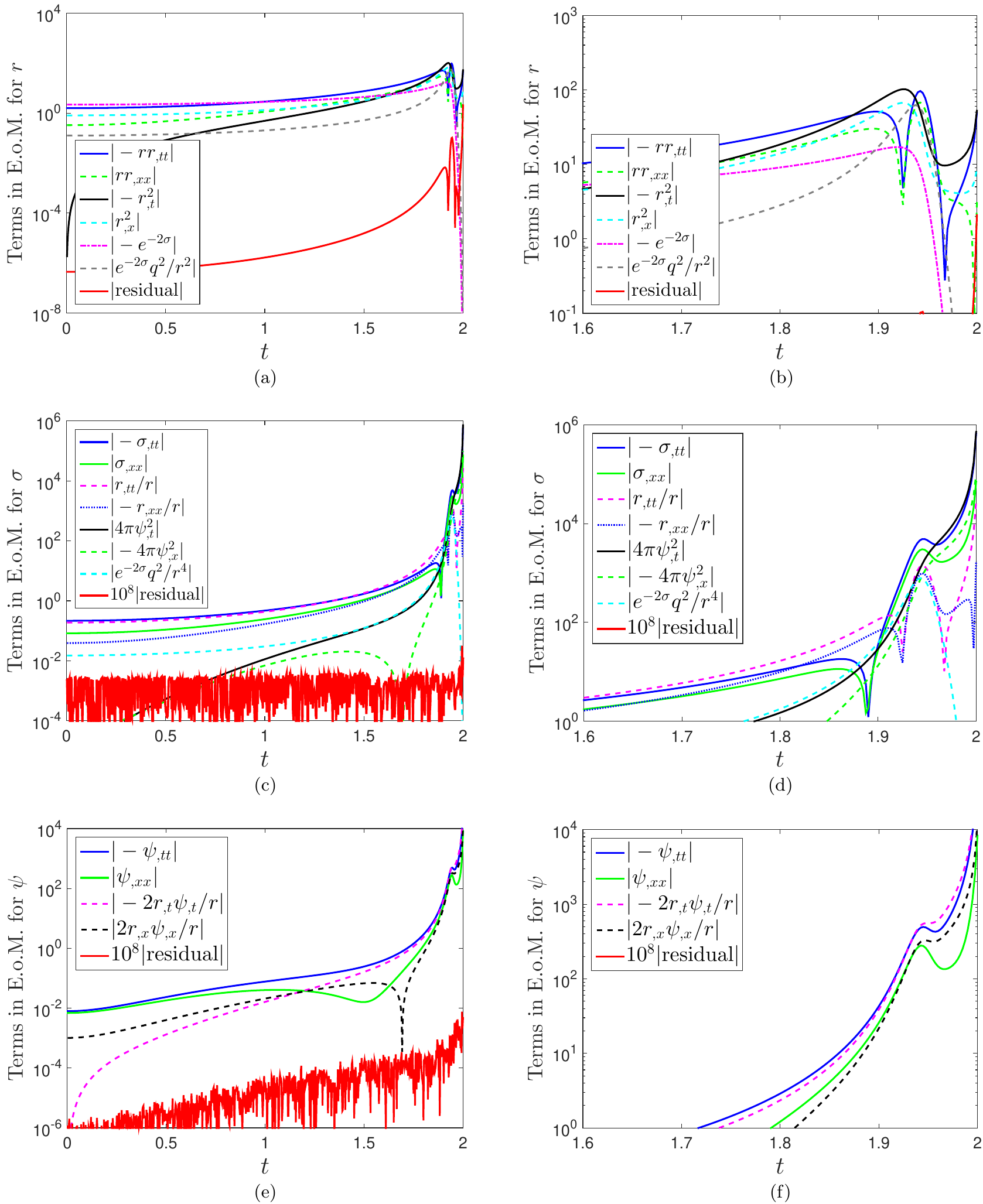, width=1\textwidth}
  \caption{(color online). Dynamics along the slice $(x=1.67,t=t)$ in spacelike charge scattering.
  (a) and (b), (c) and (d), (e) and (f): dynamical equations for $r$, $\sigma$, and $\psi$. In this configuration, the scalar field is so strong, such that $r$ does not decelerate much when it crosses the inner horizon of the given Reissner-Nordstr\"{o}m geometry.}
  \label{fig:dynamics_spacelike_singularity}
\end{figure*}

\begin{figure*}[t!]
  \epsfig{file=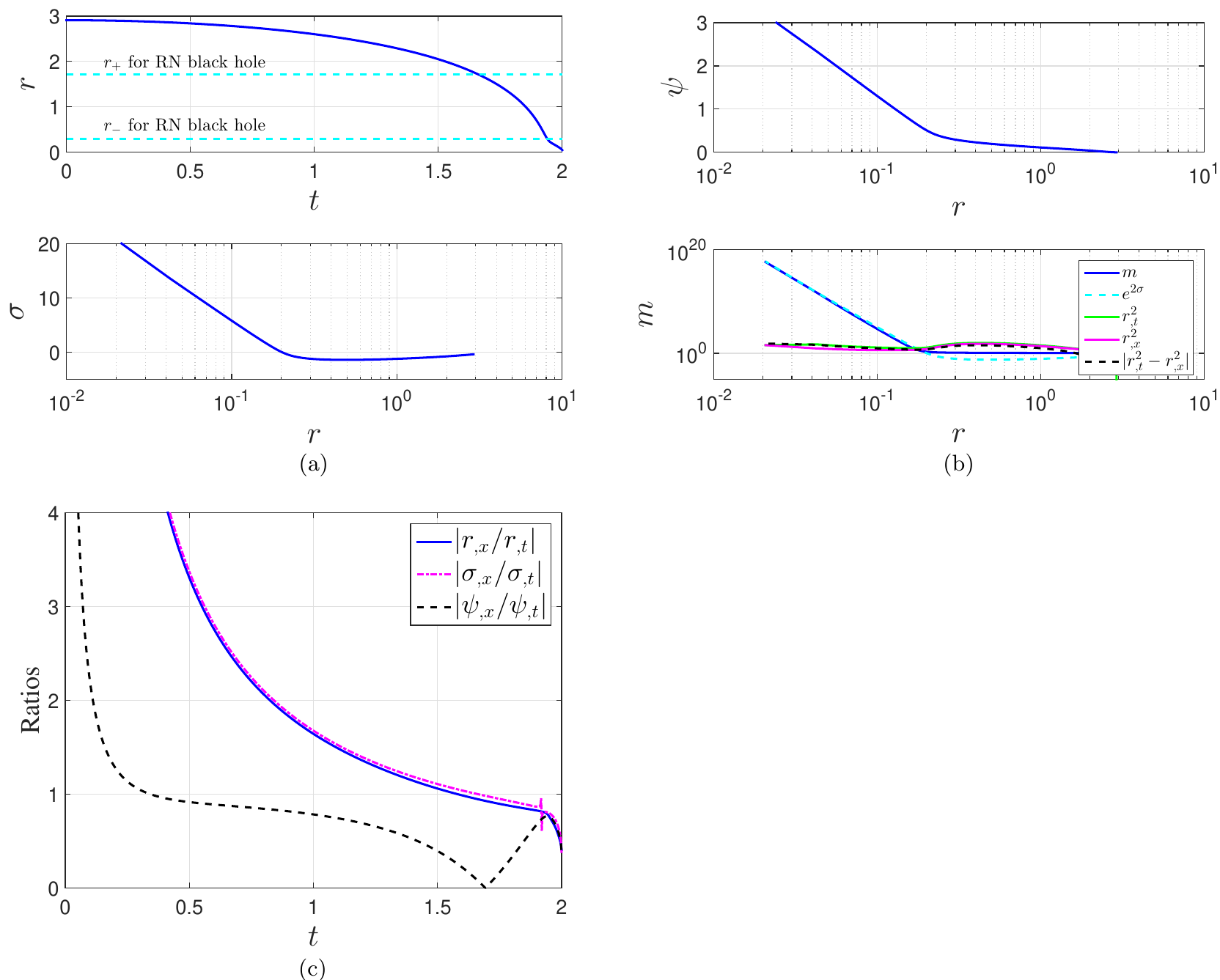, width=0.9\textwidth}
  \caption{(color online). Evolutions along the slice $(x=1.67,t=t)$ in spacelike charge scattering.
  (a) evolutions of $r$ and $\sigma$.
  (b) evolutions of $\psi$ and $m$. The mass function remains equal to the mass of the original Reissner-Nordstr\"{o}m black hole, $m=1$, until $t\approx1.9$, and by then mass inflation takes place.
  (c) ratios between spatial and temporal derivatives for $r$, $\sigma$, and $\psi$. When $t>1.9$, there is $|\psi_{,x}/\psi_{,t}|\approx|r_{,x}/r_{,t}|$.}
  \label{fig:evolutions_spacelike_singularity}
\end{figure*}

In charge scattering, the equations of motion for $\eta{\equiv}r^2$ and $\sigma$ are
\be -\eta_{,tt}+\eta_{,xx}=2e^{-2\sigma}\left(1-\frac{q^2}{r^2}\right),\label{equation_r_scattering}\ee
\be
-\sigma_{,tt}+\sigma_{,xx} + \frac{r_{,tt}-r_{,xx}}{r}+4\pi(\psi_{,t}^2-\psi_{,x}^2)+e^{-2\sigma}\frac{q^2}{r^4} = 0.
\label{equation_sigma_scattering}
\ee
At the beginning, under our initial conditions, $|\psi_{,t}|$ is less than $|\psi_{,x}|$. However, as discussed in Sec.~\ref{sec:asymptotics_collapse}, as $r$ decreases toward the central singularity, gravity becomes stronger, and $|\psi_{,t}|$ becomes greater than $|\psi_{,x}|$.
(Also see Figs.~\ref{fig:dynamics_spacelike_singularity}-\ref{fig:dynamics_null}.) Then in this case, the repulsive \lq\lq force,\rq\rq $4\pi(\psi_{,t}^2-\psi_{,x}^2)+e^{-2\sigma}q^2/r^4$, is greater than the corresponding one, $e^{-2\sigma}q^2/r^4$, in a Reissner-Nordstr\"{o}m geometry. This makes $\sigma$ accelerate faster than in the Reissner-Nordstr\"{o}m geometry. Consequently, the repulsive force from $2e^{-2\sigma}(q^2/r^2-1)$ for $\eta{\equiv}r^2$ is much weaker than the corresponding value in the Reissner-Nordstr\"{o}m geometry. As a result, near the inner horizon, $|r_{,t}|$ is much greater than the corresponding one in the Reissner-Nordstr\"{o}m case. Then $(r_{,t}^2-r_{,x}^2)$ moves from extremely tiny values in the Reissner-Nordstr\"{o}m metric case to moderate values, and $r$ crosses the inner horizon $r=r_{-}$ for the given Reissner-Nordstr\"{o}m geometry. [See Figs.~\ref{fig:dynamics_RN}(e) and \ref{fig:evolutions_spacelike_singularity}(a).] This makes the mass parameter diverge: mass inflation occurs. [See Eq.~(\ref{mass_definition_RN}).] In other words, regarding the causes of mass inflation in charge scattering, the scalar field's backreaction on $r$ is more important than that on $\sigma$. The evolutions of $r$, $\sigma$, and $\psi$ are plotted in Figs.~\ref{fig:evolutions}(a)-\ref{fig:evolutions}(d).

\subsubsection{Locations of horizons}
In charge scattering, on the horizon, there is
\be g^{\mu\nu}r_{,\mu}r_{,\nu}=e^{2\sigma}(-r_{,t}^2+r_{,x}^2){\equiv}1-\frac{2m}{r}+\frac{q^2}{r^2}=0.\ee
We locate the outer and inner horizons using this property. The results are plotted in Figs.~\ref{fig:evolutions}(e) and \ref{fig:evolutions}(f). Because of the absorption of the physical scalar field $\psi$, the outer horizon increases from the original value of $1.7$ to $2.9$. Note that the results for the inner horizon, especially at regions where $x>2$, are not that accurate. We are aware that the inner horizon is actually at infinity, while $r$ still can be very close to $r_{-}$ even when $x$ and $t$ take moderate values.

\subsection{Spacelike scattering\label{sec:spacelike}}
\begin{figure*}[t!]
  \epsfig{file=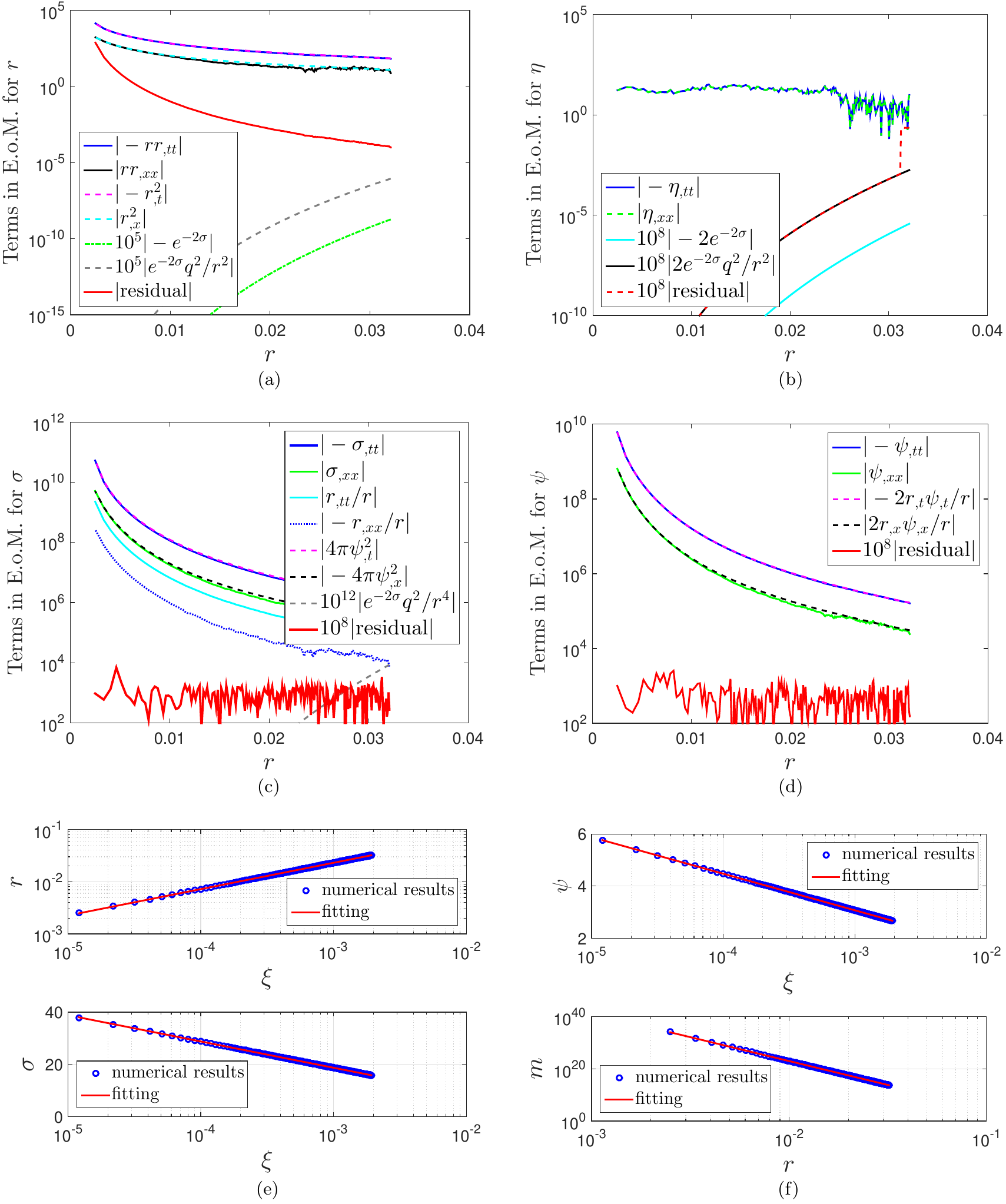, width=0.95\textwidth}
  \caption{(color online). Dynamics and solutions in the vicinity of the spacelike central singularity on the slice $(x=1.67,t=t)$ in spacelike charge scattering. In the equations of motion, in the vicinity of the singularity, the terms related to the electric field are negligible. Therefore, the equations of motion behave similarly to those in neutral scalar collapse that are plotted in Fig.~\ref{fig:dynamics_collapse_x_1}.
  (a) dynamical equation for $r$: $rr_{,tt}{\approx}-r_{,t}^2$,
  (b) for $\eta$: $\eta_{,tt}\approx\eta_{,xx}$,
  (c) for $\sigma$: $\sigma_{,tt}\approx4\pi\psi_{,t}^2$,
  (d) for $\psi$: $\psi_{,tt}\approx-2r_{,t}\psi_{,t}/r$.
  (e) $\ln r=a\ln\xi+b$, $a=0.5047\pm0.0002$, $b=-0.286\pm0.001$.
      $\sigma=a\ln\xi+b$, $a=-4.373\pm0.002$, $b=-11.50\pm0.01$.
  (f) $\psi=a\ln\xi+b$, $a=-0.6087\pm0.0002$, $b=-1.131\pm0.002$.
      $\ln m=a\ln r+b$, $a=-18.307\pm0.004$, $b=-31.39\pm0.01$.}
  \label{fig:spacelike_singularity_AMR}
\end{figure*}

In this configuration, the scalar field is so strong, such that the inner horizon can contract to zero volume rapidly, and the central singularity converts from timelike into spacelike. Taking the slice $(x=1.67,t=t)$ as a sample slice, we plot the terms in the field equations for $r$, $\sigma$, and $\psi$ in Fig.~\ref{fig:dynamics_spacelike_singularity}, and also the evolutions of $r$, $\sigma$, $\psi$, and $m$ in Fig.~\ref{fig:evolutions_spacelike_singularity}. We investigate the dynamics in the vicinity of the central singularity via mesh refinement, and plot the results in Fig.~\ref{fig:spacelike_singularity_AMR}.

The strongness of the scalar field causes several consequences as below.
\begin{enumerate}[(i)]
  \item The quantity $r$ does not decelerate much when it crosses the inner horizon of the given Reissner-Nordstr\"{o}m black hole, and it can approach the center. At the same time, the mass function grows dramatically. [See Fig.~\ref{fig:evolutions_spacelike_singularity}.]
  \item The central singularity converts from timelike into spacelike.
  \item The dynamics in spacelike scattering is similar to that in neutral, strong scalar collapse. The quantity $\sigma$ takes large positive values, such that in the vicinity of the central singularity, the term $e^{-2\sigma}q^2/r^2$ in the equation of motion for $r$~(\ref{equation_r}) and the term $e^{-2\sigma}q^2/r^4$ in the equation of motion for $\sigma$~(\ref{equation_sigma}) are negligible, compared to other terms in the equations. As a result, in the vicinity of the central singularity, the dynamics is similar to that in neutral, strong scalar collapse as expressed by Eqs.~(\ref{equation_r_asymptotic})-(\ref{equation_psi_asymptotic}).
      [See Figs.~\ref{fig:spacelike_singularity_AMR}(a)-\ref{fig:spacelike_singularity_AMR}(d).] Then the quantities $r$, $\sigma$, $\psi$, and $m$ take similar forms as those in neutral collapse and neutral scattering as expressed by Eqs.~(\ref{r_asymptotic_collapse})-(\ref{beta_asymptotic_collapse}) and (\ref{mass_analytic}). [See Figs.~\ref{fig:spacelike_singularity_AMR}(e) and \ref{fig:spacelike_singularity_AMR}(f).]
\end{enumerate}

\subsection{The late/fast stage of null scattering\label{sec:fast_stage}}
When the scalar field is less strong, the inner horizon may still contract to zero. However, in this case, the central singularity becomes null rather than spacelike. The equations of motion remain null: they have similar forms as free wave equations, e.g., $\psi_{,tt}\approx\psi_{,xx}$.

In the Reissner-Nordstr\"{o}m black hole case, near the center, the repulsive (electric) force dominates gravity, and the central singularity is timelike. In spacelike scattering as discussed in the last subsection, gravity from the scalar field and the background geometry dominates the repulsive force. As a result, the central singularity is spacelike. At the late stage of the null scattering that will be studied in this subsection, the scalar field is less strong, and the central singularity is null. Because of this, one may say that the null scattering is a critical case of the competition between repulsive and gravitational forces, in which case the two types of forces have a balance.

\begin{figure*}[t!]
  \epsfig{file=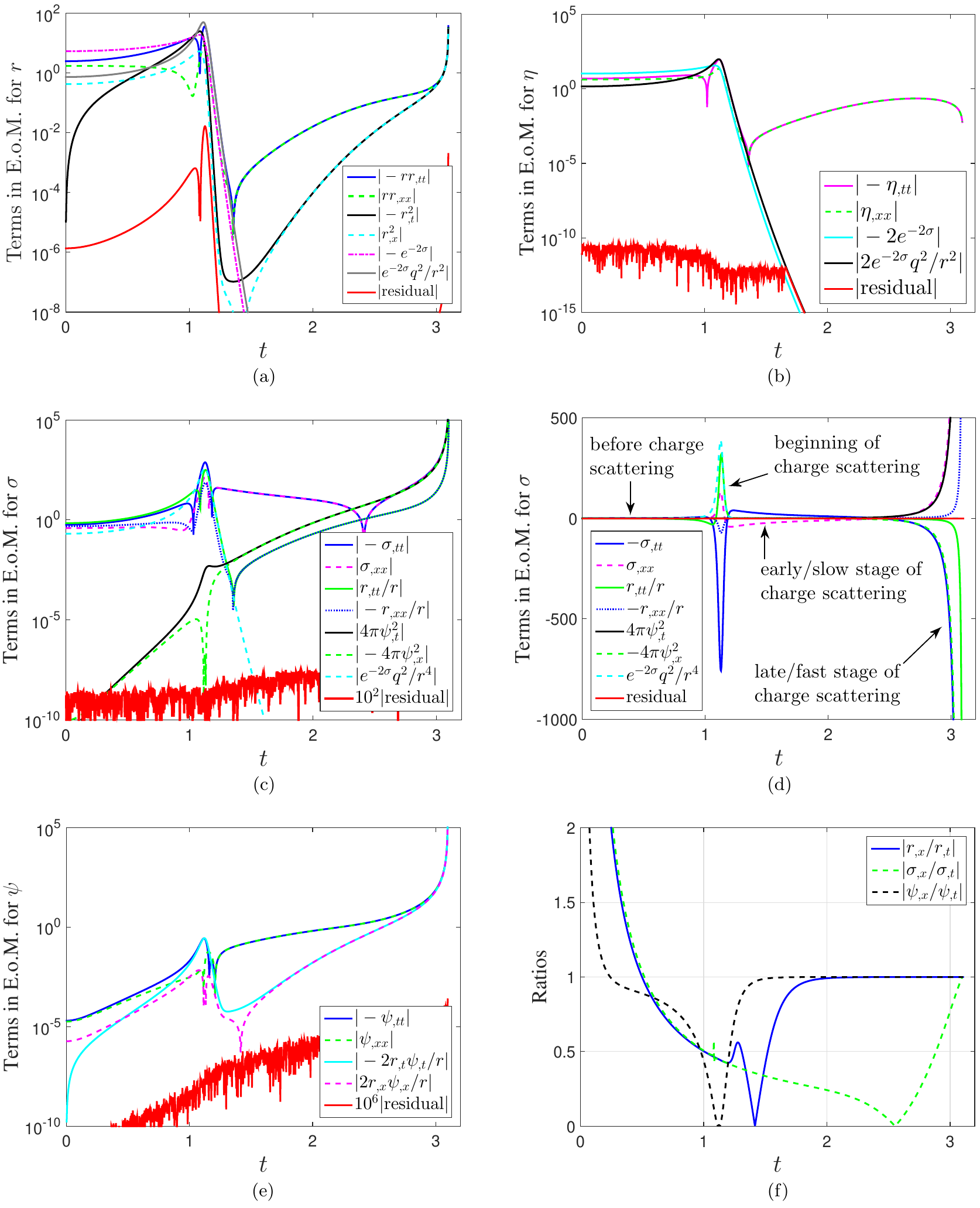, width=0.97\textwidth}
  \caption{(color online). Dynamics on the slice $(x=0.5,t=t)$ in null charge scattering. When the mass inflation happens, the field equations become null, in the sense that the temporal and spatial derivatives are almost equal. For example, in (a), $r_{,tt}{\approx}r_{,xx}$ and $r_{,t}^2{\approx}r_{,x}^2$. Near the center,
  (a) $-rr_{,tt}\approx-rr_{,xx}{\approx}r_{,t}^2{\approx}r_{,x}^2$.
  (b) $\eta_{,tt}\approx\eta_{,xx}$.
  (c) and (d): $\sigma_{,tt}\approx\sigma_{,xx}\approx4\pi\psi_{,t}^2\approx4\pi\psi_{,x}^2$.
  (e) $\psi_{,tt}\approx\psi_{,xx}\approx-2r_{,t}\psi_{,t}/r\approx-2r_{,x}\psi_{,x}/r$.
  (f) ratios between spatial and temporal derivatives for $r$, $\sigma$, and $\psi$.}
  \label{fig:dynamics_null}
\end{figure*}

\begin{figure*}[t!]
  \epsfig{file=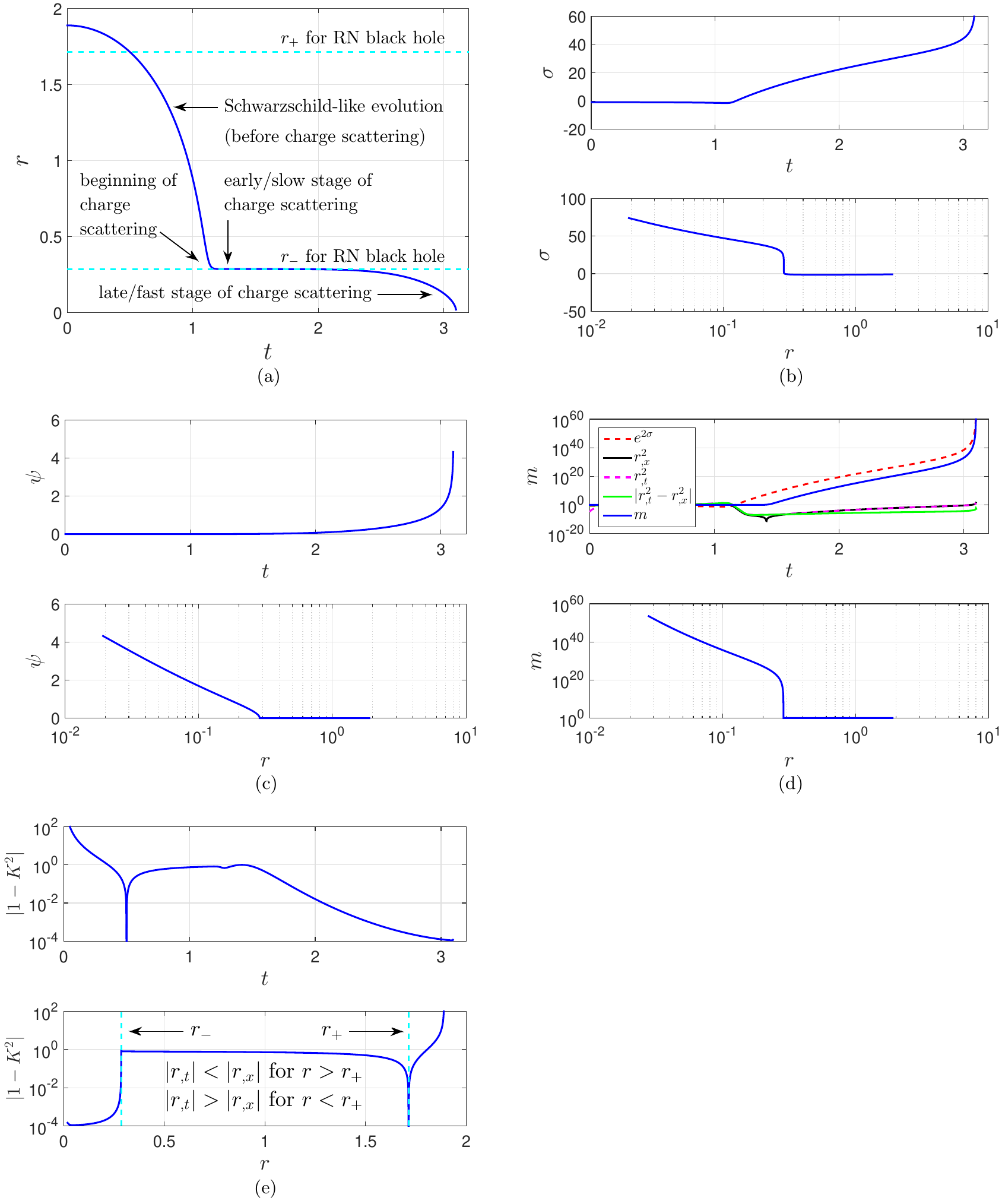, width=1\textwidth}
  \caption{Evolutions in null charge scattering on the slice $(x=0.5,t=t)$. (a)-(e): evolutions for $r$, $\sigma$, $\psi$, $m$, and $|1-K^2|$. At the early stage of mass inflation, $1.2<t<2$, $r$ varies slowly; while at the late stage, $t>2$, $r$ varies faster and faster toward a small value close to zero.}
  \label{fig:evolutions_null}
\end{figure*}

\begin{figure*}[t!]
  \epsfig{file=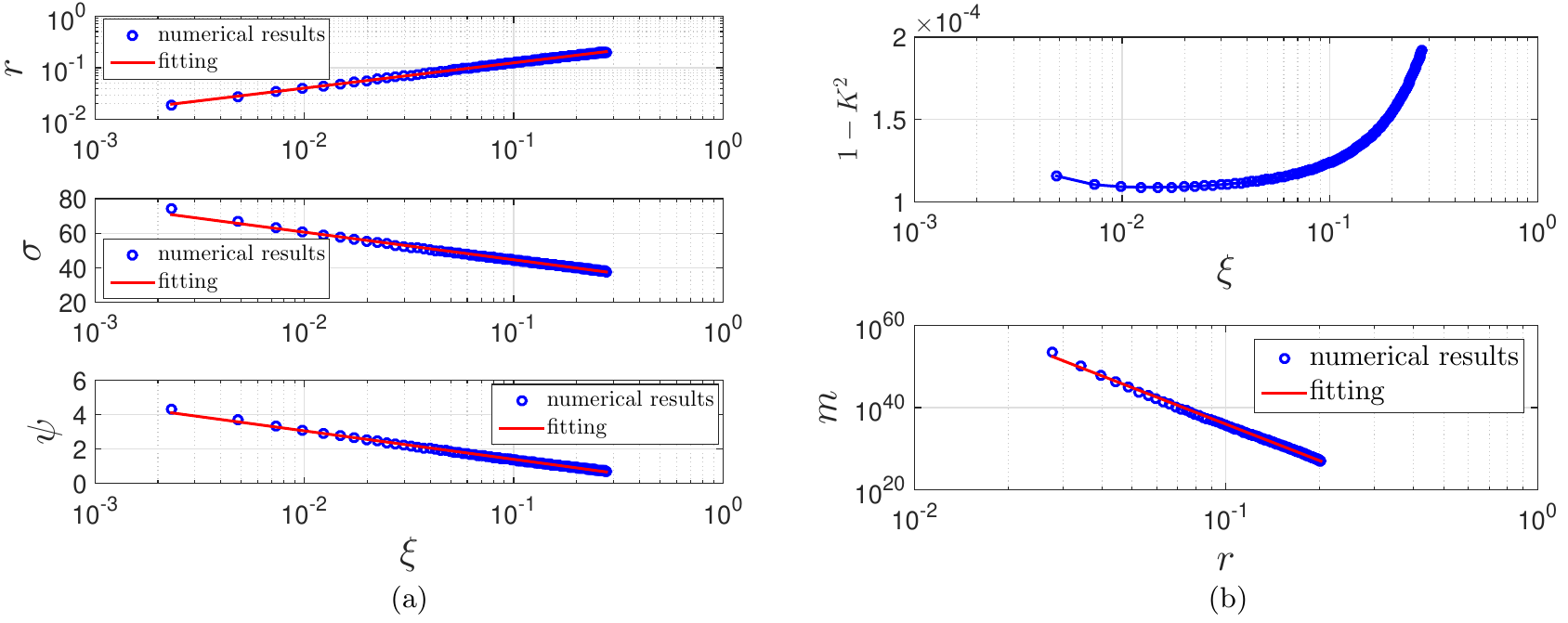, width=1\textwidth}
  \caption{Evolutions at the late/fast stage of null charge scattering on the slice $(x=0.5,t=t)$.
  (a) $\ln r=a\ln\xi+b$, $a=0.4893\pm0.0008$, $b=-0.956\pm0.002$.
      $\sigma=a\ln\xi+b$, $a=-6.95\pm0.05$, $b=28.5\pm0.1$.
      $\psi=a\ln\xi+b$, $a=-0.717\pm0.004$, $b=-0.24\pm0.01$.
  (b) $\ln m=a\ln r+b$, $a=-29.5\pm0.2$, $b=15.0\pm0.4$.}
  \label{fig:evolutions_null_fast}
\end{figure*}

Take the slice $(x=0.5,t=t)$ as a sample slice, we plot the terms in the field equations for $r$, $\sigma$, and $\psi$ in Fig.~\ref{fig:dynamics_null}, and the evolutions of $r$, $\sigma$, $\psi$, $m$, and $|1-K^2|$ in Fig.~\ref{fig:evolutions_null}. We investigate the dynamics in the vicinity of the central singularity via mesh refinement and plot the results in Fig.~\ref{fig:evolutions_null_fast}.

The null scattering has two stages: early/slow and late/fast. As shown in Figs.~\ref{fig:dynamics_null} and \ref{fig:evolutions_null}, right after the strong collision between the scalar field and the inner horizon, because of the repulsive force from the electric field and the tension force from spatial derivatives, $r$, $\sigma$, and $\psi$ change slowly. As a result, the mass function $m$ also grows slowly. We call this stage the early/slow stage. Later on, as $r$ approaches zero, maybe because of strong gravity from the central singularity as in the spacelike scattering case, these quantities change faster. We call this stage the late/fast stage.

As shown in Fig.~\ref{fig:dynamics_null}, when $r$ is very small, the equations of motion for $r$~(\ref{equation_r}), $\eta$~(\ref{equation_r_2}),
$\sigma$~(\ref{equation_sigma}), and $\psi$~(\ref{equation_psi}) can be rewritten as
\be -rr_{,tt}\approx-rr_{,xx}{\approx}r_{,t}^2{\approx}r_{,x}^2,\label{equation_r_null_late}\ee
\be \eta_{,tt}\approx\eta_{,xx},\label{equation_eta_null_late}\ee
\be \sigma_{,tt}\approx\sigma_{,xx}\approx4\pi\psi_{,t}^2\approx4\pi\psi_{,x}^2,\label{equation_sigma_null_late}\ee
\be \nonumber\ee
\be \nonumber\ee
\be \psi_{,tt}\approx\psi_{,xx}\approx-\frac{2}{r}r_{,t}\psi_{,t}\approx-\frac{2}{r}r_{,x}\psi_{,x}.\label{equation_psi_null_late}\ee
Since the above four equations have some similarities to the corresponding ones in spacelike scattering, one may guess that the quantities $r$, $\sigma$, $\psi$, and $m$ may have expressions similar to those in spacelike scattering. In fact, this guess is verified by the numerical results plotted in Fig.~\ref{fig:evolutions_null_fast}.

In the spacelike scattering case, near the central singularity, the ratio $|r_{,x}/r_{,t}|$ approaches a fixed value---the slope of the singularity curve, which is dif{}ferent from $1$. Therefore, we can obtain the approximate analytic expression for $(r_{,t}^2-r_{,x}^2)$ and then for the mass function $m$. However, in the null scattering case, $|1-K^2|$ asymptotes to zero. We plot the numerical results of $|1-K^2|$ in Fig.~\ref{fig:evolutions_null_fast}(b), while we do not derive the approximate analytic expression for $|1-K^2|$. The good thing is that in the diverging mass function, compared to the factor $e^{2\sigma}$, the term $|1-K^2|$ is a minor one.

\begin{figure*}[t!]
  \epsfig{file=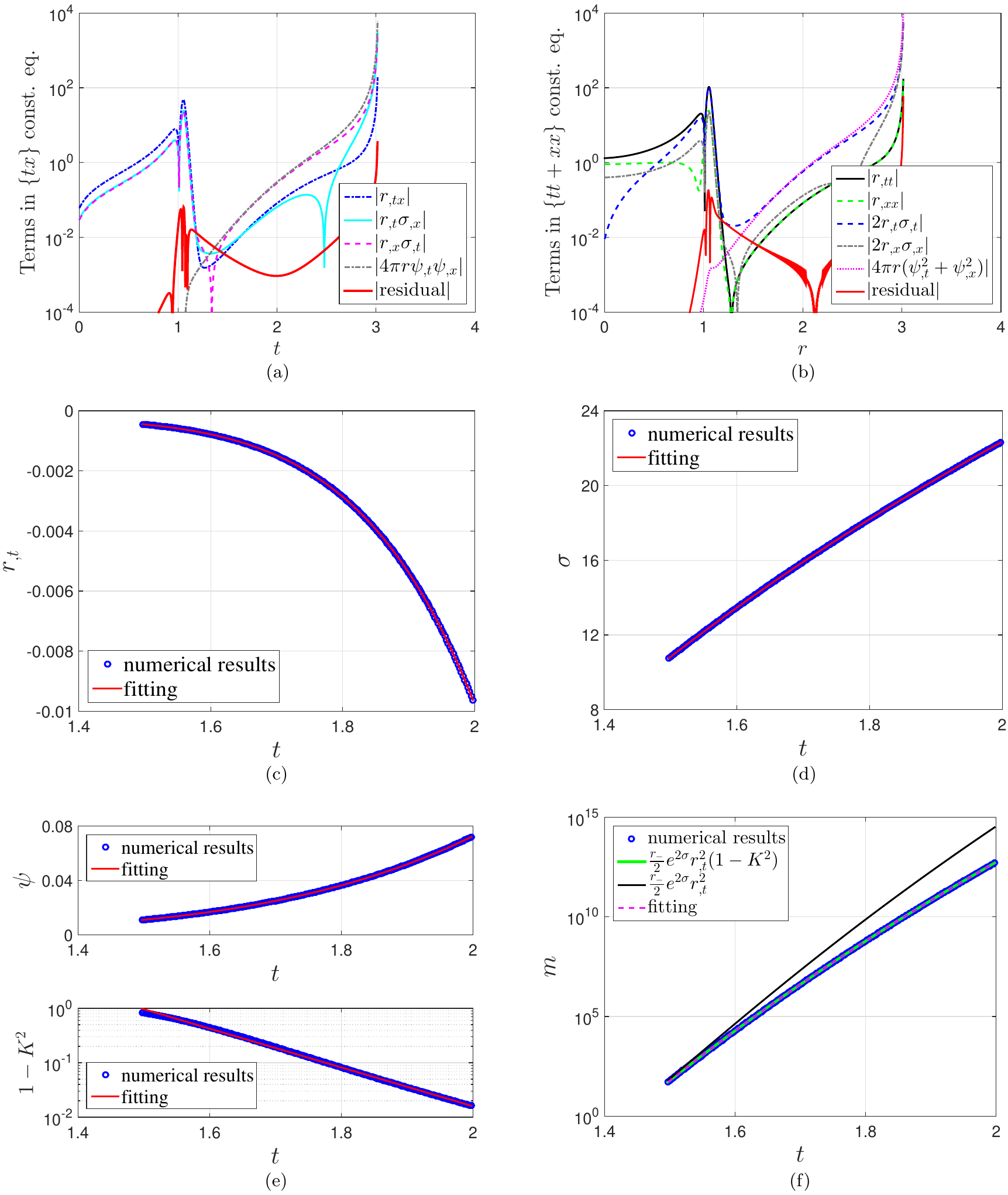, width=0.96\textwidth}
  \caption{(color online). Numerical results along the slice $(x=0.5,t=t)$ in null charge scattering.
  (a) is for the $\{tx\}$ constraint equation~(\ref{constraint_eq_xt}), and (b) is for the $\{tt\}+\{xx\}$ one~(\ref{constraint_eq_xx_tt}). (a) and (b) provide some useful information on the connections between some quantities at the early stage of charge scattering: $r_{,t}\sigma_{,t}\approx4{\pi}r_{-}\psi_{,t}^2$.
  (c)-(f): evolutions at the early/slow stage of null charge scattering.
  (c) $r_{,t}=a(t+b)^c+d$, $a=(-2.38\pm0.02)\times10^{-3}$, $b=-0.786\pm0.001$,
      $c=7.164\pm0.007$, $d=(-2.395\pm0.004)\times10^{-4}$.
  (d) $\sigma=a\ln(t+b)+c$, $a=34.11\pm0.02$, $b=-0.256\pm0.001$, $c=3.38\pm0.03$.
  (e) $\psi=a(t+b)^c+d$, $a=(2.72\pm0.06)\times10^{-3}$, $b=-0.245\pm0.004$,
      $c=5.82\pm0.01$, $d=(9.3\pm0.1)\times10^{-4}$.
      $\ln(1-K^2)=at+b$, $a=-8.21\pm0.02$, $b=12.29\pm0.04$.
  (f) $\ln m=a\ln(t+b)+c$, $a=77.52\pm0.05$, $b=-0.201\pm0.001$, $c=-16.15\pm0.08$.}
  \label{fig:evolutions_null_slow}
\end{figure*}

\begin{figure*}[t!]
  \epsfig{file=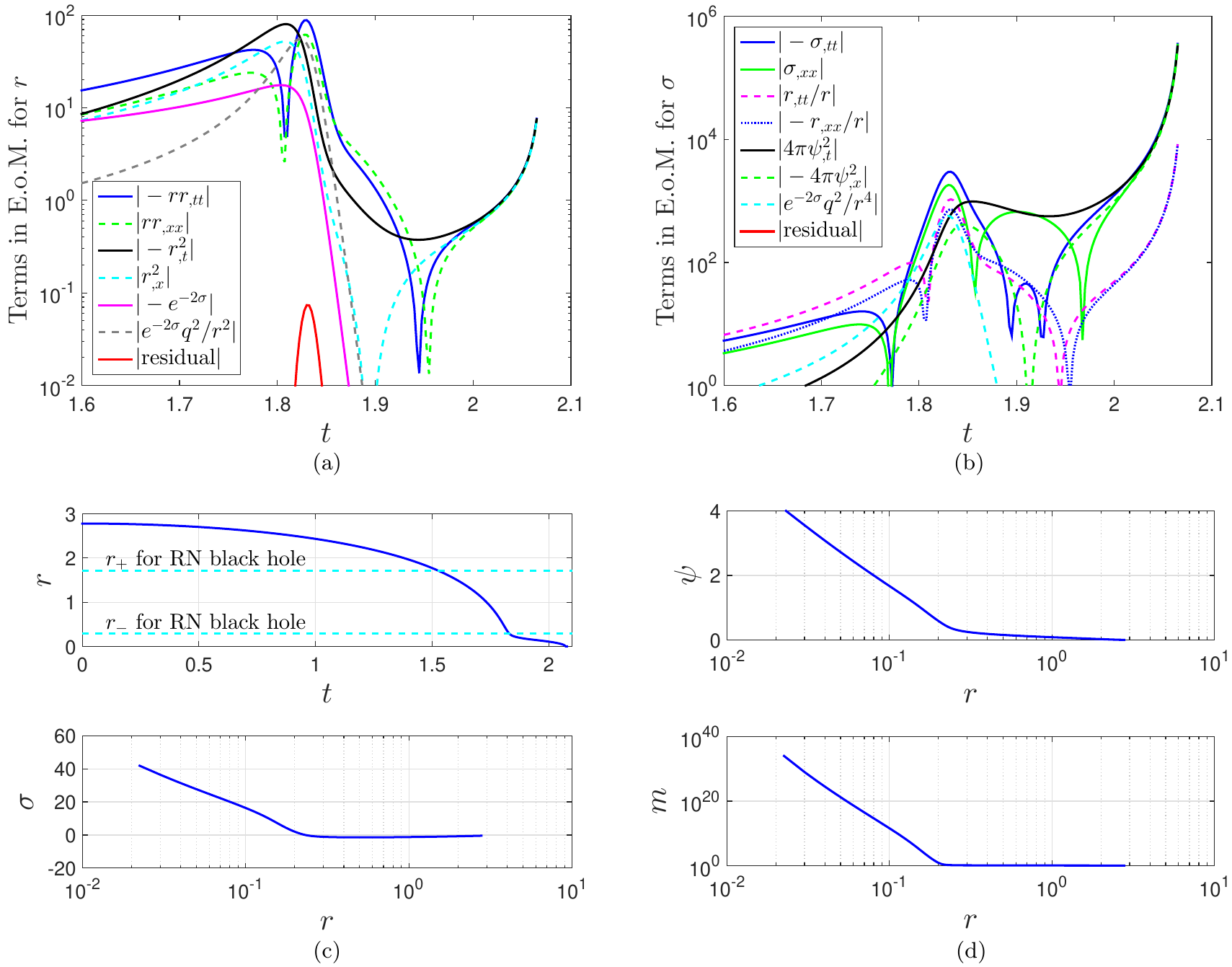, width=1\textwidth}
  \caption{(color online). Dynamics and solutions in critical charge scattering on the slice $(x=1.53,t=t)$.
  (a) and (b): dynamical equations for $r$ and $\sigma$.
  (c): evolutions of $r$ and $\sigma$.
  (d): evolutions of $\psi$ and $m$.}
  \label{fig:dynamics_critical}
\end{figure*}

\begin{figure*}[t!]
  \epsfig{file=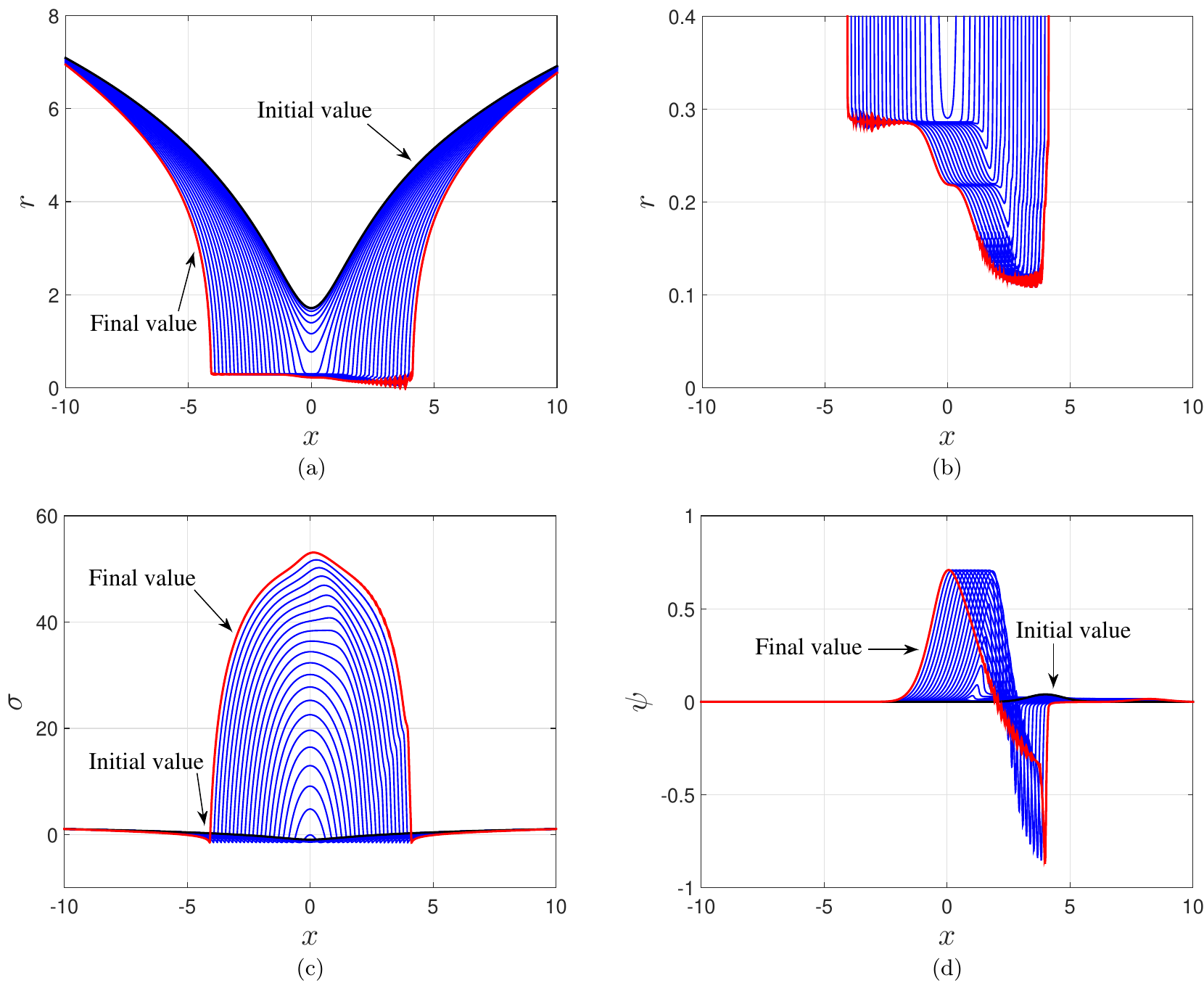, width=1\textwidth}
  \caption{Evolutions in weak scalar charge scattering. The time interval between two consecutive slices is $30{\Delta}t=0.15$.
  (a) and (b): evolutions of $r$. In the current simulations, $r$ does not approach zero. We expect this is also the case in later evolutions.
  (c) and (d): evolutions of $\sigma$ and $\psi$.}
  \label{fig:weak_scattering_evolution}
\end{figure*}

\begin{figure*}[t!]
  \epsfig{file=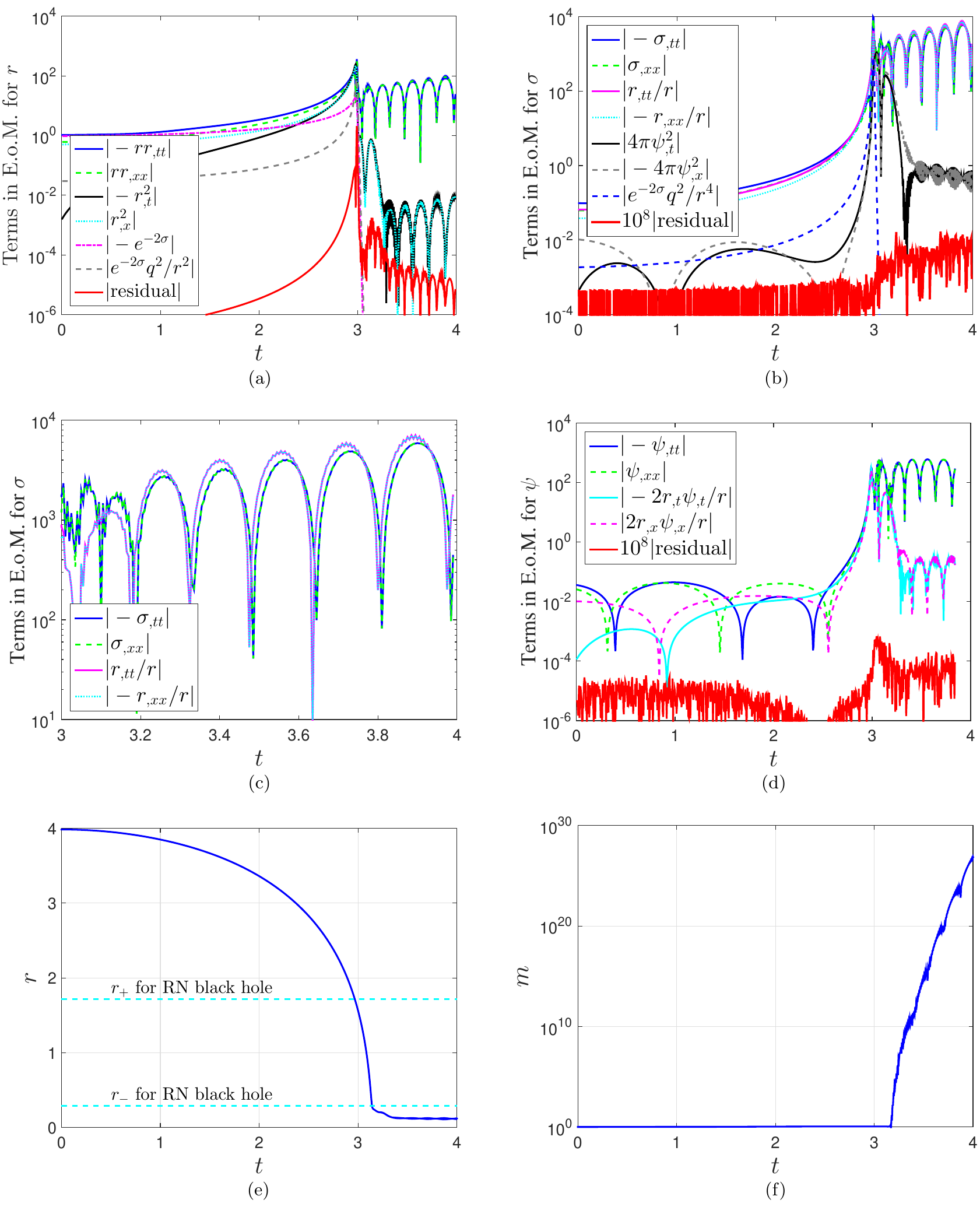, width=1\textwidth}
  \caption{(color online). Dynamics and solutions in weak scalar charge scattering on the slice $(x=3,t=t)$.
  (a)-(d): dynamical equations for $r$, $\sigma$, and $\psi$.
  (e) and (f): evolutions of $r$ and $m$.}
  \label{fig:weak_scattering_eom}
\end{figure*}

\subsection{The early/slow stage of null scattering\label{sec:slow_stage}}
As shown in Fig.~\ref{fig:dynamics_null}, at the early/slow stage of null scattering, the equations of motion
for $r$ (\ref{equation_r}), $\sigma$ (\ref{equation_sigma}), and $\psi$ (\ref{equation_psi}) can be rewritten as follows:
\be r_{,tt}{\approx}r_{,xx}, \hphantom{dd} r_{,t}^2{\approx}r_{,x}^2;\ee
\be
\sigma_{,tt}\approx\sigma_{,xx}, \hphantom{dd} \psi_{,t}^2{\approx}\psi_{,x}^2,
\hphantom{dd} r_{,tt}{\approx}r_{,xx};
\ee
\be \psi_{,tt}\approx\psi_{,xx},\hphantom{dd}r_{,t}\psi_{,t}{\approx}r_{,x}\psi_{,x}. \ee
The above equations are like free scalar wave equations in flat spacetime. The derivatives of one variable ($r$, $\sigma$, and $\psi$) are independent from the derivatives of another. Arbitrary functions of $(t+x)$ or $(t-x)$ can satisfy the above equations, and in principle, the initial conditions right after the collision between the scalar field and the
\\
\\
\\
\\
\\
\\
\\
inner horizon will decide which function each variable can take. On the other hand, we find that, as shown in Figs.~\ref{fig:evolutions_null_slow}(a) and~\ref{fig:evolutions_null_slow}(b), the constraint equations (\ref{constraint_eq_xt}) and (\ref{constraint_eq_xx_tt}) provide some useful information on the connections between $r$, $\sigma$, and $\psi$ at the early stage of charge scattering:
\be r_{,t}\sigma_{,t}\approx4{\pi}r_{-}\psi_{,t}^2.\ee

As shown in Fig.~\ref{fig:evolutions_null}, the quantities $r$, $\sigma$, $\psi$, $|1-K^2|$, and $m$ change dramatically at the beginning of charge scattering where $r{\approx}r_{-}$. Moreover, note that, near the central singularity, $r$, $\sigma$, and $\psi$ have approximate analytic expressions in terms of $\xi=t_{0}-t$, where $t_{0}$ is the time coordinate of the singularity curve. So it is natural to guess that, at the early stage of charge scattering, the above quantities may be also expressed by function of $\zeta=t-t_{s}$, where $t_{s}$ is a certain time value related to the early stage of charge scattering. We plot the evolutions of these quantities at the early stage of charge scattering in Fig.~\ref{fig:evolutions_null_slow}, from which one can see that $r$, $r_{,t}$, and $\sigma$ may have the following approximate analytic expressions:
\begin{align}
r&{\approx}r_{-},\label{r_asymptotic_slow}\\
\nonumber\\
r_{,t}&{\approx}a\zeta^{\lambda},\label{drdt_asymptotic_slow}\\
\nonumber\\
\sigma&{\approx}b\ln\zeta+\sigma_{0}.\label{sigma_asymptotic_slow}
\end{align}

In fact, a logarithmic expression for $\sigma$ is supported by its behavior near the inner horizon in the Reissner-Nordstr\"{o}m black hole case. From Eqs.~(\ref{r_RN_metric}) and (\ref{sigma_RN_metric}), one obtains that as $r$ approaches the inner horizon $r=r_{-}$, in the case of $t{\gg}x$, $\sigma$ can be approximated by a logarithmic function of $t$. As shown in Fig.~\ref{fig:dynamics_null}(c), describing the terms in the equation of motion for $\sigma$ in charge scattering, at the very early stage ($t\approx1$) of the collision between the scalar field and the inner horizon, compared to other terms, the contributions from the terms related to $\psi$ are tiny. Therefore, at this stage, the evolution of $\sigma$ should not be much dif{}ferent from the corresponding one in the Reissner-Nordstr\"{o}m geometry,

As shown in Fig.~\ref{fig:evolutions_null_slow}(e), $\psi$ can be well fitted by power law functions of $\zeta$,
\be \psi\approx c\zeta^d+\psi_{0}.\label{psi_asymptotic_slow}\ee
We also plot $(1-K^{2})$ in Fig.~\ref{fig:evolutions_null_slow}(e), and find that $\ln(1-K^{2})$ can be well fitted linearly with respect to $\zeta$,
\be \ln(1-K^{2}){\approx}f\zeta+h.\ee
Currently we do not have derivations for this linear relation. The good thing is that, in the mass function, $(1-K^2)$ is a minor factor. Therefore, as verified in Fig.~\ref{fig:evolutions_null_slow}(f), the mass function can be reduced to
\be
\begin{split}
m&=\frac{r}{2}\left[1+\frac{q^2}{r^2}+e^{2\sigma}(r_{,t}^2-r_{,x}^2)\right]\\
&{\sim}\frac{r_{-}}{2}{\cdot}e^{2\sigma}{\cdot}r_{,t}^2\\
&{\sim}\frac{r_{-}}{2}e^{2\sigma_{0}}\cdot\zeta^{2b}{\cdot}a^{2}\zeta^{2\lambda},
\end{split}
\ee
where $a$, $b$, $\lambda$, and $\sigma_{0}$ are defined in Eqs.~(\ref{drdt_asymptotic_slow}) and (\ref{sigma_asymptotic_slow}).

We list the fitting results below:
\begin{enumerate}[(i)]
  \item $r_{,t}{\approx}a(t+b)^c+d$, $a=(-2.38\pm0.02)\times10^{-3}$, $b=-0.711\pm0.001$, $c=7.164\pm0.007$, $d=(-2.395\pm0.004)\times10^{-4}$.
  \item $\sigma{\approx}a\ln(t+b)+c$, $a=34.11\pm0.02$, $b=-0.181\pm0.001$, $c=3.38\pm0.03$.
  \\
  \\
  \item $\psi{\approx}a(t+b)^c+d$, $a=(2.72\pm0.06)\times10^{-3}$, $b=-0.170\pm0.004$, $c=5.82\pm0.01$, $d=(9.3\pm0.1)\times10^{-4}$.
  \item $\ln(1-K^2){\approx}at+b$, $a=-8.21\pm0.02$, $b=11.67\pm0.03$.
  \item ${\ln}m{\approx}a{\ln}(t+b)+c$, $a=77.52\pm0.05$, $b=-0.126\pm0.001$, $c=-16.16\pm0.08$.
\end{enumerate}

It can be interesting to compare the dynamics at the early stage of null scattering with (1) that near the central singularity in spacelike scattering and (2) that at the late stage of null scattering. At the early stage of null scattering, $r$ and $\psi$ change slowly: $r{\approx}r_{-}$,
$r_{,t}\approx a\zeta^{\lambda}$ with $\lambda>1$, and $\psi\approx f\zeta^h+\psi_{0}$. Near the center, $r$ and $\psi$ change fast: $r\approx0$, $r_{,t}{\approx}a\xi^{-(1-\beta)}$ with $1-\beta\approx1/2<1$, and $\psi{\approx}C\ln\xi$. In both cases, $\sigma$ changes fast and has logarithmic expressions with respect to $\zeta$ and $\xi$, respectively.

\subsection{Critical scattering\label{sec:critical_scattering}}
In Secs.~\ref{sec:spacelike}-\ref{sec:slow_stage}, we studied spacelike and null scattering. In this subsection, we discuss critical scattering, which is on the edge between spacelike and null scattering. In fact, the dynamics of this type of mass inflation is similar to that at the late stage of null scattering.

The dynamics of $r$, $\sigma$ and the evolutions of $r$, $\sigma$, $\psi$, and $m$ in critical scattering on the slice $(x=1.53,t=t)$ are plotted in Fig.~\ref{fig:dynamics_critical}. In spacelike scattering, the terms in the field equations are grouped according to temporal and spatial derivatives. Take the field equation for $\sigma$ as an example. As shown in Fig.~\ref{fig:spacelike_singularity_AMR}(c), at small-$r$ regions, there are $\sigma_{,tt}\approx4\pi\psi_{,t}^2$ and $\sigma_{,xx}\approx4\pi\psi_{,x}^2$. At the early stage of the null scattering, the terms in the field equations are grouped according to the types of quantities. As shown in Fig.~\ref{fig:dynamics_null}(c), at the early stage of the null scattering, there are $\sigma_{,tt}\approx\sigma_{,xx}$ and $4\pi\psi_{,t}^2\approx4\pi\psi_{,x}^2$. In critical scattering, the dynamics has features from both spacelike and early stage of null scattering. As shown in Fig.~\ref{fig:dynamics_critical}(b), in this case, there are
$\sigma_{,tt}\approx\sigma_{,xx}\approx4\pi\psi_{,t}^2\approx4\pi\psi_{,x}^2$. Further details are skipped.

\section{Weak scalar charge scattering\label{sec:weak_scattering}}
In this section, we consider charge scattering with a weak scalar field. Parameter settings in this section are almost the same as those in the last section with the following exceptions:
\begin{enumerate}[(i)]
  \item Physical scalar field:

  $\psi(x,t)|_{t=0}=a\exp\left[-(x-x_{0})^2/b\right]$,

  $a=0.04$, $b=1$, and $x_{0}=4$.
  \item Grid. Spatial range: $x\in[-10~10]$. Grid spacings: ${\Delta}x={\Delta}t=0.005$.
\end{enumerate}

As discussed in the above section, in some spacetime regions where the scalar field is strong, the inner horizon can contract to zero volume, and the central singularity becomes spacelike or null. However, this does not always necessarily happen. After all, it takes energy for the inner horizon to contract. When the scalar field carries less energy, the inner horizon may only contract to a nonzero value. This is confirmed by our numerical results plotted in Fig.~\ref{fig:weak_scattering_evolution}, which are in agreement with the numerical work in Ref.~\cite{Hansen_2005} and the mathematical proof in Ref.~\cite{Dafermos_2014}. Since in this case the inner horizon is not totally destructed, one needs to reconsider whether
\\
\\
\\
the strong cosmic censorship conjecture is valid here. If not, more ef{}forts need to be spent on understanding the nature of this conjecture and the internal structure of Reissner-Nordstr\"{o}m and Kerr black holes directly.

The dynamics for the quantities $r$, $\sigma$, and $\psi$ are plotted in Figs.~\ref{fig:weak_scattering_eom}(a)-\ref{fig:weak_scattering_eom}(d). The numerical results show that at the later stage, the field equations for such quantities become null, in the sense that the temporal and spatial derivatives are almost equal, i.e., $\sigma_{,tt}\approx\sigma_{,xx}$. Moreover, the derivatives have oscillations. As shown in Figs.~\ref{fig:weak_scattering_eom}(e) and \ref{fig:weak_scattering_eom}(f), the mass function grows up continuously even as $r$ approaches a constant value. Further details are skipped.

\section{Summary\label{sec:summary}}
In this paper, we studied neutral scalar collapse, neutral scattering, and charge scattering numerically. Mass inflation was found to happen in the vicinity of the central singularity. Approximate analytic solutions were partially obtained. We summarize our work on computational and physical issues separately below.

\subsection{Computational issues}
\begin{enumerate}[(i)]
  \item \emph{Numerical vs analytic approaches.} Numerical and analytic approaches are both indispensable in gravitational physics. The results from one of the two can be enlightening for the other one; and one approach can be used to check the results from the other one. Typical examples include computations of gravitational waveforms in black hole binary systems~\cite{Boyle_2007,Hinder_2014,Ashtekar_2007} and studies of quantum evaporation of two-dimensional black holes~\cite{Ashtekar_2007}. The ef{}ficiency of combining the two approaches was displayed again in this paper.

  \item \emph{Problems with known solutions vs a new problem.} In exploring charge scattering, we tried to closely compare problems with known solutions (dynamics in Schwarzschild and Reissner-Nordstr\"{o}m geometries and neutral scalar collapse) to a problem yet to be solved (charge scattering).

  \item \emph{$dudv$ vs $(-dt^2+dx^2)$ in double-null coordinates.} In the studies of mass inflation, the $dudv$ format of the Kruskal-like coordinates, $ds^{2} = 4e^{-2\sigma}dudv+r^2d\Omega^2$, is usually used. In this paper, we used the $(-dt^2+dx^2)$ format instead, $ds^{2}=e^{-2\sigma}(-dt^2+dx^2)+r^2d\Omega^2$, with $u=(t-x)/2=\text{const}$ and $v=(t+x)/2=\text{const}$. Moreover, we set the initial conditions close to those in a Reissner-Nordstr\"{o}m geometry.

      In the $(t,x)$ line element, one coordinate is timelike, and the rest are spacelike. We are used to this setup. It is more convenient and more intuitive to use this set of coordinates. For the $(u,v)$ choice, in the field equations, many terms are mixed derivatives of $u$ and $v$, e.g., $r_{,uv}$. However, for the $(t,x)$ choice, spatial and temporal derivatives are usually separated, e.g., $(r_{,tt}-r_{,xx})$.

      In the study of charge scattering, we set the initial conditions close to those in a Reissner-Nordstr\"{o}m geometry. Consequently, removing the terms related to the scalar field, we can test our code by comparing the numerical results to the analytic ones in a Reissner-Nordstr\"{o}m geometry. Moreover, by comparing numerical results for charge scattering to the dynamics in a Reissner-Nordstr\"{o}m geometry, we can obtain intuitions as to how the scalar field af{}fects the geometry.

  \item \emph{Cauchy horizon: infinite or local regions?} As implied by Eq.~(\ref{r_RN_metric}), the exact inner horizon $r=r_{-}$ is at the regions where $uv$ and $(t^2-x^2)$ are infinite. However, $r$ still can be very close to the inner horizon even when $uv$ and $(t^2-x^2)$ take moderate values. Consequently, at regions where $uv$ and $(t^2-x^2)$ take some moderate values, the scalar field and the inner horizon still can have strong interactions, resulting in mass inflation.
\end{enumerate}

\subsection{Physical issues}
\begin{enumerate}[(i)]
  \item \emph{Interpretations of mass inflation in charge scattering: analytic computations vs numerical simulations.} According to the original papers on mass inflation~\cite{Poisson_1989,Poisson_1990}, the mechanism of mass inflation can be described below. Near the Cauchy horizon, the influx is infinitely blueshifted, while the outflux generates the crucial separation between the Cauchy and inner horizons. Then the blueshift and redshift do not cancel, and the mass parameter increase dramatically. In this paper, we explored mass inflation numerically. We examined the contributions of all the terms in the field equations and compared the dynamics in a Reissner-Nordstr\"{o}m geometry and in charge scattering. In a Reissner-Nordstr\"{o}m geometry, in the coordinates~(\ref{RN_metric_text}), $r_{,t}$ asymptotes to zero near the inner horizon, and the mass parameter remains constant. In charge scattering, a scalar field impacts the inner horizon, and makes the metric quantity $\sigma$ in the line element~(\ref{double_null_metric_dtdx}) grow faster than in the Reissner-Nordstr\"{o}m geometry, causing a smaller repulsive force for $r$. Then $r$ can cross the original inner horizon: mass inflation takes place.

  \item \emph{Contributions to mass inflation in charge scattering: $e^{2\sigma}$ vs $(r_{,t}^2-r_{,x}^2)$.} In a Reissner-Nordstr\"{o}m black hole and charge scattering, the Misner-Sharp mass function can be written as
  \be m=\frac{r}{2}\left[1+\frac{q^2}{r^2}+e^{2\sigma}(r_{,t}^2-r_{,x}^2)\right].\ee
  Near the inner horizon of a Reissner-Nordstr\"{o}m black hole, $e^{2\sigma}\gg1$. However, $r_{,t}^2{\ll}e^{-2\sigma}$. This makes $e^{2\sigma}(r_{,t}^2-r_{,x}^2)$ vanish. Then the mass function remains constant. Now we consider the early stage of null scattering that was discussed in Sec.~\ref{sec:slow_stage}. We use the example plotted in Fig.~\ref{fig:evolutions_null_slow}. At the slow stage, $e^{2\sigma}$, $r_{,t}^2$, and $(1-r_{,t}^2/r_{,x}^2)$ vary roughly from $10^{9}$ to $10^{19}$, $10^{-7}$ to $10^{-4}$, and $1$ to $10^{-2}$, respectively. So $e^{2\sigma}$ seems to play a more important role than $r_{,t}^2$ and $(1-r_{,t}^2/r_{,x}^2)$. However, comparing charge scattering and dynamics in the Reissner-Nordstr\"{o}m metric, one can see that the mass function increases dramatically not because $e^{2\sigma}$ is very large, but because $r_{,t}^2$ moves from extremely tiny values in a Reissner-Nordstr\"{o}m geometry to a small number in charge scattering. Namely, $r_{,t}^2$ is more important than $e^{2\sigma}$ in the growth of the mass function.

  \item \emph{Gravity vs repulsive (electric) force $\Leftrightarrow$ Nature of the central singularity (timelike, spacelike, or null).} In charge scattering, the quantities in the field equations can be separated into two sides: the gravitating side ($r$, $\sigma$, and $\psi$) and the repulsive side (electric field). The dynamics in charge scattering mainly describes how these quantities interact, including how the gravitating and repulsive sides compete.

  In the Reissner-Nordstr\"{o}m geometry, at small-$r$ regions, the electric field dominates gravity. As a result, the central singularity is timelike. In charge scattering, when the scalar field is strong enough, the total gravity from the black hole and the scalar field dominates the electric field. The inner horizon contracts to zero volume rapidly, and the central singularity becomes spacelike. When the scalar field is intermediate, the inner horizon contracts to a small or zero value. The equations of motion become null, in the sense that in the equations, the temporal and spatial derivatives are almost equal, e.g., $\sigma_{,tt}\approx\sigma_{,xx}$. In the case of the inner horizon contracting to zero, the central singularity becomes null.

  \item \emph{Compare neutral collapse and dif{}ferent types of charge scattering.}  In the late stages of strong and intermediate scalar charge scattering, the dynamics and hence the solutions in the vicinity of the central singularity are similar to those in neutral scalar collapse.

      Regarding the early stage of intermediate scalar charge scattering, we noticed that, in the mass function, $e^{2\sigma}$ is the dominant factor, and hence the mass function can be well fitted by a logarithmic function of the coordinate time $t$.

  \item \emph{The inner horizon in a Reissner-Nordstr\"{o}m black hole vs the central singularity in a Schwarzschild black hole.} These two share some similarities as below.

  For Reissner-Nordstr\"{o}m and Schwarzschild black holes, throughout the whole spacetime, the Misner-Sharp mass function is constant. When a scalar field impacts the inner horizon of a Reissner-Nordstr\"{o}m black hole, the scalar field can modify the geometry in the vicinity of the inner horizon significantly, especially on $r_{,t}$. The inner horizon contracts and mass inflation takes place. In neutral scalar collapse toward a Schwarzschild black hole formation, the scalar field can also modify the geometry in the vicinity of the central singularity dramatically, especially on the metric component $\sigma$~\cite{Guo_1312}. Then mass inflation also happens.

  The BKL conjecture is an important result on the dynamics in the vicinity of a spacelike singularity~\cite{Belinskii_1970,Belinskii_1973,Belinski_1404,einstein_online}. The first statement of this conjecture is that as the singularity is approached, the dynamical terms dominate the spatial terms in the field equations. In other words, the way gravity changes over time is more important than the variation of the gravitational field from one location to the next~\cite{einstein_online}. We would like to say that, to a large extent, later evolutions in a strong gravitational field largely erase away the initial information on the connections between neighboring points. As discussed in Ref.~\cite{Guo_1312} and also in this paper, in double-null coordinates, using the above arguments, one can interpret the following behaviors displayed in numerical simulations: near the central singularity of a Schwarzschild black hole and the inner horizon of a Reissner-Nordstr\"{o}m black hole, there is
  \be \bigg|\frac{\psi_{,x}}{\psi_{,t}}\bigg|\sim\bigg|\frac{r_{,x}}{r_{,t}}\bigg|<1.\label{ratio_derivatives}\ee
  In this paper, it was found that Eq.~(\ref{ratio_derivatives}) can interpret how mass inflation takes place.

  The second and third statements of the BKL conjecture are that i) the metric terms dominate the matter field terms, while the matter field may not be negligible if it is a scalar field; ii) the dynamics of the metric components and the matter fields is described by the Kasner solution. These two statements were confirmed in simulations of neutral scalar collapse in $f(R)$ gravity in Ref.~\cite{Guo_1312} and in general relativity in this paper. The second statement was also verified in charge scattering in this paper. However, the third statement on Kasner solution may not apply to charge scattering.

  Near the central singularity in a Schwarzschild black hole, the asymptotic solutions are $r{\approx}A\xi^{\frac{1}{2}}$, $\sigma{\approx}B{\ln}\xi+\sigma_0$, and $\psi{\approx}C\ln\xi$, where $\xi{\equiv}t_0-t$, and $t_0$ is the time coordinate on the singularity curve $r(x,t)=0$. Near the inner horizon of a Reissner-Nordstr\"{o}m black hole, the asymptotic solutions are $r{\approx}a\zeta^{b}+c$ with $b>1$, $\sigma{\approx}d{\ln}\zeta+f$, and $\psi{\approx}h\zeta^{j}+\psi_{0}$, where $\zeta{\equiv}t-t_{s}$, and $t_{s}$ is the coordinate time $t$ at the beginning of charge scattering.

  \item \emph{Strong cosmic censorship conjecture.} On the Cauchy horizon of a Reissner-Nordstr\"{o}m black hole, the predicability is violated. Based on first-order calculation, the inner (Cauchy) horizon appears unstable under perturbations~\cite{Simpson_1973}. Then it was conjectured that for generic asymptotically flat initial data, the maximal Cauchy development is future inextendible. In this paper, it was found that for weak scalar field perturbation, the inner (Cauchy) horizon only contracts to a nonzero value. This is also in agreement with the numerical work in Ref.~\cite{Hansen_2005} and the mathematical proof in Ref.~\cite{Dafermos_2014}. Therefore, one needs to reconsider the validity of the strong cosmic censorship.

  \item \emph{Inside vs outside black holes: local vs global.} Throughout the whole spacetime of stationary Schwarzschild and Reissner-Nordstr\"{o}m black holes, the Misner-Sharp mass function is equal to the black hole mass. For a gravitational collapsing system, at asymptotic flat regions, the mass function describes the total mass of the dynamical system. However, in this system, near the central singularity of a Schwarzschild black hole or near the inner horizon of a Reissner-Nordstr\"{o}m black hole, the dynamics is local. Then the mass function does not provide global information on the mass of the collapsing system.
\end{enumerate}

In summary, in this paper, we studied neutral scalar collapse, neutral scattering, and charge scattering. Regarding charge scattering, for convenience and intuitiveness, Kruskal-like coordinates were used, and initial conditions were set up to be close to those in a Reissner-Nordstr\"{o}m geometry. Mass inflation was also found to happen near the central singularity in neutral scalar collapse. Approximate analytic solutions for mass inflation were partially obtained. Connections between Schwarzschild black holes, Reissner-Nordstr\"{o}m black holes, neutral collapse, neutral scattering, and charge scattering were explored.

\section*{Acknowledgments}
The authors are grateful to Andrei V. Frolov, Jos\'{e} T. G\'{a}lvez Ghersi, and Ken-ichi Nakao for useful discussions. JQG would like to thank Simon Fraser University where part of this work was done.

\begin{appendix}
\section{Reissner-Nordstr\"{o}m metric in Kruskal-like coordinates\label{sec:appendix_Kruskal_RN}}
The Reissner-Nordstr\"{o}m metric in Kruskal-like coordinates was obtained in Ref.~\cite{Graves_1960} (also see Ref.~\cite{Reall_2015}). For reference concern, we list the derivations here via analog to the Schwarzschild black hole case in the form of Ref.~\cite{Reall_2015}.

We start from the Reissner-Nordstr\"{o}m metric in the conventional form,
\be ds^2=-{\triangle}dt^2+{\triangle^{-1}}dr^2+r^2d\Omega^2,\label{RN_metric}\ee
where
\begin{align}
\triangle&=1-\frac{2m}{r}+\frac{q^2}{r^2}=\frac{(r-r_{+})(r-r_{-})}{r^2},\label{triangel_define}\\
\nonumber\\
r_{\pm}&=m\pm\sqrt{m^2-q^2}.
\end{align}
Considering the radial null curves along which $ds^2=0$, we have
\be \frac{dt}{dr}=\pm\left(1-\frac{2m}{r}+\frac{q^2}{r^2}\right).\ee
The solution to the above equation can be expressed as
\be t={\pm}r^{*}+\text{const},\ee
where
\begin{align}
r^{*}&=r+\frac{1}{2k_{+}}\ln\Big|1-\frac{r}{r_{+}}\Big|+\frac{1}{2k_{-}}\ln\Big|1-\frac{r}{r_{-}}\Big|,\label{r_star}\\
\nonumber\\
k_{\pm}&=\frac{r_{\pm}-r_{\mp}}{2r_{\pm}^2}.
\end{align}
Define
\be u=t-r^{*}, \hphantom{hh} v=t+r^{*},\label{u_and_v_appendix}\ee
and then the line element (\ref{RN_metric}) becomes
\be
\begin{split}
ds^2&=\triangle(-dt^2+d{r^{*}}^2)+r^2d\Omega^2\\
&=-\frac{1}{2}\triangle(dudv+dvdu)+r^2d\Omega^2.
\end{split}
\label{RN_metric_uv}\ee

\subsection{Patch I: $r_{-}<r<+\infty$}
\noindent{(i) \emph{$r_{+}<r<+\infty$ of Patch I}}

As implied in Eq.~(\ref{r_star}), the inner and outer horizons in the metric (\ref{RN_metric_uv}) are pushed to infinity. To compact them into finite space, we define
\be U^{+}=-e^{-k_{+}u}, \hphantom{hh} V^{+}=e^{k_{+}v}. \hphantom{dd} \mbox{for $r>r_{+}$.}\label{U_V_part1}\ee
Rewrite Eq.~(\ref{r_star}) as
\be
\begin{split}
r^{*}=
&r+\frac{1}{2k_{+}}\ln\left[\Big|1-\frac{r}{r_{+}}\Big|\left(\frac{r}{r_{-}}-1\right)\right]\\
&-\frac{1}{2k_{+}}\left(1+\frac{k_{+}}{|k_{-}|}\right)\ln\Big|1-\frac{r}{r_{-}}\Big|.
\end{split}
\label{r_star_patch1}
\ee

Using Eqs.~(\ref{triangel_define}), (\ref{u_and_v_appendix}), (\ref{U_V_part1}), and (\ref{r_star_patch1}), the line element~(\ref{RN_metric_uv}) can be converted into the following format
\be ds^2=-\frac{r_{+}r_{-}}{k_{+}^2r^{2}}e^{-2k_{+}r}\left(\frac{r}{r_{-}}-1\right)^{1+\frac{k_{+}}{|k_{-}|}}dU^{+}dV^{+}+r^2d\Omega^2.
\label{RN_metric_UV}\ee
Note that we get used to coordinate systems in which one coordinate is timelike and the rest are spacelike. So defining
\begin{align}
T&\equiv\frac{V^{+}+U^{+}}{2}=e^{k_{+}r^{*}}\sinh(k_{+}t),\label{T_part1}\\
\nonumber\\
X&\equiv\frac{V^{+}-U^{+}}{2}=e^{k_{+}r^{*}}\cosh(k_{+}t),\label{X_part1}
\end{align}
one can rewrite the metric (\ref{RN_metric_UV}) as
\be ds^2=\frac{r_{+}r_{-}}{k_{+}^2r^{2}}e^{-2k_{+}r}\left(\frac{r}{r_{-}}-1\right)^{1+\frac{k_{+}}{|k_{-}|}}(-dT^2+dX^2)+r^2d\Omega^2.
\label{RN_metric_TX}\ee
Moreover, using Eqs.~(\ref{r_star}), (\ref{T_part1}), and (\ref{X_part1}), $r$ and $t$ can be expressed as
\begin{align}
T^2-X^2&=e^{2k_{+}r}\left(1-\frac{r}{r_{+}}\right)\left(\frac{r}{r_{-}}-1\right)^{-\frac{k_{+}}{|k_{-}|}},\label{r_TX_1}\\
\nonumber\\
\frac{T}{X}&=\tanh(k_{+}t).
\end{align}

\noindent{(ii) \emph{$r_{-}<r<r_{+}$ of Patch I}}

Define
\be U^{+}=-e^{-k_{+}u}, \hphantom{hh} V^{+}=-e^{k_{+}v}. \hphantom{dd} \mbox{for $r_{-}<r<r_{+}$.}\label{U_V_part2}\ee
Similar to the case of $r_{+}<r<+\infty$, one can obtain
\begin{align}
T&=\frac{V^{+}+U^{+}}{2}=e^{k_{+}r^{*}}\cosh(k_{+}t),\label{T_part2}\\
\nonumber\\
X&=\frac{V^{+}-U^{+}}{2}=e^{k_{+}r^{*}}\sinh(k_{+}t),\label{X_part2}\\
\nonumber\\
\frac{T}{X}&=\coth(k_{+}t).
\end{align}
The line element is the same as Eq.~(\ref{RN_metric_TX}), and the expression for $r$ is the same as Eq.~(\ref{r_TX_1}).

\subsection{Patch II: $0<r<r_{+}$}
\noindent{(i) \emph{$r_{-}<r<r_{+}$ of Patch II}}

Define
\be U^{-}=-e^{-k_{-}u}, \hphantom{hh} V^{-}=-e^{k_{-}v}. \hphantom{dd} \mbox{for $r_{-}<r<r_{+}$.}\label{U_V_part1_patchII}\ee
In a similar routine as in Patch I, the metric for the spacetime of $r_{-}<r<r_{+}$ can be obtained as
\be ds^2=\frac{r_{+}r_{-}}{k_{-}^2r^{2}}e^{2|k_{-}|r}\left(1-\frac{r}{r_{+}}\right)^{1+\frac{|k_{-}|}{k_{+}}}(-dT^2+dX^2)+r^2d\Omega^2,
\label{RN_metric_TX_2}\ee
with
\begin{align}
T^2-X^2&=e^{-2|k_{-}|r}\left(\frac{r}{r_{-}}-1\right)\left(1-\frac{r}{r_{+}}\right)^{-\frac{|k_{-}|}{k_{+}}},\label{r_TX_2}\\
\nonumber\\
\frac{T}{X}&=\coth(k_{-}t).
\end{align}

\noindent{(ii) \emph{$0<r<r_{-}$ of Patch II}}

Define
\be U^{-}=-e^{-k_{-}u}, \hphantom{hh} V^{-}=e^{k_{-}v}. \hphantom{dd} \mbox{for $0<r<r_{-}$.}\label{U_V_part2_patch_II}\ee
Then we have
\be \frac{T}{X}=\tanh(k_{-}t).\ee
The line element is the same as Eq.~(\ref{RN_metric_TX_2}), and the expression for $r$ is the same as Eq.~(\ref{r_TX_2}).

\section{Einstein tensor and Energy-momentum tensor of a massive scalar field\label{sec:appendix_tensors}}
In this appendix, we give specific expressions of the Einstein tensor and the energy-momentum tensor of a massive scalar field. In double-null coordinates expressed by Eqs.~(\ref{double_null_metric_dudv}) and (\ref{double_null_metric_dtdx}), some components of the Einstein tensor can be expressed as follows:
\begin{align}
G^{t}_{t}=&\frac{2e^{2\sigma}}{r^2}\bigg[r(r_{,t}\sigma_{,t}+r_{,x}\sigma_{,x})+rr_{xx}\\
&+\frac{1}{2}(-{r_{,t}}^2+{r_{,x}}^2)-\frac{1}{2}e^{-2\sigma}\bigg],\\
\nonumber\\
G^{x}_{x}=&\frac{2e^{2\sigma}}{r^2}\bigg[-r(r_{,t}\sigma_{,t}+r_{,x}\sigma_{,x})-rr_{tt}\\
          &+\frac{1}{2}(-{r_{,t}}^2+{r_{,x}}^2)-\frac{1}{2}e^{-2\sigma}\bigg],\\
\nonumber\\
G^{\theta}_{\theta}=G^{\phi}_{\phi}&=\frac{e^{2\sigma}}{r}\left[-r_{,tt}+r_{,xx}-r(-\sigma_{,tt}+\sigma_{,xx})\right],\\
\nonumber\\
G_{uu}&=-\frac{2}{r}(r_{,uu}+2\sigma_{,u}r_{,u}),\\
\nonumber\\
G_{vv}&=-\frac{2}{r}(r_{,vv}+2\sigma_{,v}r_{,v}).
\end{align}

For a massive scalar field with energy-momentum tensor
\be T_{\mu\nu}=\psi_{,\mu}\psi_{,\nu}-\left[\frac{1}{2}g^{\alpha\beta}\psi_{,\alpha}\psi_{,\beta}+V(\psi)\right]g_{\mu\nu},
\label{T_mu_nu}\ee
there are
\begin{align}
T^{t}_{t}&=-e^{2\sigma}\left[\frac{1}{2}(\psi_{,t}^2+\psi_{,x}^2)+e^{-2\sigma}V(\psi)\right],\label{T_tt}\\
\nonumber\\
T^{x}_{x}&=e^{2\sigma}\left[\frac{1}{2}(\psi_{,t}^2+\psi_{,x}^2)-e^{-2\sigma}V(\psi)\right],\label{T_xx}\\
\nonumber\\
T^{\theta}_{\theta}=T^{\phi}_{\phi}&=e^{2\sigma}\left[\frac{1}{2}(\psi_{,t}^2-\psi_{,x}^2)-e^{-2\sigma}V(\psi)\right],\\
\nonumber\\
T_{uu}&=\psi_{,u}^2,\\
\nonumber\\
T_{vv}&=\psi_{,v}^2,\\
\nonumber\\
T{\equiv}T^{\alpha}_{\alpha}&=e^{2\sigma}(\psi_{,t}^2-\psi_{,x}^2)-4V(\psi).
\end{align}
The equations obtained in this appendix can be used to derive the field equations as discussed in Sec.~\ref{sec:field_eqs}.

\section{Schwarzschild geometry in Kruskal coordinates\label{sec:appendix_singularity_Schw}}
In this appendix, we derive the analytic expressions for the spatial and temporal derivatives near the singularity curve for a Schwarzschild black hole in Kruskal coordinates, and verify the mass formula~(\ref{mass_function}). The Schwarzschild metric in Kruskal coordinates is
\be ds^2=\frac{32m^3}{r}e^{-\frac{r}{2m}}(-dt^2+dx^2)+r^2d\Omega^2,\label{Kruskal_coordinate}\ee
with
\be t^2-x^2=\left(1-\frac{r}{2m}\right)e^{\frac{r}{2m}},\label{radius_schw_BH}\ee

\be e^{2\sigma}=\frac{r}{32m^3}e^{\frac{r}{2m}}.\label{e_2sigma}\ee
The solution to Eq.~(\ref{radius_schw_BH}) is
\be \frac{r}{2m}=1+W(z),\label{r_W}\ee
where
\be z=\frac{x^2-t^2}{e},\nonumber\ee
and $W$ is the Lambert $W$ function defined by \cite{Corless}
\be Y=W(Y)e^{W(Y)}. \label{lambertW_definition} \ee
$Y$ can be a negative or a complex number.

The first- and second-order derivatives of $W$ are
\be \frac{dW}{dz}=\frac{W}{z(1+W)}, \hphantom{ddd} \mbox{for } z\neq \left\{0, -\frac{1}{e}\right\}, \label{W_z}\ee
\be \frac{d^{2}W}{dz^2}=-\frac{W^{2}(2+W)}{z^2(1+W)^{3}}, \hphantom{ddd} \mbox{for } z\neq \left\{0, -\frac{1}{e}\right\}.\label{W_zz}\ee
Consequently, with Eqs.~(\ref{r_W}), (\ref{W_z}), and (\ref{W_zz}), one obtains the first- and second-order derivatives of $r$ with respect to $x$:
\begin{align}
\frac{1}{2m}\cdot\frac{dr}{dx}&=\frac{dW}{dz}\cdot \frac{2x}{e},\label{r_x}\\
\nonumber\\
\frac{1}{2m}\cdot\frac{d^{2}r}{dx^2}&=\frac{d^{2}W}{dz^2}\left(\frac{2x}{e}\right)^{2}+\frac{dW}{dz}\cdot\frac{2}{e}.\label{r_xx}
\end{align}
Near the singularity curve, $z[=(x^2-t^2)/e]$ approaches $-1/e$, and $W$ asymptotes to $-1$. Consequently, the second-order derivative of $r$ with respect to $x$ can be approximated as follows:
\be
\frac{1}{2m}\cdot\frac{d^{2}r}{dx^2}\approx-\frac{4x^2}{(1+W)^3}\approx\frac{d^{2}W}{dz^2}\left(\frac{2x}{e}\right)^{2}.
\label{r_xx_v2}
\ee
Similarly, one obtains the first- and second-order derivatives of $r$ with respect to $t$ near the singularity curve:
\be \frac{1}{2m}\cdot\frac{dr}{dt}=-\frac{dW}{dz}\cdot \frac{2t}{e},\label{r_t}\ee
\be
\frac{1}{2m}\cdot\frac{d^{2}r}{dt^2}\approx-\frac{4t^2}{(1+W)^3}\approx\frac{d^{2}W}{dz^2}\left(\frac{2t}{e}\right)^{2}.
\label{r_tt}
\ee
Therefore, with Eqs.~(\ref{r_x}) and (\ref{r_xx_v2})-(\ref{r_tt}), the ratios between the spatial and temporal derivatives can be expressed by the slope of the singularity curve, $K$,
\be \frac{r_{,x}}{r_{,t}}=-\frac{x}{t}=-K,\ee
\be \frac{r_{,xx}}{r_{,tt}}\approx\left(\frac{x}{t}\right)^2=K^2.\ee
As discussed in Sec.~\ref{sec:neutral_collapse}, in spherical scalar collapse in double-null coordinates, in the vicinity of the singularity curve of the formed black hole, for the metric components and scalar field, the ratios between the spatial and temporal derivatives are also defined by $K$.

Using Eqs.~(\ref{radius_schw_BH}), (\ref{e_2sigma}), (\ref{r_W}), (\ref{r_x}), and (\ref{r_t}), one obtains
\be
\begin{split}
r_{,x}^2-r_{,t}^2&=64m^{4}\frac{\left(\frac{r}{2m}-1\right)^2}{(x^2-t^2)r^2}\\
&=\frac{32m^3}{r}e^{-\frac{r}{2m}}\left(1-\frac{2m}{r}\right)\\
&=e^{-2\sigma}\left(1-\frac{2m}{r}\right).
\end{split}
\ee
This implies that the definition of mass function (\ref{mass_function}) for a Schwarzschild black hole is valid throughout the spacetime of the black hole, inside and outside the horizon, and also in the vicinity of the central singularity,
\be g^{\mu\nu}r_{,\mu}r_{,\nu}=e^{2\sigma}(-r_{,t}^2+r_{,x}^2){\equiv}1-\frac{2m}{r}.\nonumber\ee
In other words, the Misner-Sharp mass function is equal to the black hole mass everywhere.
\end{appendix}



\begin{thebibliography}{99}
\bibitem{Burko_1997_book}
{\it Internal Structure of Black Holes and Spacetime Singularities,}
edited by L. M. Burko and A. Ori
(Institute of Physics Publishing, Bristol, UK, and The Israel Physical Society, Jerusalem, Israel. 1997).

\bibitem{Brady_1999}
P. R. Brady,
{\it ``The Internal Structure of Black holes,''}
Prog. Theor. Phys. Suppl. {\bf 136}, 29 (1999).

\bibitem{Berger_2002}
B. K. Berger,
{\it ``Numerical Approaches to Spacetime Singularities,''}
Living Rev. Relativity {\bf 5}, 1 (2002).
[\arXiv{gr-qc/0201056}]

\bibitem{Joshi_2007}
P. S. Joshi,
{\it Gravitational Collapse and Spacetime Singularities}
(Cambridge University Press, Cambridge, UK, 2007).

\bibitem{Henneaux_2008}
M. Henneaux, D. Persson, and P. Spindel,
{\it ``Spacelike Singularities and Hidden Symmetries of Gravity,''}
Living Rev. Relativity {\bf 11}, 1 (2008).
[\arXiv[hep-th]{0710.1818}]

\bibitem{Price}
R. H. Price,
{\it ``Nonspherical Perturbations of Relativistic Gravitational Collapse. I. Scalar and Gravitational Perturbations,''}
Phys. Rev. D {\bf 5}, 2419 (1972).

\bibitem{Simpson_1973}
M. Simpson and R. Penrose,
{\it ``Internal instability in a Reissner-–Nordstr\"{o}m black hole,''}
Int. J. Theor. Phys. {\bf 7}, 183 (1973).

\bibitem{Dafermos_2003}
M. Dafermos,
{\it ``Stability and instability of the Cauchy horizon for the spherically symmetric Einstein-Maxwell-scalar field equations,''}
Ann. Math. (N. Y.) {\bf 158}, 875 (2003).

\bibitem{Dafermos_2014}
M. Dafermos,
{\it ``Black holes without spacelike singularities,''}
Commun. Math. Phys. {\bf 332}, 729 (2014).
[\arXiv[gr-qc]{1201.1797}]

\bibitem{Ringstrom_2015}
H. Ringstr\"{o}m,
{\it ``Origins and development of the Cauchy problem in general relativity,''}
Classical Quantum Gravity {\bf 32}, 124003 (2015).

\bibitem{Isenberg_2015}
J. Isenberg,
{\it ``On Strong Cosmic Censorship,''}
\arXiv[gr-qc]{1505.06390}.

\bibitem{Belinskii_1970}
V. A. Belinskii, I. M. Kalathnikov, and E. M. Lifshitz,
{\it ``Oscillatory Approach to a Singular Point in the Relativistic Cosmology,''}
Adv. Phys. {\bf 19}, 525 (1970) [Sov. Phys. Usp. {\bf 13}, 745 (1971)].

\bibitem{Belinskii_1973}
V. A. Belinskii and I. M. Khalatnikov,
{\it ``Ef{}fect of scalar and vector fields on the nature of the cosmological singularity,''}
Zh. Eksp. Teor. Fiz. {\bf 63}, 1121 (1972) [Sov. Phys. JETP {\bf 36}, 591 (1973)].

\bibitem{Belinski_1404}
V. A. Belinskii,
{\it ``On the cosmological singularity,''}
Int. J. Mod. Phys. D {\bf 23}, 1430016 (2014).
[\arXiv[gr-qc]{1404.3864}]

\bibitem{einstein_online}
BKL conjecture.
\url{http://www.einstein-online.info/dictionary/bkl-conjecture}

\bibitem{Berger}
B. K. Berger, D. Garfinkle, J. Isenberg, V. Moncrief, and M. Weaver,
{\it ``The Singularity in Generic Gravitational Collapse Is Spacelike, Local, and Oscillatory,''}
Mod. Phys. Lett. A \textbf{13}, 1565 (1998).
[\arXiv{gr-qc/9805063}]

\bibitem{Garfinkel_1}
D. Garfinkle,
{\it ``Numerical Simulations of Generic Singularities,''}
Phys. Rev. Lett. \textbf{93}, 161101 (2004).
[\arXiv{gr-qc/0312117}]

\bibitem{Garfinkel_2}
R. Saotome, R. Akhoury, and D. Garfinkle,
{\it ``Examining Gravitational Collapse With Test Scalar Fields,''}
Classical Quantum Gravity \textbf{27}, 165019 (2010).
[\arXiv[gr-qc]{1004.3569}]

\bibitem{Ashtekar}
A. Ashtekar, A. Henderson, and D. Sloan,
{\it ``A Hamiltonian Formulation of the BKL Conjecture,''}
Phys. Rev. D \textbf{83}, 084024 (2011).
[\arXiv[gr-qc]{1102.3474}]

\bibitem{Guo_1312}
J.-Q. Guo, D. Wang, and A. V. Frolov,
{\it ``Spherical collapse in $f(R)$ gravity and the Belinskii-Khalatnikov-Lifshitz conjecture,''}
Phys. Rev. D {\bf 90}, 024017 (2014).
[\arXiv[gr-qc]{1312.4625}]

\bibitem{Poisson_1989}
E. Poisson and W. Israel,
{\it ``Inner-horizon instability and mass inflation in black holes,''}
Phys. Rev. Lett. {\bf 63}, 1663 (1989).

\bibitem{Poisson_1990}
E. Poisson and W. Israel,
{\it ``Internal structure of black holes,''}
Phys. Rev. D {\bf 41}, 1796 (1990).

\bibitem{Barrabes_1990}
C. Barrabes, W. Israel, and E. Poisson
{\it ``Collision of light-like shells and mass inflation in rotating black holes,''}
Classical Quantum Gravity {\bf 7}, L273 (1990).

\bibitem{Gnedin_1991}
N. Yu. Gnedin and M. L. Gnedina,
{\it ``Instability of the internal structure of a Reissner-Nordstr\"{o}m black hole,''}
Sov. Astron. {\bf 36}, 296 (1992) [Astron. Zh. {\bf 69}, 584 (1992)].

\bibitem{Gnedin_1993}
M. L. Gnedin and N. Yu. Gnedin,
{\it ``Destruction of the Cauchy horizon in the Reissner-Nordstrom black hole,''}
Classical Quantum Gravity {\bf 10}, 1083 (1993).

\bibitem{Brady_1995}
P. R. Brady and J. D. Smith,
{\it ``Black hole singularities: a numerical approach,''}
Phys. Rev. Lett. {\bf 75}, 1256 (1995).
[\arXiv{gr-qc/9506067}]

\bibitem{Burko_1997}
L. M. Burko,
{\it ``Structure of the Black Hole's Cauchy Horizon Singularity,''}
Phys. Rev. Lett. {\bf 79}, 4958 (1997).
[\arXiv{gr-qc/9710112}]

\bibitem{Burko_1997b}
L. M. Burko and A. Ori,
{\it ``Late-time evolution of nonlinear gravitational collapse,''}
Phys. Rev. D {\bf 56}, 7820 (1997).
[\arXiv{gr-qc/9703067}]

\bibitem{Hansen_2005}
J. Hansen, A. Khokhlov, and I. Novikov,
{\it ``Physics of the interior of a spherical, charged black hole with a scalar field,''}
Phys. Rev. D {\bf 71}, 064013 (2005).
[\arXiv{gr-qc/0501015}]

\bibitem{Hod_1997}
S. Hod and T. Piran,
{\it ``Mass Inflation in Dynamical Gravitational Collapse of a Charged Scalar Field,''}
Phys. Rev. Lett. {\bf 81}, 1554 (1998).
[\arXiv{gr-qc/9803004}]

\bibitem{Oren_2003}
Y. Oren and T. Piran,
{\it ``Collapse of charged scalar fields,''}
Phys. Rev. D {\bf 68}, 044013 (2003).
[\arXiv{gr-qc/0306078}]

\bibitem{Kommemi_2011}
J. Kommemi,
{\it ``The Global Structure of Spherically Symmetric Charged Scalar Field Spacetimes,''}
Commun. Math. Phys. {\bf 323}, 35 (2013).
[\arXiv[gr-qc]{1107.0949}]

\bibitem{Avelino_2009}
P. P. Avelino, A. J. S. Hamilton, and C. A. R. Herdeiro,
{\it ``Mass inflation in Brans-Dicke gravity,''}
Phys. Rev. D {\bf 79}, 124045 (2009).
[\arXiv[gr-qc]{0904.2669}]

\bibitem{Borkowska_2011}
A. Borkowska, M. Rogatko, and R. Moderski,
{\it ``Collapse of charged scalar field in dilaton gravity,''}
Phys. Rev. D {\bf 83}, 084007 (2011).
[\arXiv[hep-th]{1103.4808}]

\bibitem{Hwang}
D.-i. Hwang, B.-H. Lee, and D.-h. Yeom,
{\it ``Mass inflation in f(R) gravity: A conjecture on the resolution of the mass inflation singularity,''}
J. Cosmol. Astropart. Phys. {\bf 12} (2011) 006.
[\arXiv[astro-ph]{1110.0928}]

 analytic work
\bibitem{Burko_1998}
L. M. Burko and A. Ori,
{\it ``Analytic study of the null singularity inside spherical charged black holes,''}
Phys. Rev. D {\bf 57}, R7084 (1998).
[\arXiv{gr-qc/9711032}]

\bibitem{Burko_1999}
L. M. Burko,
{\it ``Strength of the null singularity inside black holes,''}
Phys. Rev. D {\bf 60}, 104033 (1999).
[\arXiv{gr-qc/9907061}]

\bibitem{Guo_1508}
J.-Q. Guo and P. S. Joshi,
{\it ``Interior dynamics of neutral and charged black holes in f(R) gravity,''}
Universe {\bf 1}, 239 (2015).
[\arXiv[gr-qc]{1508.00461}]

\bibitem{Poisson_2004}
E. Poisson,
{\it A Relativist's Toolkit: The Mathematics of Black-Hole Mechanics}
(Cambridge University Press, Cambridge, UK, 2004).

\bibitem{Frolov_2004}
A. V. Frolov,
{\it ``Is It Really Naked? On Cosmic Censorship in String Theory,''}
Phys. Rev. D {\bf 70}, 104023 (2004).
[\arXiv{hep-th/0409117}]

\bibitem{Graves_1960}
J. C. Graves and D. R. Brill
{\it ``Oscillatory Character of Reissner-Nordstr\"{o}m Metric for an Ideal Charged Wormhole,''}
Phys. Rev. {\bf 120}, 1507 (1960).

\bibitem{Reall_2015}
H. Reall,
{\it Lecture Notes on Black Holes,}
\url{http://www.damtp.cam.ac.uk/user/hsr1000/black_holes_lectures_2015.pdf}

\bibitem{Pretorius}
F. Pretorius,
{\it ``Numerical relativity using a generalized harmonic decomposition,''}
Classical Quantum Gravity {\bf 22}, 425 (2005).
[\arXiv{gr-qc/0407110}]

\bibitem{Sorkin}
E. Sorkin and T. Piran,
{\it ``Ef{}fects of pair creation on charged gravitational collapse,''}
Phys. Rev. D {\bf 63}, 084006 (2001).
[\arXiv{gr-qc/0009095}]

\bibitem{Golod}
S. Golod and T. Piran,
{\it ``Choptuik's critical phenomenon in Einstein-Gauss-Bonnet gravity,''}
Phys. Rev. D {\bf 85}, 104015 (2012).
[\arXiv[gr-qc]{1201.6384}]

\bibitem{Misner}
C. W. Misner and D. H. Sharp,
{\it ``Relativistic Equations for Adiabatic, Spherically Symmetric Gravitational Collapse,''}
Phys. Rev. {\bf 136}, B571 (1964).

\bibitem{Hayward}
S. A. Hayward,
{\it ``Gravitational energy in spherical symmetry,''}
Phys. Rev. D {\bf 53}, 1938 (1996).
[\arXiv{gr-qc/9408002}]

\bibitem{Baumgarte}
T. W. Baumgarte and S. L. Shapiro,
{\it Numerical Relativity: Solving Einstein's Equations on the Computer}
(Cambridge University Press, Cambridge, UK, 2010).

\bibitem{Csizmadia}
P. Csizmadia and I. R\'{a}cz,
{\it ``Gravitational collapse and topology change in spherically symmetric dynamical systems,''}
Classical Quantum Gravity {\bf 27}, 015001  (2010).
[\arXiv[gr-qc]{0911.2373}]

\bibitem{Garfinkel}
D. Garfinkle,
{\it ``Choptuik scaling in null coordinates,''}
Phys. Rev. D {\bf 51}, 5558 (1995).
[\arXiv{gr-qc/9412008}]

\bibitem{Boyle_2007}
M. Boyle, D. A. Brown, L. E. Kidder, A. H. Mrou\'{e}, H. P. Pfeiffer, M. A. Scheel, G. B. Cook, and S. A. Teukolsky,
{\it ``High-accuracy comparison of numerical relativity simulations with post-Newtonian expansions,''}
Phys. Rev. D {\bf 76}, 124038 (2007).
[\arXiv[gr-qc]{0710.0158}]

\bibitem{Hinder_2014}
I. Hinder \emph{et al.} (The NRAR Collaboration),
{\it ``Error-analysis and comparison to analytical models of numerical waveforms produced by the NRAR Collaboration,''}
Classical Quantum Gravity {\bf 31}, 025012 (2014).
[\arXiv[gr-qc]{1307.5307}]

\bibitem{Ashtekar_2007}
A. Ashtekar, F. Pretorius, and F. M. Ramazano\v{g}lu,
{\it ``Evaporation of two-Dimensional Black Holes,''}
Phys. Rev. D {\bf 83}, 044040 (2011).
[\arXiv[gr-qc]{1012.0077}]

\bibitem{Corless}
R. M. Corless, G. H. Gonnet, D. E. G. Hare, D. J. Jef{}frey, and D. E. Knuth,
{\it ``On the Lambert W function,''}
Adv. Comput. Math. {\bf 5}, 329 (1996).
\end{thebibliography}
\end{document}